\documentclass[onecolumn]{emulateapj}

\RequirePackage{xspace}
\usepackage{graphicx}
\usepackage{float}
\usepackage{amsmath}
\usepackage{mathtools}

\begin{document}

\newcommand\xtenexp[2]{$#1 \times 10^{#2}$}
\newcommand\tenexp[1]{$10^{#1}$}

\slugcomment{Accepted for publication by the Astrophysical Journal}

\shorttitle{Interaction of Type Ia Supernovae With The Circumstellar Environment}
\shortauthors{Paul Dragulin, Peter Hoeflich}

\title{Type Ia Supernovae and their Environment:
Theory \& Applications to SN~2014J}
\author{Paul Dragulin\altaffilmark{1}, Peter Hoeflich\altaffilmark{2}}

\altaffiltext{1}{Department of
  Physics, Florida State University, Tallahassee, FL 32306, USA;
  pd09@my.fsu.edu}

\altaffiltext{2}{Department of
  Physics, Florida State University, Tallahassee, FL 32306, USA;
  phoeflich77@gmail.com}
  
\keywords{Supernovae: Type Ia: Circumstellar: Environment: Interaction : SN2014J}

\begin{abstract}
We present theoretical semi-analytic models for the interaction of stellar 
winds with the interstellar medium (ISM) or prior mass loss implemented in our code SPICE 
\footnote{Supernovae Progenitor Interaction Calculator for parameterized 
Environments, available on request}, assuming spherical symmetry and power-law 
ambient density profiles and 
using the $\Pi$-theorem. This allows us to test a wide variety of 
configurations, their functional dependencies, and to find classes of solutions 
for given observations.  

Here, we study Type Ia Supernova (SN~Ia) surroundings of single and 
double degenerate systems, and their observational signatures. Winds may 
originate from the progenitor prior to the white dwarf (WD) stage, the WD, 
a donor star, or an accretion disk (AD).
For $M_{Ch}$ explosions, 
the AD wind dominates and produces a low-density void several light 
years across { surrounded by a dense shell}. The bubble explains the lack of 
observed interaction in late time SN light curves for, at least, several 
years. {The shell produces narrow ISM lines Doppler shifted 
by 10-100 $km/s$, and equivalent widths of 
$\approx 100~m\mbox{\normalfont{\AA}}$ and $\approx 1~m\mbox{\normalfont{\AA}}$
 in case of ambient environments with constant density and produced by 
prior mass loss, respectively. } For SN 2014J, both mergers and $M_{Ch}$ mass 
explosions have been suggested based on radio and narrow lines.
 {As a consistent and most likely solution}, we find an AD wind running into
 an environment produced by the RG wind of the progenitor during the pre-WD 
stage, {and a short delay, $0.013$ to $1.4~Myr$, between the WD formation and the explosion.} 
Our framework may be applied more generally to stellar winds and star-formation 
feedback in large scale galactic evolution simulations.
\end{abstract}

\section{Introduction}\label{Introduction}
  Type Ia supernovae (SNe Ia) allow us to study the Universe at large and have proven invaluable in 
cosmological studies, the understanding of the  origin of elements, and they 
are laboratories to study physics such as: hydrodynamics, radiation transport, non-equilibrium 
systems, nuclear and high energy physics. 
 The consensus picture is that SNe Ia result from a degenerate C/O white dwarf (WD) undergoing a 
thermonuclear runaway \citep{hf60}, and that they originate from  close binary stellar systems.
 Potential progenitor systems may either consist of two WD, the so called double degenerate system (DD), 
or a single WD along with a Main Sequence (MS) or Red Giant (RG) star, the so called single degenerate system (SD).

{ Within this general picture for progenitors, four classes of explosion scenarios are discussed which 
are distinguished by the mechanism which triggers the thermonuclear explosion: (1) Within the Chandraskhar mass
$M_{Ch}$ models, the explosion is triggered by compressional heat close to the center when the WD approaches $M_{Ch}$.
 The accretion material
may originate either from a RG or MS companion via Roche-lobe overflow, or from tidal disruption of another WD in an DD system.
(2) In a second class, the explosion is triggered  by heat released during the dynamical merging (DM) of two WDs of a DD system.
(In some cases, the combination may result in what is known as an accretion induced collapse).  For overviews, see
\citet{Branch1995,Nomoto2003,Di_Stefano2011,Di_Stefano2012,Wang2012,hoeflich2013}. (3) Recently, a third trigger mechanism has been revived known as double  detonation scenario or He-detonation
 in a  sub-$M_{Ch} $ mass WD \citep{n82,livne1990,woosley94,hk96,Kromer2010,WK2011}. In this picture, a C/O WD star accretes He-rich material at a low rate to prevent 
 burning. Explosions are triggered from ignition of the surface He layer with masses of about a few
hundredths to 0.1 $M_\odot$. The resulting strong shock wave may
trigger a detonation of underlaying C/O. Previous calculations produced a few $ 1/100^{th}~ M_\odot$ of $^{56}\rm{Ni}$ which is inconsistent with the early time spectra
when the photosphere is formed within $10^{-3...-2} M_\odot$
\citep{het97plp}.  Modern recalculations
have utilized a smaller He shell mass and obtain better agreement with
observations \citep{Kromer2010,WK2011}, though, the problem with the outer
layers still persists. Recent studies of helium detonations including curvature and expansion effects may be in better
agreement with the observations \citep{Moore2013,Ruiter2014,Zhou2014}. For this class of explosions, we may expect
accretion disk winds similar to those in $M_{Ch} $ scenarios as discussed below. 
(4) Another explosion path, also recently set forth by \citet{Soker}, is known as the 
core-degenerate (CD) scenario. 
They suggested that the merger between a WD and the hot core 
of an AGB star could take place within a common-envelope (CE) phase of binary evolution.  
The rotation of the core slows down by emitting magnetic dipole radiation until 
the angular momentum is insufficient to prevent collapse and susequent SN~Ia. 
Binary/envelope interaction is expected to produce wind during a very short phase, 
similar to a planetary nebula (PN) mass loss phase. It is not obvious that 
the properties would resemble a typical SNe~Ia. 
Depending on the delay time from merging to explosion, the merged pair 
might explode within the PN shell or in a lower density ISM \citep{SNIPS}.

 The stellar environment will shed light on the evolutionary history of the progenitor, supernovae
light curves (LCs), and spectra, with X-rays and radio emission being the probes (e.g. \citet{Chugai1994,Dwarkadas1998,Chevalier2012,Chandra2012,Fransson2014}).
 As discussed below, the density limits for the environment of a typical SNe~Ia are well below those of the solar
neighborhood and one of the goals is to probe whether SD and/or DD systems may create this environment.

 { In the case of DD progenitors, we may expect long evolutionary time scales after the formation of the WDs 
compared to the accretion phase in $M_{Ch}$ and the double detonation scenarios. 
 The time scales depend on the unknown initial separation and mass of the binary WDs and the decay of the 
orbits due to gravitational waves (possibly modified during a common envelope phase). The time scale of angular momentum loss
by gravitational waves scales with the fourth power of the separation \citep{landau71}.  
For example, the orbits of two 1 $M_\odot$ WDs at $1~R_\odot$ will decay in 
$\approx 6 \times 10^8~yr$ representing a period with no or little mass loss.
 However, we may expect wind just prior to the explosion when the WDs fill their
 Roche lobe. The size of the Roche lobe corresponds to a separation of 
$\approx 13,000~km$ \citep{eggleton83} which translates to mass loss at most 
some months prior to the merging \citep{hanweb99}, and material close to the 
system which will be quickly overrun by the SN material. In DD, therefore, we 
expect no ongoing wind with the exception of a brief period just prior to the 
dynamical merging. Thus, the environment of a DD system may be dominated by the 
ISM the system has moved into which depends on its peculiar velocity and the 
delay between WD formation and explosion. 
 \citet{Mannucci2006} argued that the observed evidence of SNe~Ia rates favors a bimodial distribution of the delay times between
star formation and explosion with about 50 \% of all explosions take place after 
$\approx 0.1~Gyr$ and  $ 3~Gyr$, respectively.  Using the star formation rates and assuming that all SNe~Ia originate 
from DD systems, \citet{PiersantiTornambe09} concluded that 50 \% of all DD systems explode within $4 \times 10^8~yr$, 
with a long tail to about $14~Gyr$.  Recent studies show that the distribution of delay times is more continuous (see \citet{Maoz2014} and references therein).} 
 It is likely that the DD system moved far away from its birthplace and that the explosion happens in the (constant density) ISM . 
 In most cases, we may expect a low density environment consistent with observations.

 The environment of SD  systems can be expected to consist of three main components:
1) Some matter bound in the progenitor system at the time of the explosion that may originate from the 
accretion disk or be shed from the donor star; 2) the wind from the WD, accretion disk or donor star; and 3) the interstellar medium (ISM).

 Within the scenario of $M_{Ch}$ WD explosions,
 hydrodynamic calculations have shown that the expanding supernova ejecta wraps around the companion star and may pull off 
 several tenths of a solar mass of material in case of a RG donor \citep{Marietta,kasen10}. Besides the donor star, another source of matter is the accretion disk
material \citep{gerardy03du04} lifted during a pulsational phase during the explosion, or debris from the merging of two WDs \citep{hk96,q07}.
  There has been some observed evidence for interaction between the explosion and the immediate environment. 
 Although H-lines like in SN~2002ic are rare, a common feature is a   
high-velocity $Ca II$ line which, first, was prominently seen in events like SN1995D, SN 2001el, SN2003du, and SN 2000cx, a
feature present in almost all SNe~Ia \citep{hatano99,fisher00,wang2003,folatelli2013a,Silverman2015}. This line may be attributed
to the material even of solar metallicity bound or in close proximity to the progenitor system \footnote{ $H_\alpha$ emission is suppressed by collisions, charge exchange and low ionization of hydrogen in absense of a strong, ongoing interaction with a Red Giant wind, see also Sect. \ref{Discussion}.}. 
\citep{Gerardy,Quimby}.}

  At intermediate distances of up to several light years, in the case of $M_{Ch}$ explosions, the environment may be
  dominated by the wind from the donor star,
 the accretion disk or, for high accretion rates, the wind from the WD, or the interstellar material (ISM).
  A number of possible interaction signatures
has been studied, including X-rays, Radio, and narrow H and He lines, but no evidence has been found, with
 an upper limit of $10^{-5} M_\odot$ for the mass loss \citep{Chugai86,Schlegel93,schlegel95,Cumming,Chomiuk12}. In late-time
light curves, interaction should result in excess luminosity but, in general, is not seen. No sign of an interaction has 
been found even in SN1991T, which has been observed up to day 1000. This implies
 particle densities less than $\approx 10^{-3}~cm^{-3} $\citep{Schmidt}.

At large distances, from few tenths to several light years, the environment is determined by the ISM.
 It is known that Type Ia SNe generally explode away from star forming regions
\citep{Wang}. This can be partly attributed to the long stellar evolutionary 
lifetimes of the low-mass stars in the progenitor systems, allowing them sufficient time to move away from 
 their place of birth. It is also known that SNe~Ia occur in elliptical and spiral galaxies, including galactic disks, 
the bulge and the halo. One may expect the explosion to occur in ISM particle densities of $\approx 10^{-3...1} 
~cm^{-3}$ \citep{Ferri01}. Light echos from SNe~Ia have been used to probe their environments, 
 and showed that many SNe~Ia have circumstellar dust shells at distances ranging from a few up to several hundred parsecs
\citep{Hamy00,rest05,aldering06,patat07,crotts08,wang08,rest08}. 

Most evidence for a link between SNe~Ia and their   
environment comes from the observations of narrow, time-dependent, blue-shifted NaID and KI absorption lines 
which, for a significant fraction of all SNe~Ia, indicates outflows \citep{Odorico89,Patat07b,Blondin09,Foley11,Sternberg2011,Phillips2013}.
  In addition, extinction laws derived from SNe~Ia seem to be different from the interstellar medium
in our galaxy, suggesting a component linked to the environment of SNe~Ia rather than the general host galaxy 
\citep{cardelli89,kevin00,elias66,nordin11,pastorello11}. Possibly, the  
hydrodynamical impact of the SN ejecta will produce additional emission    
and may modify the outer structure of the envelope and, thus, the Doppler shift of spectra features. Light emitted 
from the photosphere of the supernovae may heat up matter in the environment, which, in turn may change the ionization 
balance or the dust properties \citep{raymond13,kruegel15,slavin15,patat15}.
We note that this effect for different dust formation in the host galaxy may lead to extinction laws
different from the Milky Way as commonly observed in SNe~Ia \citep{goobar2008,kevin2013,folatelli2014,Kawara2014,burns2014}. 
The dust properties may effect the light echoes which could in turn change the extinction laws \citep{Wang2014}.

 The following picture of the environment emerges: SNe~Ia are surrounded by a cocoon with a much lower environment
than the ISM. Sometimes, narrow ISM lines indicate clumps or surrounding shells.
However, the diversity of supernova and progenitor channels leaves open a huge parameter space which cannot
be covered by numerical simulations. To cover the parameter space, we use a semi-analytic approach 
similar to those developed  by \citet{Weaver} { for stellar wind/ISM 
interaction, \citet{Chevalier&Imamura} for stellar wind/stellar wind interaction,} and \citet{Chevalier} for supernova remnants. 
In our study, we  
make use of  the $\Pi $ theorem \citep{Buckingham1914,Sedov1959} to study the classes of self-similar solutions for the environment of SNe~Ia. 

 The current state of the research leaves some important questions unresolved. How can we understand the ubiquitously  low density 
environment, their general structure, and their link to the progenitor systems? Do SNe~Ia all originate from merging WDs?
  Which of the wide variety of progenitor systems are compatible with the  
observations and the range of parameters?  What other possible signatures might be seen due to the interaction of the explosion 
within the possible progenitor systems? For SN~2014J, can we find a class of progenitor systems which is consistent with the lack of 
X-rays and radio \citep{Margutti, Perez-Torres}, which favor dynamical merging scenarios, and the narrow ISM lines, which favor $M_{Ch}$ mass explosions \citep{Graham}?

 To address the questions, we developed a parameterized model in chapter 2 using fluid mechanics.
  In chapter 3, we present the application of semi-analytic models and our code SPICE as an analysis tool 
in the framework of environment of SNe~Ia. We evaluate the imprint of different environments and wind properties.
 In chapter 4, we apply the framework to SN 2014J as an example { (see Fig. 17)} and discuss the results.
In chapter 5, final discussions and conclusions are presented.

\section{Theory and Assumptions}\label{Theory}

\begin{figure}
\begin{center}$
\begin{array}{cc}
\includegraphics[width=0.48\textwidth]{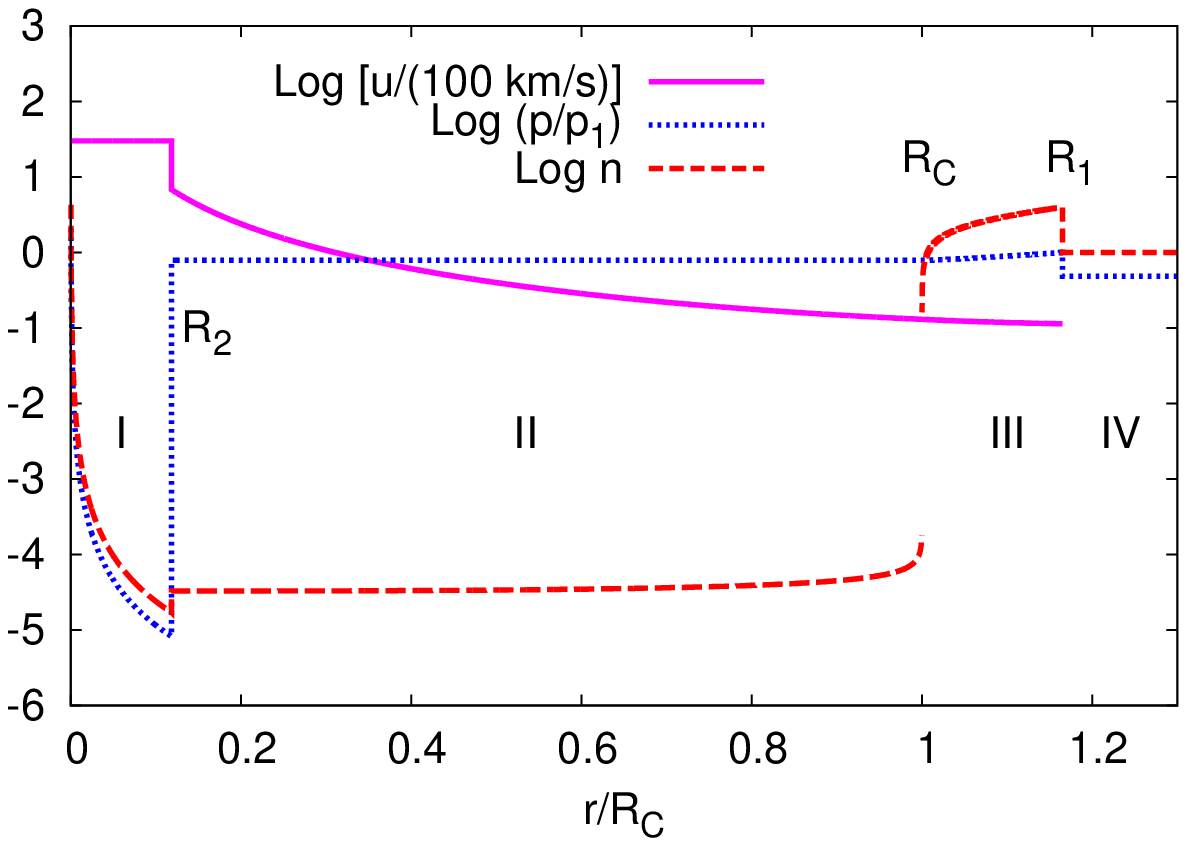}&
\includegraphics[width=0.48\textwidth]{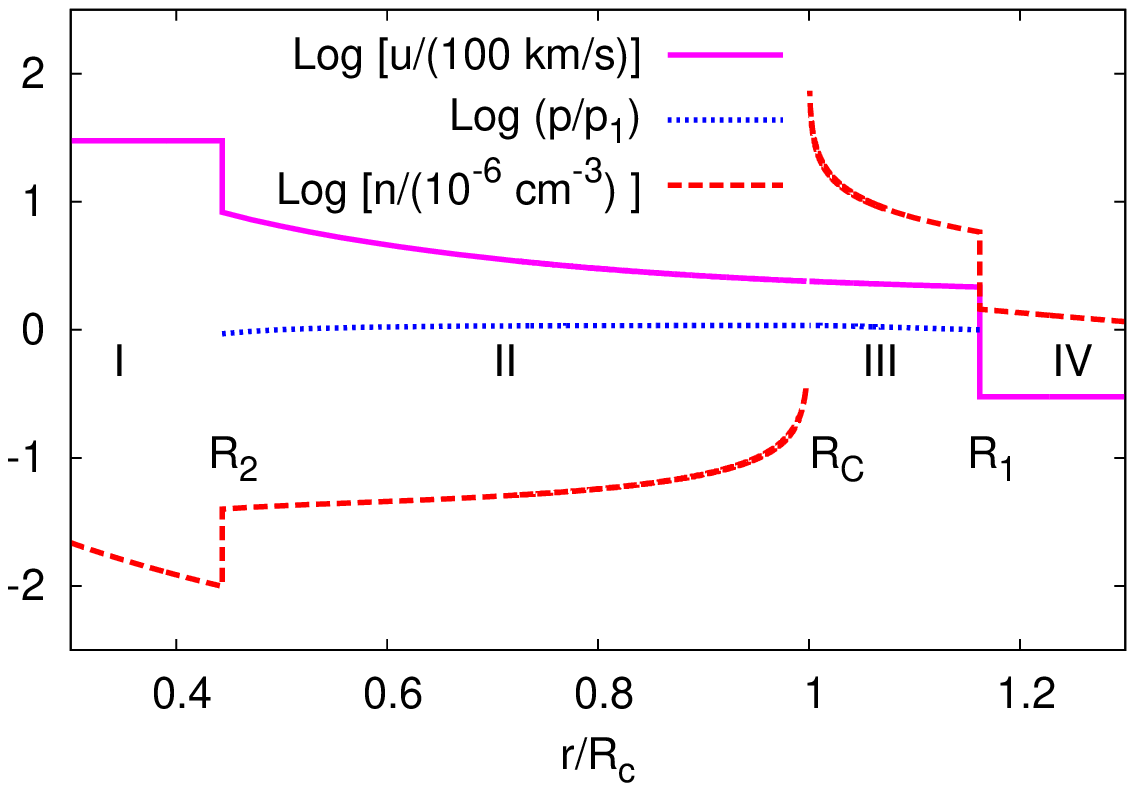}
\end{array}$
\caption{Velocity, pressure and density structure of a typical model shows four distinct regions dominated by  
the wind (I), the reversed shock (II), a shell (III), and the environment (IV) both for constant density environments (left) and    
environments produced by prior mass loss (right). In the figures of this work, density 
is shown in particle density $n = \rho N_A/\overline{\mu}$, where $N_A$ is Avogadro's number and
$\overline{\mu}$ is the mean molecular weight. In this work, $\overline{\mu}$ is set equal 
to 1. For different $\overline{\mu}$, $\rho \rightarrow \overline{\mu} \rho$ in all equations.
}
\label{example}
\end{center}
\end{figure}

We develop here a model for wind-environment interaction from basic fluid mechanics.
In the case of spherical symmetry and adiabatic flows, the hydrodynamic equations take the form:
\begin{equation}\frac {\partial \rho}{\partial t} + \frac{1}{r^2}\frac{\partial}{\partial r}(r^2 \rho u) = 0 \label{CE}\end{equation}
\begin{equation}\frac{\partial u}{\partial t} + u\frac{\partial u}{\partial r} + \frac{1}{\rho}\frac{\partial p}{\partial r} =  0 \label{NSE}\end{equation}
\begin{equation}\frac{\partial p}{\partial t} + u\frac{\partial p}{\partial r} + \frac{\gamma p}{r^2}\frac{\partial}{\partial r}(r^2 u) = 0 \label{PE}\end{equation}
where $u$ is the fluid velocity, $\rho$ is the mass density, $p$ is the pressure, and $\gamma = 5/3$ is the adiabatic index. 
 The structures have four characteristic regions (Fig. \ref{example}): I) 
undisturbed wind emanating from the source between $r=0$ and an inner shock front $R_2$, II) the inner shocked region of accumulated 
wind matter between $R_2$ and the contact discontinuity $R_{\rm{C}}$, III) an outer region of swept-up interstellar gas between $R_{\rm{C}}$ 
and the outer shock $R_1$, and IV) the outermost, undisturbed, ambient medium.

The solutions for regions I and IV are trivial. 
The solution for region III was found by \citet{Parker} who implimented a self-similar ansatz and the following transformations:

\begin{equation}u = \frac{r}{t} U(\eta) \label{UT}\end{equation}
\begin{equation}\rho = \Omega(\eta)\rho_c r^{-s} \label{RT}\end{equation}
\begin{equation}p = r^{2-s} t^{-2} \rho_c P(\eta) \label{PT}\end{equation} 
\begin{equation}\eta = tr^{-\lambda} \label{SV}\end{equation}

\noindent
where $\rho_0 (r)=\rho_c r^{-s}$ for the outer density. The similarity exponent $\lambda$ can be found by dimensional analysis.

\begin{equation} R_{\rm{C}} \propto (\dot{m}{v_w}^2/2)^{\alpha}(\rho_c)^{\beta}t^{1/\lambda}\label{s0_Rc}\end{equation}
  
\noindent
$\lambda=5/3$ for $s = 0$ and $\lambda=1$ for $s = 2$ (See Eqs. \ref{WRC} and 
\ref{Pi_K2c} for exact expressions of $R_{\rm{C}}$). Substituting these into equations \ref{CE}, \ref{NSE} 
and \ref{PE}, making the substitution  $\chi=\gamma \frac{P(\eta)}{\Omega(\eta)}$, and re-arranging gives us the following:

\begin{equation} 
\frac{d\chi}{d \log(\eta/\eta_1)} = \frac{\gamma \chi(1-\lambda U) 
[2+U(1-3\lambda -3\gamma +\lambda \gamma)+2\lambda \gamma U^2 ] + \lambda \chi^2 
(s\gamma-2\gamma+2-s- 2\lambda+ 2\lambda \gamma U)} {\gamma(1-\lambda U)[(1-\lambda U)^2-\lambda^2\chi]} 
\label{CDE}
\end{equation}

\begin{equation}
 \frac{dU}{d \log(\eta/\eta_1)} = \frac{\gamma U(1-U)(1-\lambda U)+ \chi
(2\lambda +s -2 -3\lambda \gamma U)}{\gamma(1-\lambda U)^2-\gamma \lambda^2\chi} 
\label{EDE}
\end{equation}

\begin{equation} 
\frac{d\log(P/P_1)}{d \log(\eta/\eta_1)} = \frac{ 2+U(s-2-2\lambda 
+\lambda\gamma -3\gamma)  +\lambda U^2 (2-s +2\gamma)+\lambda 
\chi (s-2)}{(1-\lambda U)^2-\lambda^2\chi} 
\label{PDE}
\end{equation}

%
%

The outer boundary conditions are obtained from the Rankine-Hugoniot jump conditions for large Mach numbers: 
\begin{equation} \chi_1=\frac{2\gamma(\gamma-1)}{\lambda^2 (\gamma+1)^2 } \label{RHC1}\end{equation} 
\begin{equation} P_1=\frac{2}{\lambda^2(\gamma+1)} \label{RHP1}\end{equation} 
\begin{equation} U_1=\frac{2}{\lambda(\gamma+1)} \label{RHU1}.\end{equation}  

Integration is carried out with respect to $U$ from the outer shock $R_1$ where $U = U_1$ to the contact discontinuity $R_{\rm{C}}$ 
where the velocity $u$ is equal to $dR_{\rm{C}}/dt=\frac{R_{\rm{C}}}{\lambda t}$ and hence $U_c = 1/\lambda$. 
	
        The solution for region III is self-similar because the only relevant parameters are the mechanical luminosity 
of the wind $ \frac 1 2 \dot{m} {v_w}^2 $ emanating from the origin and the outer density constant $\rho_0$. There is no way to 
obtain a parameter with dimension of length or time using those parameters. This is not the case in region II, where
the relevant parameters are $\rho_c$, as well as $\dot{m}$ and $v_w$, individually. Therefore, in the case of $s = 0$ a self-similar 
solution is not possible in region II, as was shown by \citet{Weaver}. They did however obtain useful analytic relations directly 
from the hydrodynamic equations by assuming that region II was isobaric. Their results for the locations of the inner shock, contact 
discontinuity, outer shock, the velocity, pressure, and density are the following:

\begin{equation} R_2(t) = 0.748\left (\dot{m}/\rho_0 \right )^{3/10}v_w^{1/10}t^{2/5} \label{WR2}\end{equation}
\begin{equation} R_{\rm{C}}(t) = 0.660\left (\dot{m}v_w^2/\rho_0 \right )^{1/5}t^{3/5}  \label{WRC}\end{equation}
\begin{equation} R_1(t) = 0.769\left (\frac{\dot{m}v_w^2} {\rho_0} \right )^{1/5}t^{3/5} \label{WR1}\end{equation}
\begin{equation} u(r,t)=\frac{11}{25}\frac{R_{\rm{C}}(t)^3}{r^2t}+\frac{4}{25}\frac{r}{t}  \label{WU}\end{equation}
\begin{equation} p(t) = 0.126\left ( \dot{m}^2 v_w^4 \rho_0^3 \right )^{1/5}t^{-4/5} \label{WP}\end{equation}
\begin{equation} \rho(r,t) = 0.628\left(\frac{\dot{m}^2 \rho_0^3}{v_w^6}\right )^{1/5}t^{-4/5}\left(1-\frac{r^3}{R_{\rm{C}}(t)^3} \right )^{-8/33} \label{WR}\end{equation}

\noindent
after correcting a small typographical error in their given expression for the velocity. This is the solution for the structure of the 
interaction region for the $s = 0$ (constant IS density) case as the Mach number goes to infinity. 
	
        One notable feature about these structures is that the density goes to zero as $r$ approaches $R_{\rm{C}}$ from above 
but diverges to infinity upon approach from below. Pressure and fluid velocity are finite and continuous 
across the boundary. Using the analytic expression for $\rho(r)$ (equation \ref{WR}) we may define a characteristic 
width of the density peak in the inner region by $\Delta r_{\rho} \equiv R_{\rm{C}}-r(2\rho_2 )=R_{\rm{C}} [1-2^{-33/8} (1-{R_2}^3/{R_{\rm{C}}}^3 )]^{1/3}$ , where 
$r(\rho)$ is the inverse of equation $\rho(r)$ as given in \ref{WR}. The width of region III is given by the integration of the equations 
\ref{CDE} and \ref{EDE}; it is $\Delta r_{III}= 0.165 R_{\rm{C}}$ . The temperature varies with the inverse of the density. In 
reality the extreme temperature discontinuity will smooth out due to finite heat conduction.

\subsection{Self-similar solutions for s=2}
	
        The case for $s=2$ allows us self-similar solutions because all characteristic scales  $R$ are proportional to $ t$
 (Eq. 8) and, thus, the time-dependence cancels out. \citet{Chevalier&Imamura} found self similar solutions for the 
interaction regions of colliding winds. Their work is similar to what we do here.
  The density in region IV is of the form 
$\rho_0 (r)=\frac{\dot{m}_1}{4\pi v_{w,1} r^2}$ assuming it is of a prior stellar wind with parameters $\dot{m}_1$ and $v_{w,1}$. 
The boundary conditions become 
	 
\begin{equation} \chi_i=\frac{2\gamma(\gamma-1)}{(\gamma+1)^2} \left (\frac{v_{w,i} t}{R_i} -1 \right )^2 \label{RHC2} \end{equation} 
\begin{equation} P_i=\frac{2}{\gamma+1} \left (\frac{v_{w,i} t}{R_i} -1 \right )^2 \label{RHP2} \end{equation}
\begin{equation} U_i=\frac{1}{\gamma+1}\left [2+(\gamma-1) \frac{v_{w,i} t}{R_i} \right ] \label{RHU2} \end{equation}
	
\noindent	 
where the subscript i is either 1 or 2, referring to the outer and inner shock front boundaries, respectively
($v_{w,2} = v_w$).
 Using the Buckingham Pi theorem \citep{Buckingham1914, Sedov1959}, we obtain 
the following expression for $R_{\rm{C}}$:

\begin{equation} R_{\rm{C}} = K_{2C}(\Pi_{\dot{m}}, \Pi_{v_w})v_{w}t   \label{Pi_K2c}\end{equation}
\begin{equation} R_1 = K_{21}(\Pi_{\dot{m}}, \Pi_{v_w})v_{w}t   \label{Pi_K21}\end{equation}
\begin{equation} R_2 = K_{22}(\Pi_{\dot{m}}, \Pi_{v_w})v_{w}t   \label{Pi_K22}\end{equation}

with 

\begin{equation} \Pi_{\dot{m}} = \dot{m}/\dot{m}_1 \label{Pi_m}\end{equation}
\begin{equation} \Pi_{v_w} = \frac{v_{w}}{v_{w,1}}-1 \label{Pi_v}.\end{equation}

$K_{2C,1,2}$ are functions to be determined numerically.
By requiring pressure continuity across $R_{\rm{C}}$, we acquire an analytic expression for the inner shock radius as a function 
of time: 
\begin{equation} R_2=v_{w} t \left[1+\left(1-\frac{tv_{w,1}}{R_1}\right)\sqrt{\left(\frac{P_2}{P_c^- }\right)\left(\frac{P_c^+}{P_1}\right)\frac{\rho_{c,1}}{\rho_{c,2}}}\right]^{-1}\label{EQ1}. \end{equation}
The outer shock radius can be given by
\begin{equation} R_1=(\eta_c^+/\eta_1 )(\eta_2/\eta_c^-)R_2, \end{equation} 
and likewise 
\begin{equation} R_{\rm{C}}=(\eta_1/\eta_c^+ )R_1. \end{equation} The quantities  $(P_2/P_c^- )$,  $(P_c^+/P_1 )$, $(\eta_c^+/\eta_1 )$ and $(\eta_c^-/\eta_2 )$ 
are found from integrating equations \ref{CDE}, \ref{EDE} and \ref{PDE} in either region III ($+$) or II ($-$).  However, in order to calculate them, initial guesses 
of $R_1$, $R_2$ and $R_{\rm{C}}$ are required. A consistent solution is obtained by damped fixed-point iteration. 
As Fig. \ref{example} shows, the structures are qualitatively different for $s=2$ than for when $s = 0$. The density goes to infinity as one 
approaches $R_{\rm{C}}$ from either side while the pressure is finite. Formally, the temperature therefore goes to zero.

\subsection{Self-similar Solutions for s=0 and Boundary Conditions}\label{Boundary}

As discussed above, the self-similar solution depends on the ambient density, $\rho_0$, and the kinetic energy flux, $1/2\,\dot{m}{v_w}^2$ at the
inner boundary. For constant density ISM (s=0), the solutions are not valid for $t \rightarrow 0$ and $t \rightarrow \infty$. 
 This is because physical assumptions break down and the results are unphysical solutions.

 For small times, this can be seen as follows:
 As shown in Table \ref{eqntable1}, $R_{\rm{C}} \propto t^{3/5}$ and $R_2 \propto t^{2/5}$. 
This implies the velocity  of the reverse shock  would go against infinity. No interaction is possible if the wind cannot overrun $R_2$, 
 therefore the description becomes unphysical. Also, the reverse shock $R_2$ is greater than the contact discontinuity $R_{\rm{C}}$ for small enough $t$. 
Therefore, no self-similar solutions exist for $\dot{R_2} \approx v_w$ and $R_2 \approx R_{\rm{C}}$, i.e. times shorter than  
 $t \sim \sqrt{\frac{\dot{m}}{\rho_0 {v_w}^3}}$. For 
our application to SN environments at times greater than when interaction takes place within the progenitor system, this hardly poses a limitation.
For our reference model in the accretion disk wind case, the MS wind, and the RG-like wind  (see tables \ref{s0_table1a},\ref{s0_table1b} \& \ref{s0_table1c})
the critical times are  $3.8$ $yr$, $0.17$ $yr$, and  $1.2 \times 10^4$ $yr$, respectively. 
{ The times where no self-similar solutions exist are short compared to the duration of the winds from the progenitor system.}

\begin{deluxetable}{lcr}
\tablecaption{For constant ISM and $t \lesssim t_p$, the relations for the radius of the contact discontinuity $R_{\rm{C}}$, 
the forward and reverse shock, $R_1$ and $R_2$, the fluid velocity at the shell $u_c$, its mass column density $\tau_m$, and the density of the inner void, $n_2$,
all as a funciton of the density of the environment $n_0$, the mass loss $\dot{m}$,  its  wind velocity $v_w$, and duration $t$.
For $n_2$, we assumed $(R_2/R_{\rm{C}})^3 \ll 1$ (see Eq. \ref{WR}). Asymptotic ($t \rightarrow \infty$) values of the reverse shock $R_{2,\infty}$, and the particle density $n_{2,\infty}$ are 
obtained from the FAP model and given below. $p_0$ is the ambient pressure.}
\tablewidth{20pc}
\tablehead{}
\startdata
$R_{\rm{C}}$         & $\propto$ & $n_0^{-\frac{1}{5}}  v_w^{\frac{2}{5}}  \dot{m}^{\frac{1}{5}}  t^{\frac{3}{5}}    $ \\ \vspace{0.1cm}
$R_1$         & $\propto$ & $n_0^{-\frac{1}{5}}  v_w^{\frac{2}{5}}  \dot{m}^{\frac{1}{5}}  t^{\frac{3}{5}}    $ \\ \vspace{0.1cm}
$R_2$         & $\propto$ & $n_0^{-\frac{3}{10}} v_w^{\frac{1}{10}} \dot{m}^{\frac{3}{10}} t^{\frac{2}{5}}    $ \\ \vspace{0.1cm}
$u_c$         & $\propto$ & $n_0^{-\frac{1}{5}}  v_w^{\frac{2}{5}}  \dot{m}^{\frac{1}{5}}  t^{-\frac{2}{5}}   $ \\ \vspace{0.1cm}
$n_2$         & $\propto$ & $n_0^{\frac{3}{5}}   v_w^{-\frac{6}{5}} \dot{m}^{\frac{2}{5}}  t^{-\frac{4}{5}}   $ \\ \vspace{0.1cm}
$\tau_{m}$    & $\propto$ & $n_0^{\frac{4}{5}}   v_w^{\frac{2}{5}}  \dot{m}^{\frac{1}{5}}  t^{\frac{3}{5}}    $ \\ \vspace{0.2cm}
$t_p$         & $\propto$ & $n_0^{\frac{3}{4}}   v_w                \dot{m}^{-\frac{1}{2}} p_0^{-\frac{5}{4}} $ \\ \hline \vspace{-0.1cm}\\

$R_{2,\infty}$ & $\approx$ & $0.30 \sqrt{\frac{\dot{m} v_w}{p_0}}$ \\ \vspace{0.1cm}
$n_{2,\infty}$ & $\approx$ & $3.9~  p_0/v_w^2$ \vspace{0.0cm}  
\enddata
\label{eqntable1}
\end{deluxetable}

For large times, the solution depends on the outer boundary condition, namely the pressure of the ambient medium.
In the following, we want to consider the validity of solutions at large times, and develop approximations which allow us to study 
environmental properties. We will compare solutions with and without ambient pressure. We will refer to those as
zero-ambient pressure (ZAP) and finite-ambient pressure models (FAP), respectively. 

 For $t \rightarrow \infty $, $R_{\rm{C}}$ goes out indefinitely according to the self-similar solution as external pressure is neglected. In reality, the outer pressure 
will increasingly confine the expansion of the structure and, thus, $R_{\rm{C}}$. In the self-similar solution without ambient pressure,       
 $p_1 $ decreases with ${R_1}^2/t^2 \propto t^{-4/5}$ (Eq. 39) and, eventually, it will drop below the ambient pressure of the physical medium. 
As reference, we define the pressure-equilibration time $t_p$ as the time at which the pressure
just inside $R_1$ equals the ambient (constant) pressure, $p_0$. It is given by 

\begin{equation} t_p = 1.23\sqrt{\frac{\dot{m}{v_w}^2}{\rho_0}} \times \left[ \lambda^2(\gamma+1)\frac{p_0}{\rho_0} \right]^{-5/4}.\label{t_p}\end{equation} 

\begin{figure}
\begin{center}
$\begin{array}{ccc}
\includegraphics[width=0.3\textwidth]{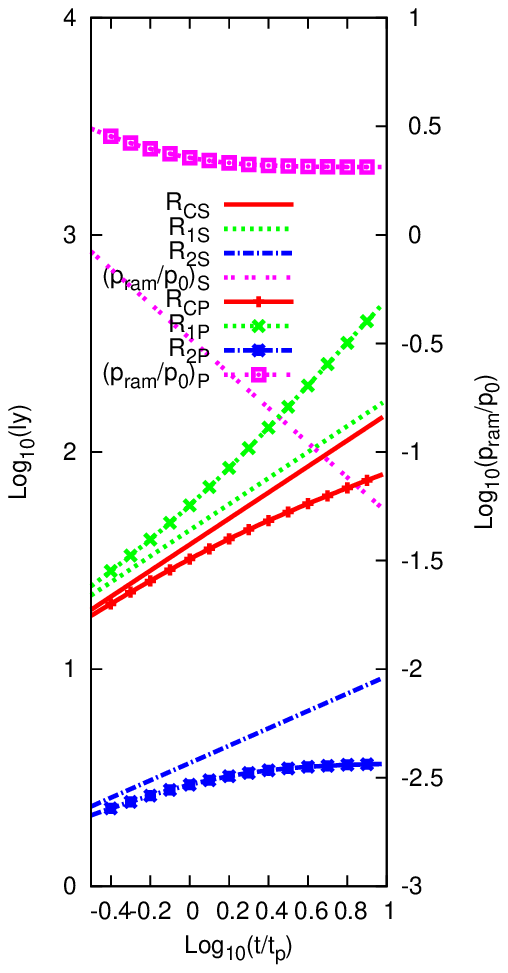} &
\includegraphics[width=0.3\textwidth]{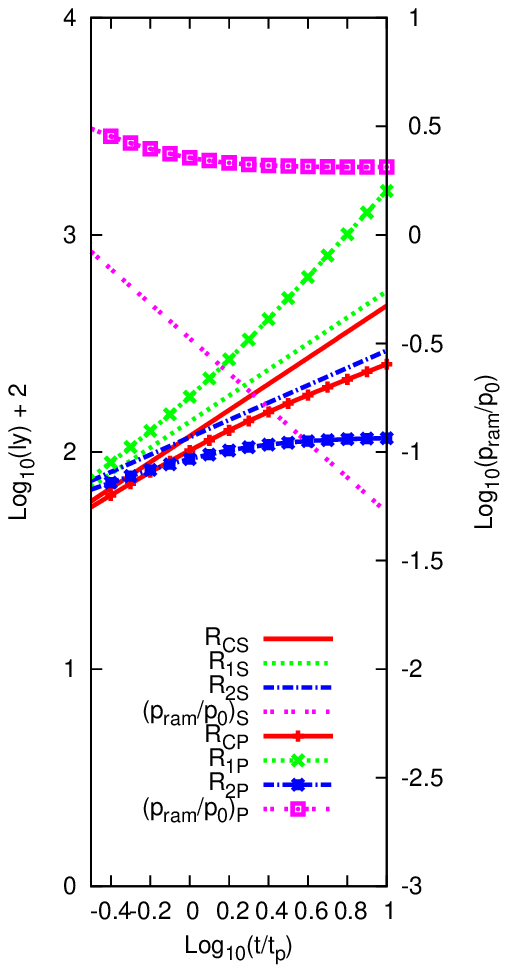} &
\includegraphics[width=0.3\textwidth]{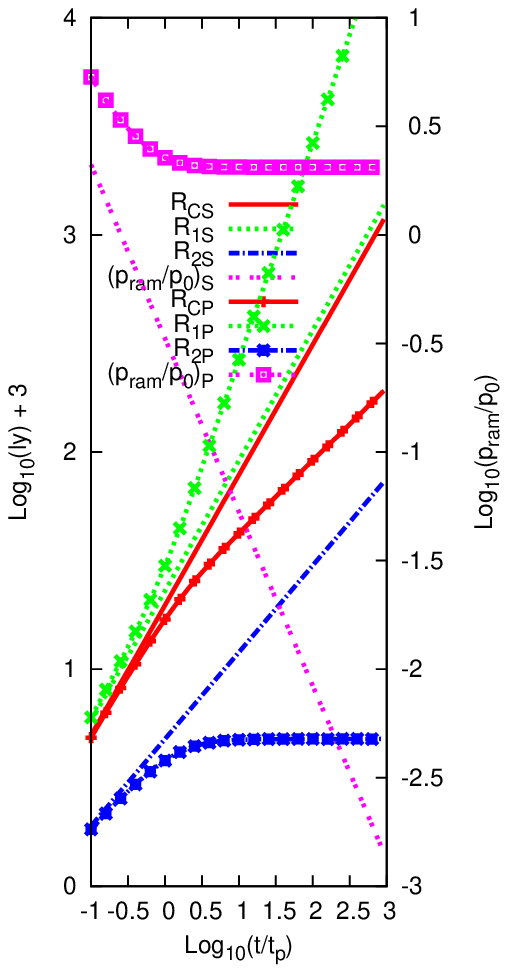} 
\end{array}$
\caption{ Structure feature comparisons of ZAP and FAP models for given sets of parameters as a function of $t/t_p$ for the reference models 
of the AD-, RG-like and MS wind (right to left).
Dependence of the radii of the outer and inner shock $R_{1,2}$, of the contact discontinuity $R_{\rm{C}}$, and the ram pressure $p_{ram}/p_0$ are shown as a 
function of the duration of the wind $t$ normalized to the time $t_p$ at which
the inner and outer pressure are equal. We show the functions for  the zero ambient  and finite pressure model indicated by the small and large symbols, respectively.
}
\label{ComparePlots}
\end{center}
\end{figure}

 In Fig. \ref{ComparePlots}, we give the evolution of the basic physical quantities as a function of time
for models with parameters typically for AD-, RG-like and MS-star winds.
 For our reference models (see Tables \ref{s0_table1a}, \ref{s0_table1b} and \ref{s0_table2}), assuming an ideal gas ambient pressure
given by temperature $T_0 = 10^4$ $K$, we obtain in the AD wind case, the RG-like wind, and the MS wind
 $t_p = 7.45 \times 10^5$ $yr$,  $t_p = 2.3 6\times 10^5$ $yr$, $t_p = 124$ $yr$, respectively.

 As an extreme case and benchmark for modifications, we use a MS wind with
parameters similar to the Sun and an evolutionary time $t=4 \times 10^7 yr$ (see reference model in Table \ref{s0_table2}). 
$t_p$ is some 124 years only, i.e. smaller than $t$ by a factor of $3 \times 10^5$. In the ZAP model, the contact discontinuity  $R_{\rm{C}}$ and
the location of the reversed shock are about 23 and 20 ly, respectively. 
The solar wind has similar properties but the termination shock is at about 75 to 95 AU based on Voyager 1 
\citep{Shiga2007}. The discrepancies can be understood due to the ambient pressure not being taken into account.
 Moreover, for times much larger than $t_p$, $\dot{R_{\rm{C}}} \ll c_s$ where $c_s$ is the ambient sound speed. 
We would therefore expect turbulent instabities which results in mixing. 
Ignoring ambient pressure for the MS (solar) model results in the contact discontinuity overrunning the  
the heliopause in about 10 years. Therefore, it is imperative that we consider how to 
account for finite ambient density in order to get realistic solutions.

 A first order estimate for the solution may be obtained by stopping the time integration 
at $t_p$ for a model without ambient pressure, and neglecting the further evolution.
 For $R_{\rm{C}}$ and $R_2$ and with this crude approximation, we obtain the right order of magnitude 
with $R_2 = 190~AU$ compared to the solar value of 75 to 95 AU.

In the following, we will construct physically motivated boundary conditions for moderate $log(t/t_p) <$ 1 ... 3, and
discuss the uncertainties estimated by a comparison between the finite ambient pressure model (FAP) and
zero ambient pressure model (ZAP). 

Besides simply truncating the solution at $t_p$, there is a way to approximately incorporate the ambient
pressure in a way that retains the self similar solution at any time, although with a modified ambient pressure
profile. In order to see how, we first notice the Rankine-Hugoniot jump conditions (in shock rest frame):

\begin{equation} p' = \frac{2 \rho_0 {u_0}^2 - (\gamma - 1)p_0}{\gamma + 1} \label{RHGC1}\end{equation} 
\begin{equation} u' = \frac{2 \gamma p_0 + (\gamma - 1)\rho_0 {u_0}^2}{(\gamma + 1) \rho_0 u_0} \label{RHGP1}\end{equation} 
\begin{equation} \rho' = \frac{(\gamma+1) {\rho_0}^2 {u_0}^2}{2 \gamma p_0 + (\gamma-1) \rho_0 {u_0}^2}\label{RHGU1}\end{equation}  

\noindent
where the 0 subscripts denote pre-shock and the primed variables are post-shock quantities. After applying the 
substitutions in \ref{UT}, \ref{RT}, and \ref{PT}, we see that the boundary conditions can remain constant with
respect to space and time if $p_0$ follows a spacial power law with $r^{-4/3}$ (or,
equivalently, a time power-law of $t^{-4/5}$. This is because $R_1 \propto t^{3/5}$.). 
An effective pressure power law environment can be defined by
requiring that, at a certain final time $t_f$, the thermal energy contained within $R_1(t_f)$ in our effective
environment is equal to the thermal energy in the physical, constant ambient density at the same radius. 
We find it by volume integration along $R_1(t)$: $p_{\rm{eff}}(t) = \frac{5 p_0}{9} \left (\frac{t_f}{t} \right )^{4/5}$.
Although $t_f$ can be thought of as a constant parameter used to define the environment, in practice $t_f = t$ and
$p_{\rm{eff}}(t) = \frac{5 p_0}{9}$. However, we note that, as the solution advances forward in time, this means that $t_f$ and the 
boundary condition will vary as well, meaning the solution will not be truly ``self-similar,'' ie the morphology changes with time. 
The solution obtained in this way is, in fact, a series of snapshots of self-similar solutions where $t_f$ is equal to the instantaneous time $t$.
Using this parameterisation, the Buckingham $\Pi$ theorem gives us the following:

\begin{equation} R_{\rm{C}} = K_{0\rm{C}} \left( \Pi_0 \right )\left( \frac{\dot{m} v_w^2 t^3}{\rho_0}\right)^{1/5}\label{Pi_R0C}\end{equation}
\begin{equation} R_1 = K_{01} \left( \Pi_0 \right )\left( \frac{\dot{m} v_w^2 t^3}{\rho_0}\right)^{1/5}\label{Pi_R01}\end{equation}
\begin{equation} \Pi_0 = \left( \frac{\dot{m} v_w^2}{\rho_0(R_s T_0)^{5/2}t^2}\right)^{1/5}\label{Pi_0}\end{equation}

\noindent
where the $K$s are to be determined numerically and the ideal gas relation was used: 
$R_s$ is the gas constant divided by the mean molecular weight and $T_0$ is the 
physical ambient, constant temperature. $T_0$ will be taken to be $10^4$ $K$ typical for ISM gas \citep{Osterbrock}.
Note that, using Eqs \ref{Pi_0} and \ref{t_p} { (for $\gamma = 5/3$), we have $\Pi_0 = 2.50(t/t_p)^{-2/5}$.}

The outer boundary conditions of the ODEs are then given as:

\begin{equation} \chi_1= \frac{(2\gamma-(\gamma - 1)/M^2)(2/M^2 + \gamma - 1)}{\lambda^2 (\gamma + 1)^2 } \label{RHCM1}\end{equation} 
\begin{equation} P_1= \frac{(2 \gamma - (\gamma-1)/M^2)}{\lambda^2 \gamma (\gamma+1)} \label{RHPM1}\end{equation} 
\begin{equation} U_1= 2 \frac{ 1 - 1/M^2 }{ \lambda ( \gamma + 1 )} \label{RHUM1}\end{equation}  

\noindent
where the Mach number $M$ is given by:

\begin{equation} M = {\frac{R_1(t)}{\lambda t}  \sqrt{\frac{9 \rho_c}{5\gamma p_0}}} \label{mach_no}. \end{equation}

An initial guess of $R_1$ is required in order to numerically solve the ODEs and obtain the structure,
therefore iteration is necessary in order to obtain a consistent solution. Following the method of
\citet{Weaver} using with our modified boundary conditions, we obtain the following expressions for the radii and inner
structure profile:

\begin{equation} R_2(t) = 1.13(\beta \alpha)^{3/2} \left( \frac{\dot{m}}{\rho_0} \right)^{3/10} {v_w}^{1/10} t^{2/5} \label{R2M}\end{equation}
\begin{equation} R_1(t) = \alpha/2^{1/5} \left( \frac{\dot{m}{v_w}^2}{\rho_0}\right)^{1/5} t^{3/5} \label{R1M}\end{equation}
\begin{equation} R_{\rm{C}}(t) = \beta R_1 \label{RCM}\end{equation}
\begin{equation} u(r,t) = \frac{11}{25}\frac{{R_{\rm{C}}}^3}{r^2 t} + \frac{r}{25t}\label{UM}\end{equation}
\begin{equation} p(r,t) = \frac{5}{22 \pi (\beta \alpha)^3}(1/4 \dot{m}^2 {v_w}^4 {\rho_0}^3)^{1/5}t^{-4/5} \label{PM}\end{equation}
\begin{equation} \rho(r,t) = \frac{0.274}{(\alpha \beta)^3} \left( \frac{\dot{m}^2 {\rho_0}^3}{{v_w}^6}\right)^{1/5} t^{-4/5} 
  \left( 1 - \frac{r^3}{{R_{\rm{C}}}^3}\right)^{-8/33}\label{RH0M}\end{equation}

\noindent
where $\alpha = \frac{1}{\beta} \left( \frac{5}{22 \pi P_c^+} \right)^{1/5}$, $\beta = \left( \frac{\eta_c}{\eta_1} 
\right)^{-1/\lambda}$, and $P_c^+$ and $\eta_c$ are evaluated at the contact discontinuity. Comparison of eqs. \ref{R1M}, \ref{RCM} 
with eqs. \ref{Pi_R01}, \ref{Pi_R0C} give $K_{01} =$ $\alpha \over 2^{1/5}$ and $K_{0\rm{C}} =$ $\beta \alpha \over 2^{1/5}$. We can then
define a proportionality for the reverse shock: 
\begin{equation} R_2 = K_{02}(\Pi_0) \left(\frac{\dot m}{\rho_0}\right)^{3/10} v_w^{1/10} t^{2/5}\label{Pi_R02}\end{equation}
where
\begin{equation} K_{02} = 1.39 K_{0\rm{C}}^{3/2}. \label{K_02}\end{equation}

\begin{figure}
\begin{center}
\includegraphics[width=0.8\textwidth]{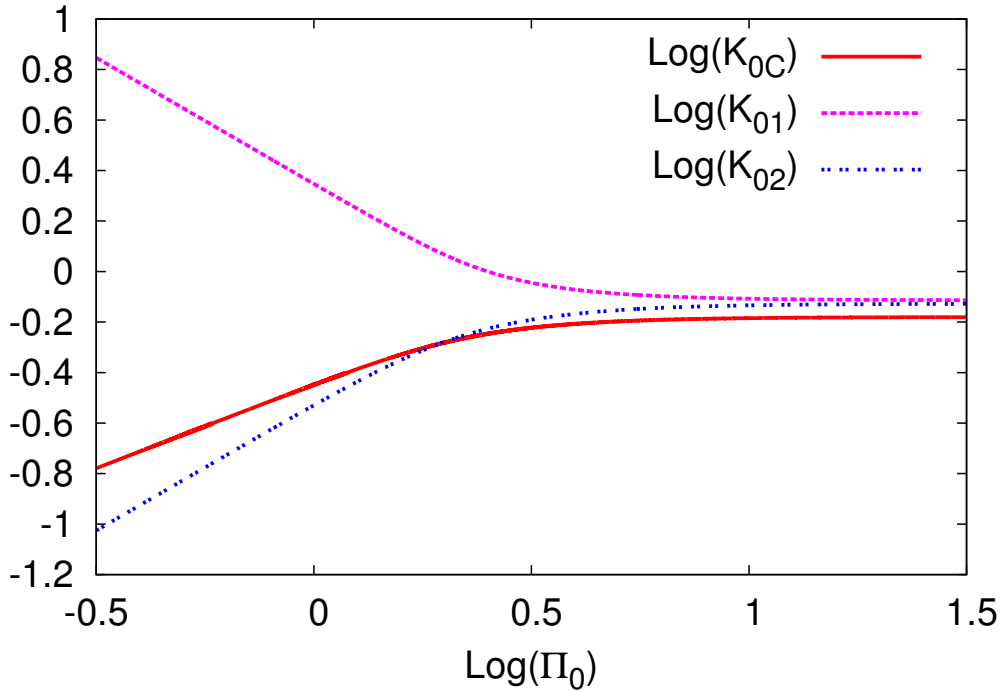}
\caption{Proportionality factors for the characteritic distances as a function of $\Pi_0$
 for constant density environments and the FAP model. Note that all factors are constant  for ZAP models with values corresponding to large $\Pi_0$ (see Sect.2).
}
\label{K0_Plots}
\end{center}
\end{figure}

\begin{figure}
\begin{center}
$\begin{array}{cc}
\includegraphics[width=0.45\textwidth]{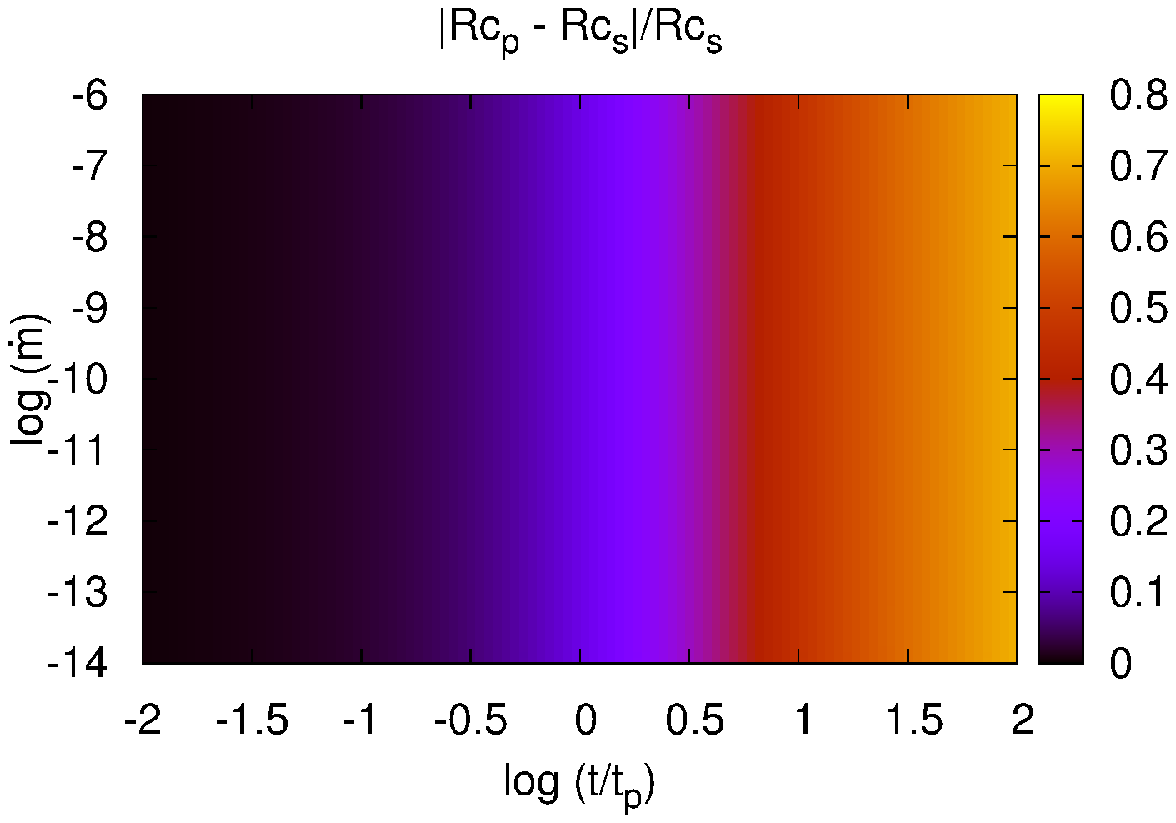} &
\includegraphics[width=0.45\textwidth]{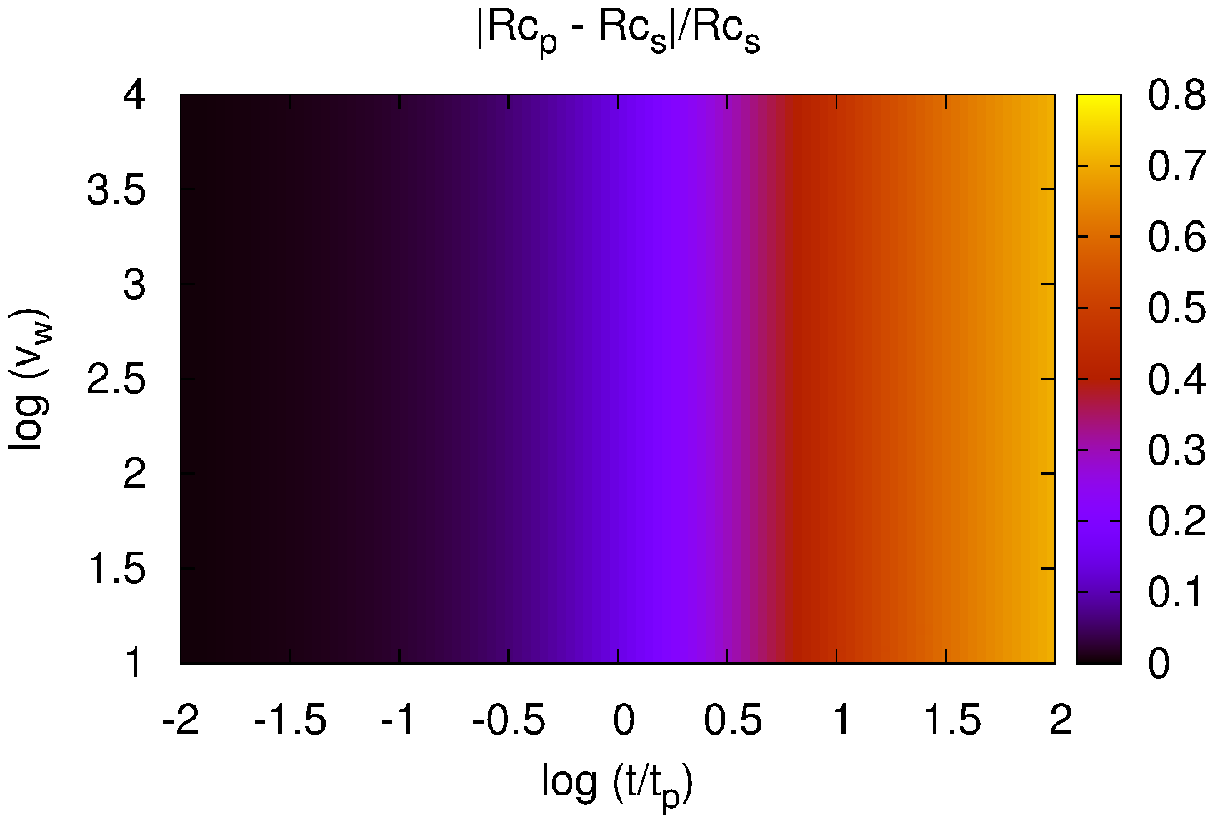} \\
\includegraphics[width=0.45\textwidth]{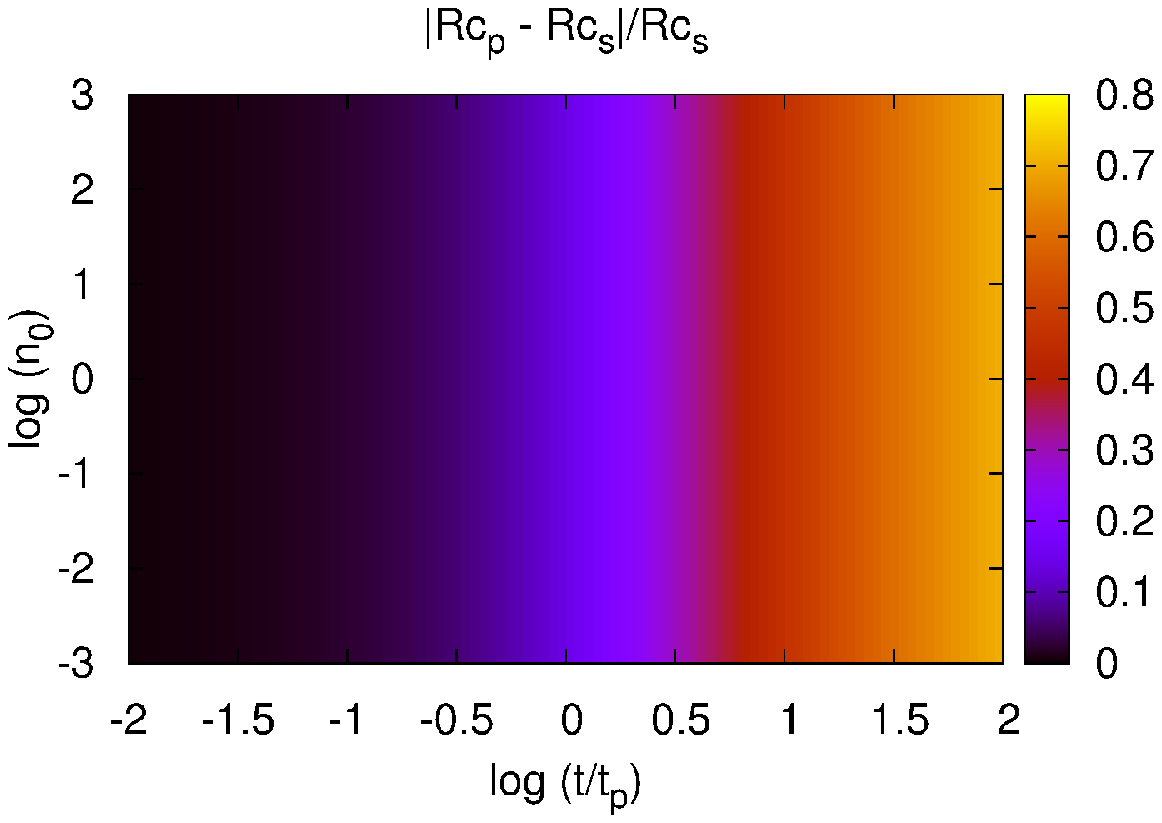} &
\includegraphics[width=0.45\textwidth]{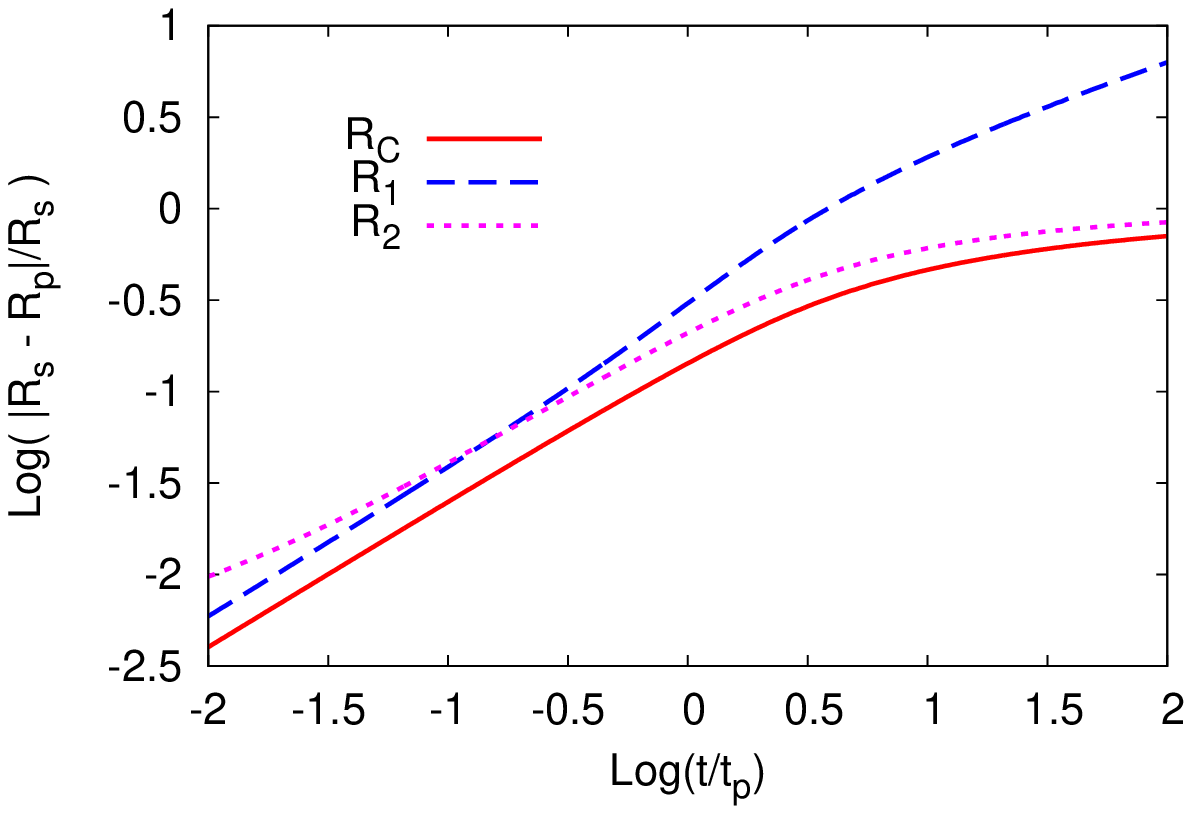} 
\end{array}$
\caption{Fractional difference in $R_{\rm{C}}$  between models of zero ambient pressure ($R_{C,s}$) and finite pressure models ($R_{C,p}$) 
in the parameter space of the mass loss $\dot{m}$, the wind velocity $v_w$, and the environment density $n_0$ as a function of $log(t/t_p)$. 
The plots visualize the $\Pi$-theorem as discussed in Sect. 3:  The difference depends on $(t/t_p)$ only. The ratios between scale-free variables 
are constant throughout the parameter space. The $\Pi$-theorem applies also to $R_1$. In the lower right, we show the fractional difference of 
$R_{\rm{C}}$ (red), $R_1$ (blue) and $R_2$ (magenta) between the ZAP and FAP models as a function of $log(t/t_p)$.
}
\label{color_plots}
\end{center}
\end{figure}

 For $p_0 \rightarrow 0$, $M \rightarrow \infty$, $\alpha \rightarrow 0.88$, $\beta \rightarrow 0.86$, and eqns \ref{WR2}-\ref{WR} 
from the last section are reproduced.
 For the reference models for an AD-, RG-like and MS-wind,  a comparison of the basic properties between the ZAP and FAP models 
as a function of $t/t_p$ is shown in Fig. \ref{ComparePlots}. 
Qualitatively, the main differences are as follows: For FAP models,  $R_2$ and $R_{\rm{C}}$ are smaller and  $R_1$ is larger than for ZAPs,
and $R_2$ goes to a constant value for large $t/t_p$. For the range shown and for the RG-like wind, $R_2$ becomes larger 
than $R_{\rm{C}}$ at about $log(t/t_p) \approx -0.5$ marking the regime of ``unphysical'' solutions already discussed above for the
ZAP model. The functional relations appear to be similar and, in fact, they are identical as a consequence of the $\Pi $ theorem.
 The relative shifts between the various quantities are given by proportionality factors which, in turn, depend only on the basic
parameters, namely $v_w$, $\dot{m}$ and $n_0$. For the FAP model, the proportionality constants have to be determined 
numerically (Fig. \ref{K0_Plots}). A further consequence of the $\Pi$ theorem is that the differences are only a  
function of $t/t_p$ and do not depend individually on $\dot{m}$, $v_w$ and $n_0$ (Fig. \ref{color_plots}). For a constant density medium,
the characteristic parameters can be directly obtained using Fig. \ref{K0_Plots}.

\begin{figure}
\begin{center}$
\begin{array}{ccc}
\includegraphics[width=0.30\textwidth]{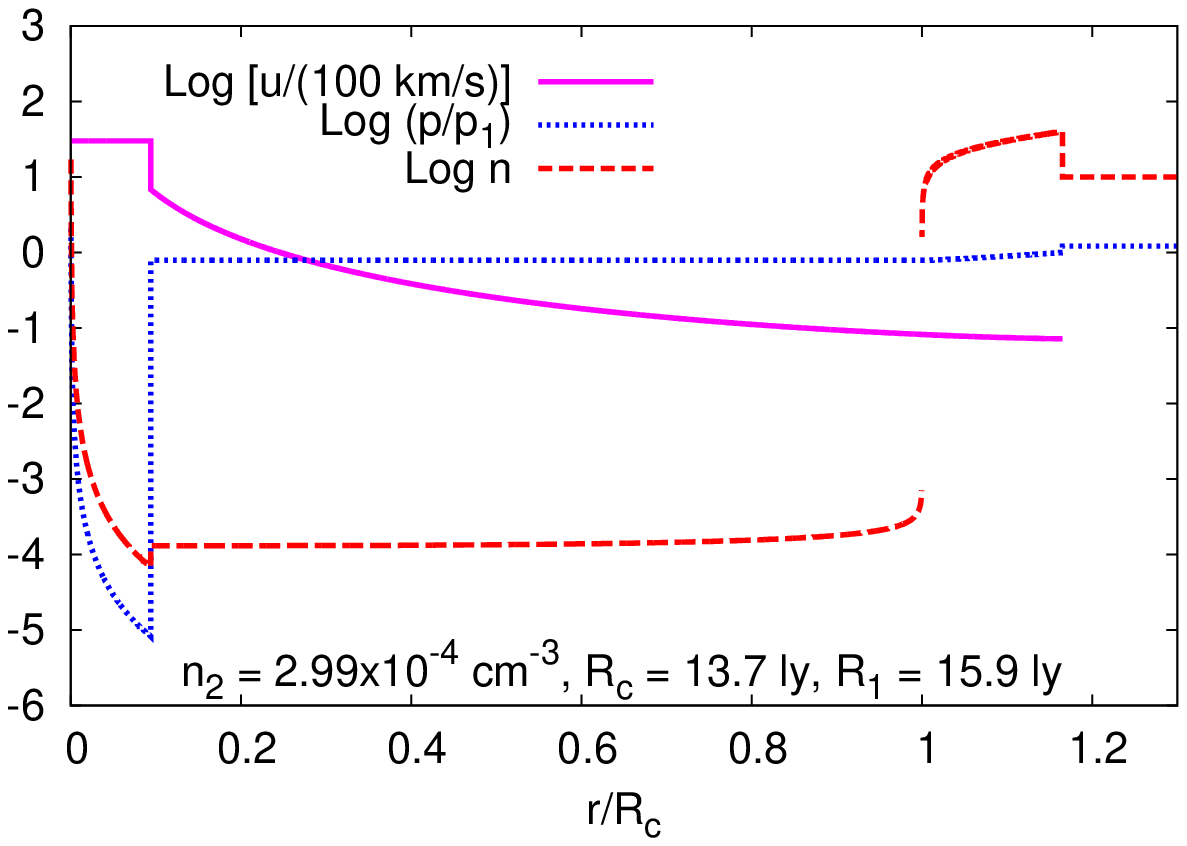} &
\includegraphics[width=0.30\textwidth]{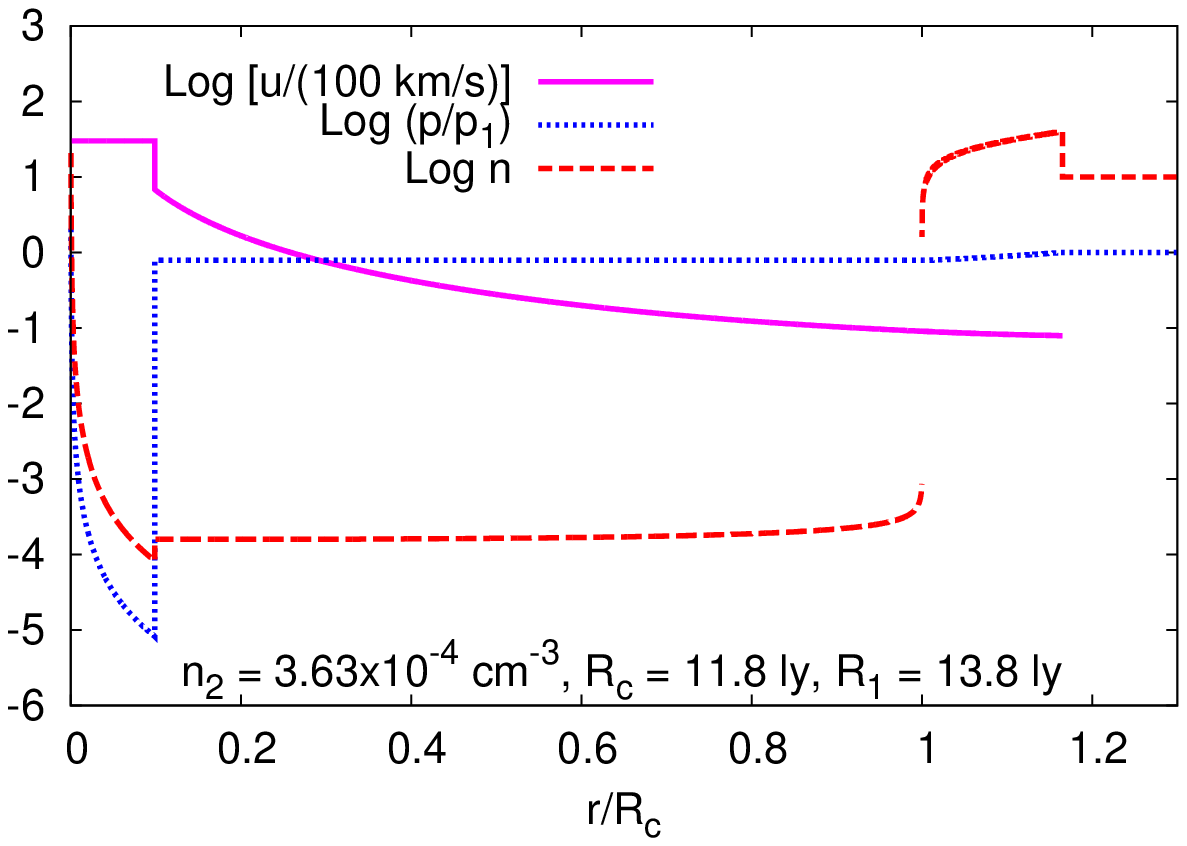} &
\includegraphics[width=0.30\textwidth]{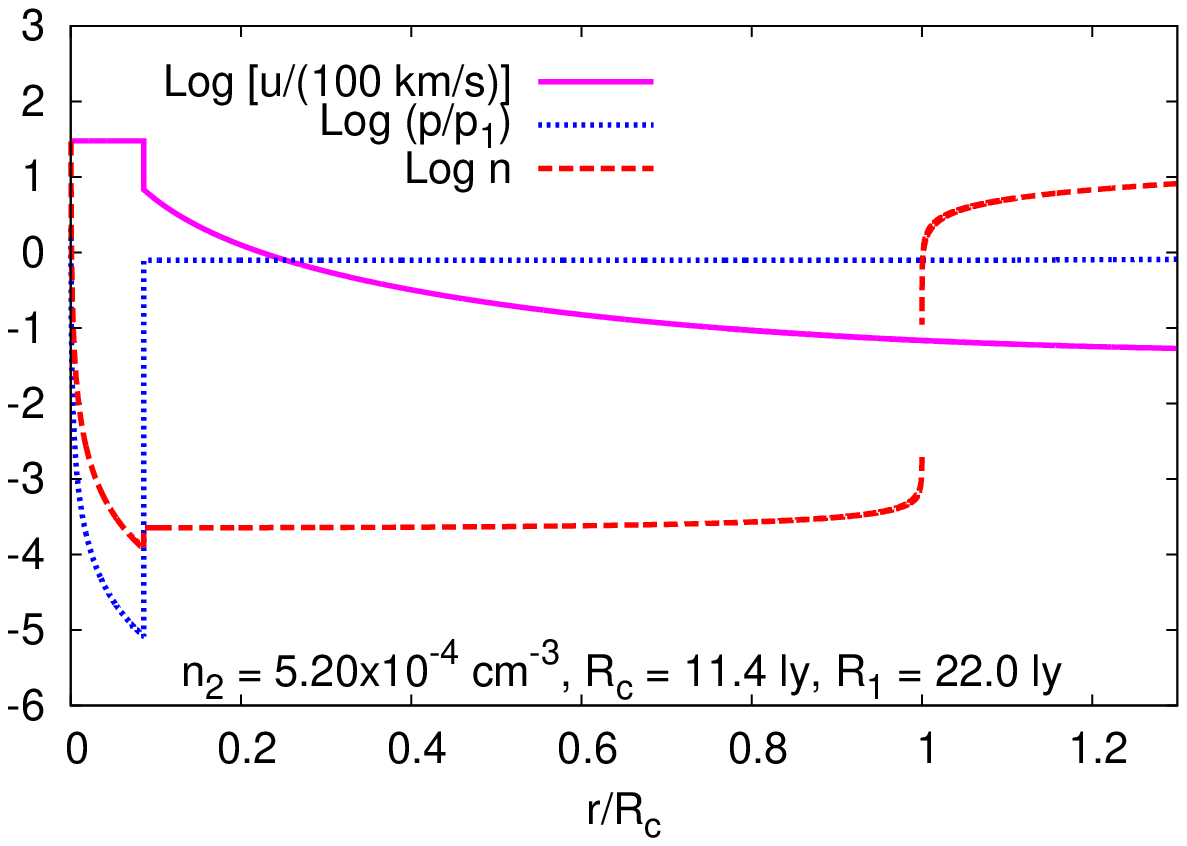}
\end{array}$
\caption{Structure of model 3 for AD wind (see Table \ref{s0_table1a}).
We show the ZAP (left) with $t=3 \times 10^5~yr$ and at $t=t_p$ (middle),
and the FAP model at $t=3 \times 10^5~yr$. The overall structure is similar
within the parameter range.}
\label{BoundaryFigure1}
\end{center}
\end{figure}

\begin{figure}
\begin{center}$
\begin{array}{ccc}
\includegraphics[width=0.30\textwidth]{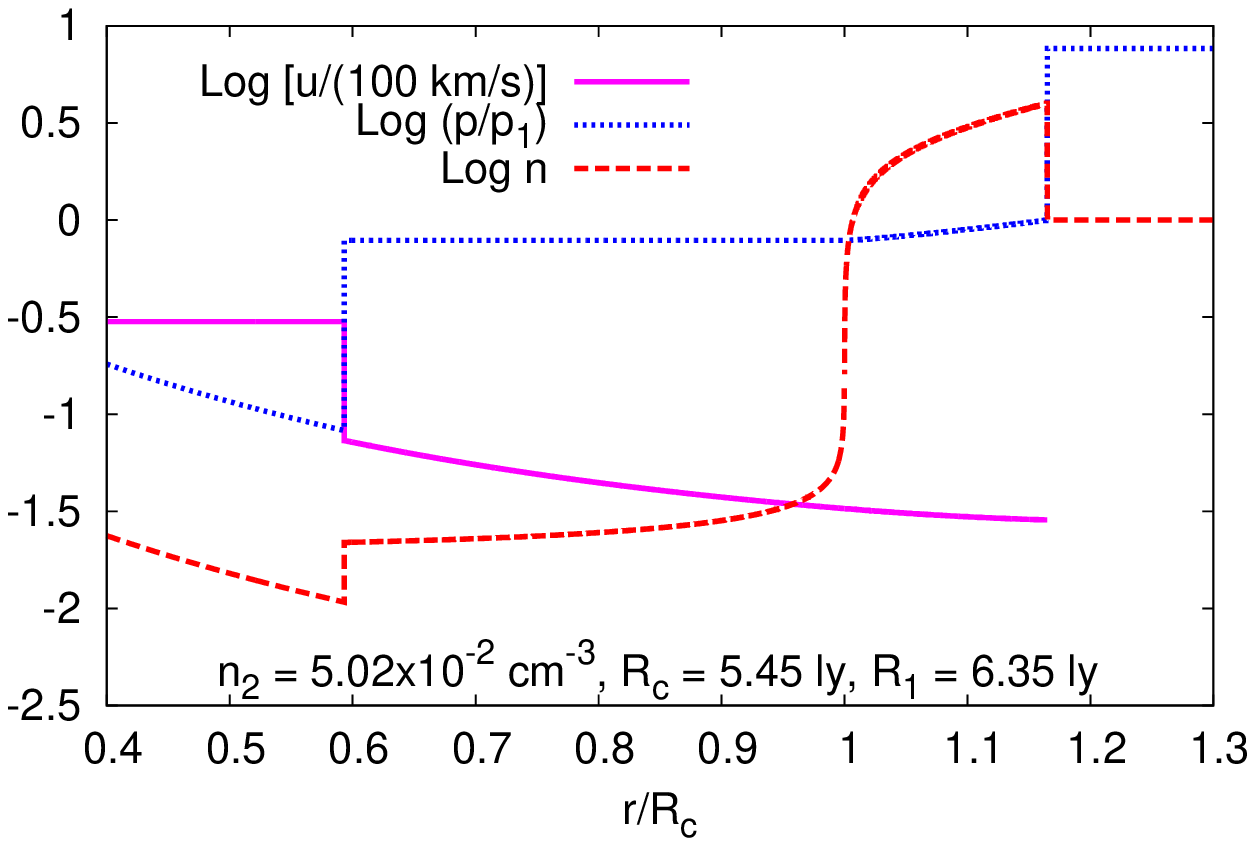} &
\includegraphics[width=0.30\textwidth]{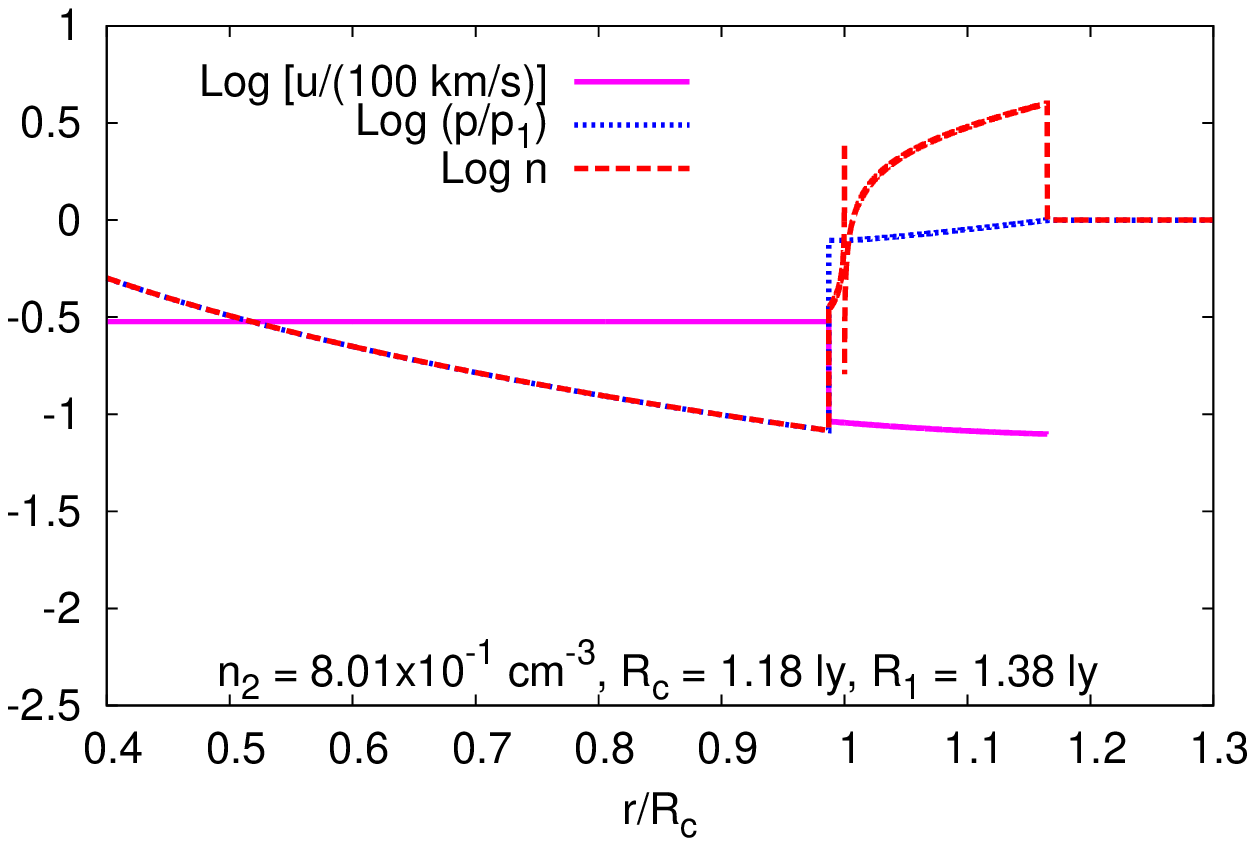} &
\includegraphics[width=0.30\textwidth]{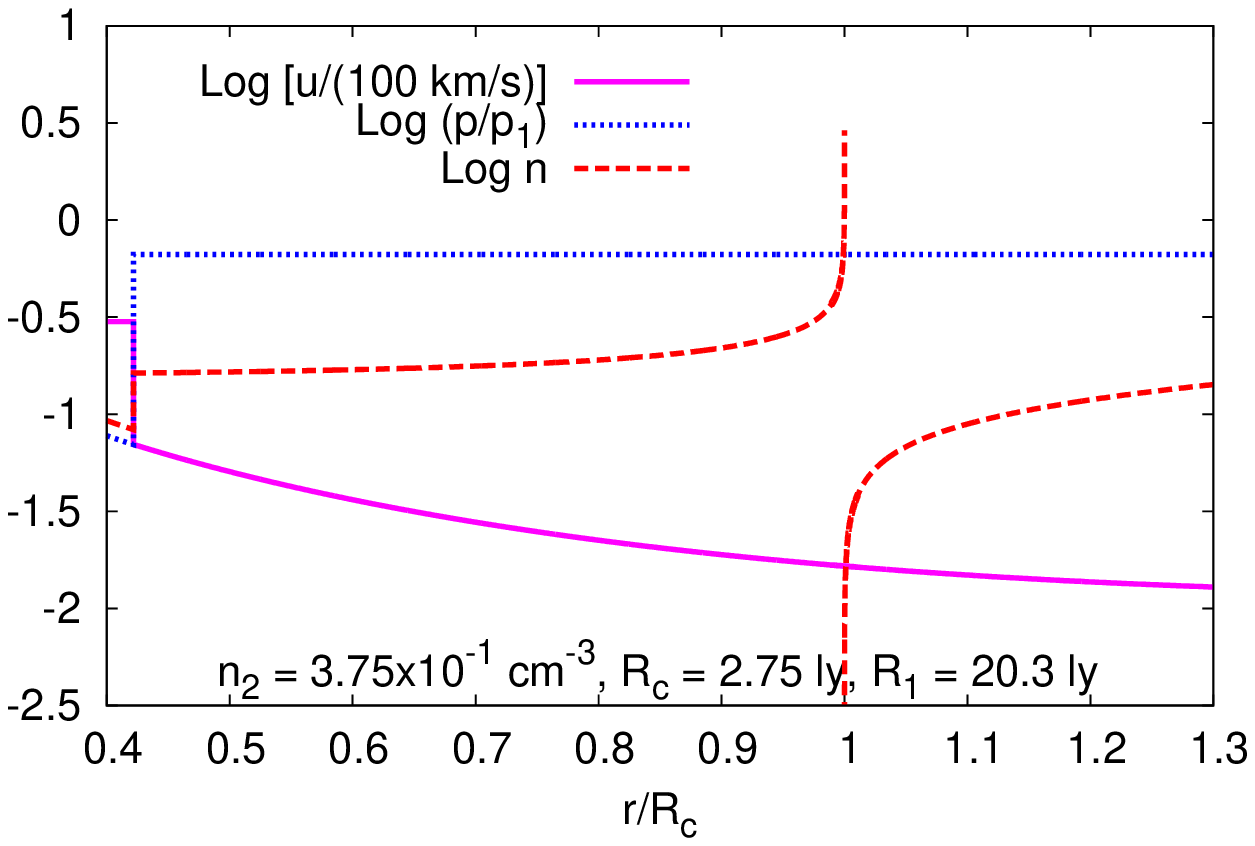} 
\end{array}$
\caption{Same as Fig. \ref{BoundaryFigure1} but for RG-like winds (see Table \ref{s0_table1b}). }  
\label{BoundaryFigure2}
\end{center}
\end{figure}

The detailed solutions for our reference models are shown in  Figs. \ref{BoundaryFigure1} \& \ref{BoundaryFigure2}.
The morphology of the envelopes does not change for a wide range of parameters
and time.  As discussed above in case of the MS star wind, however, we must expect strong mixing for $t/t_p \gg 1$ in FAP models. 
The solution becomes ``un-physical'' for $r \gtrsim R_{\rm{C}}$ in the regime of a weak shock.
 
\subsection{Existence of Solutions}\label{Solutions}
 Here, we want to provide the range for which self-similar solutions exists
using the $\Pi  $ theorem.

\noindent
Case I (s=0) : For FAP and s=0, there is one $\Pi $ group given by Eq \ref{Pi_0}.
Solutions do not exist for Mach numbers less than 1.3849.

\begin{figure}
\includegraphics[width=1\textwidth]{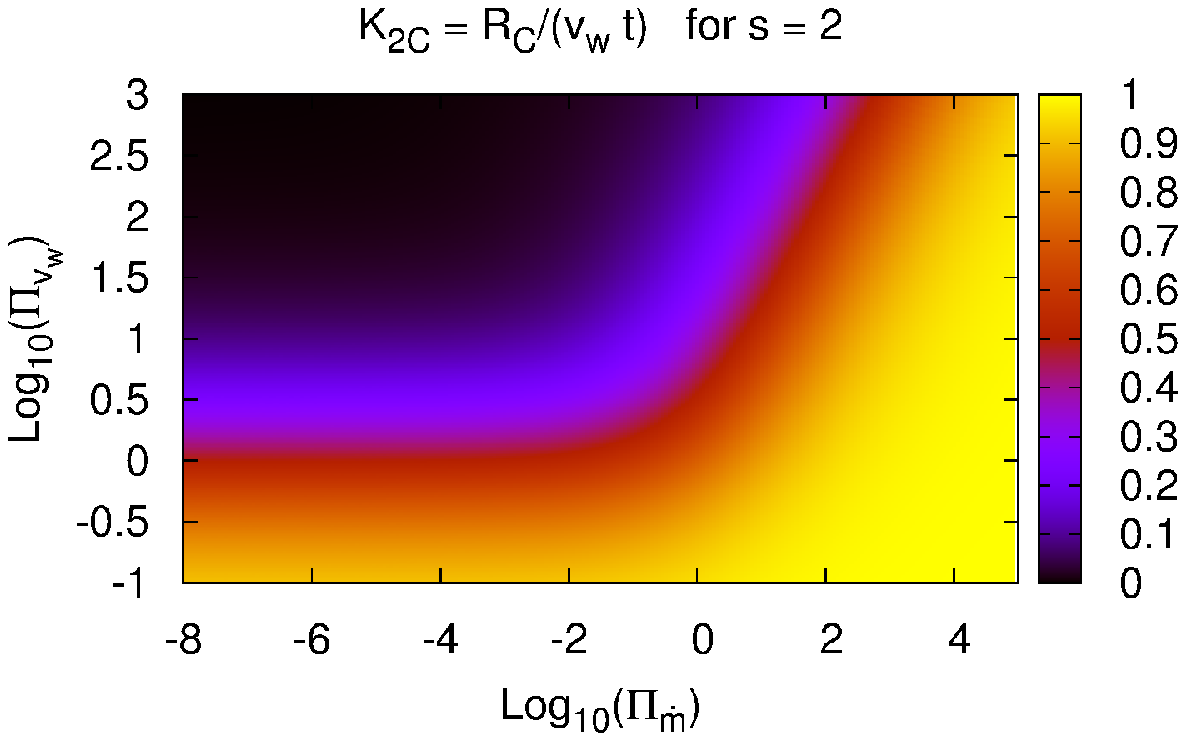}
\caption{Value of $K_{2\rm{C}}$ as a function of $v_w$ and log $K_2 (\Pi_{\dot m}, \Pi_{v_w})$
where $\Pi_{v_w}=(v_{w}-v_{w,1})/v_{w,1})$ and $\Pi_{\dot{m}}=(\dot m/\dot m_1)$. $K_{2\rm{C}} $ 
is close to constant in the regime of low mass winds running 
into high mass loss wind, i.e. $log(\Pi_{\dot m}) < 0$. \vspace{0.25cm}}
\label{Rc_lookup2}
\end{figure}

\begin{figure}
\includegraphics[width=1\textwidth]{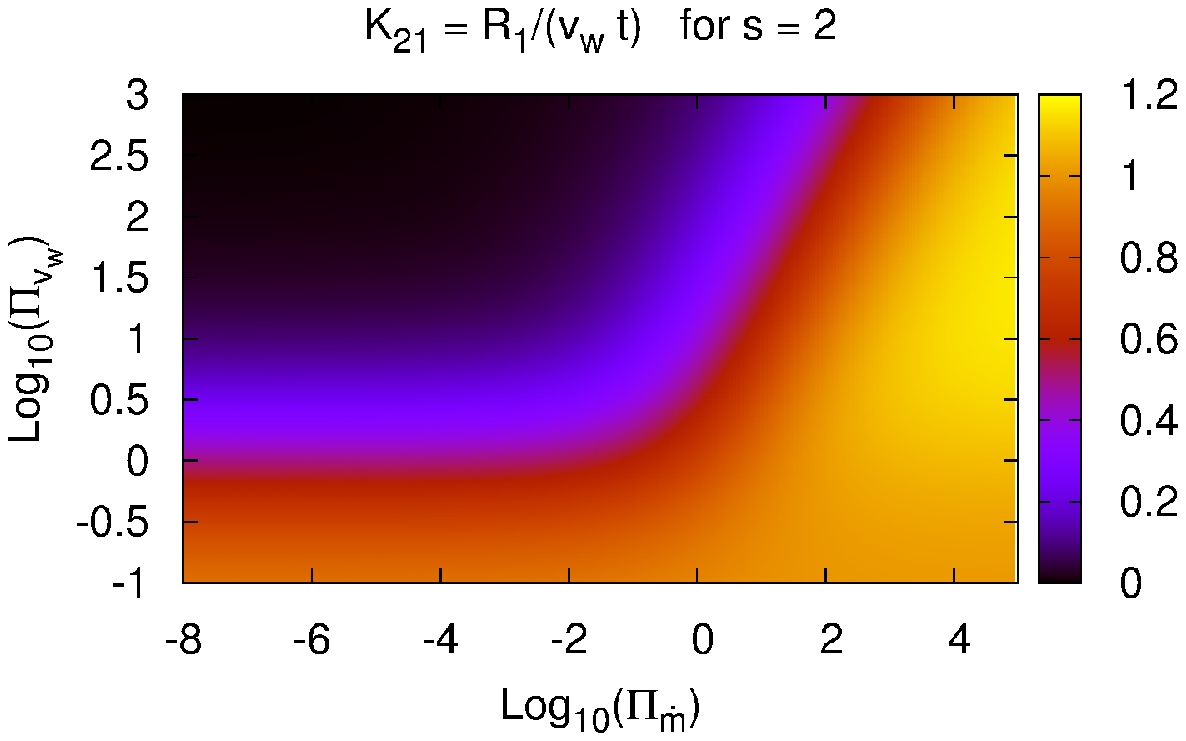}
\caption{Value of $K_{21}$ as a function of $v_w$ and log $K_2 (\Pi_{\dot m}, \Pi_{v_w})$
where $\Pi_{v_w}=(v_{w}-v_{w,1})/v_{w,1})$ and $\Pi_{\dot{m}}=(\dot m/\dot m_1)$. $K_{21} $ is close to constant in the regime of low mass winds running
into high mass loss wind, i.e. $log(\Pi_{\dot m}) < 0$. \vspace{0.9cm}}
\label{R1_lookup2}
\end{figure}

\begin{figure}
\includegraphics[width=1\textwidth]{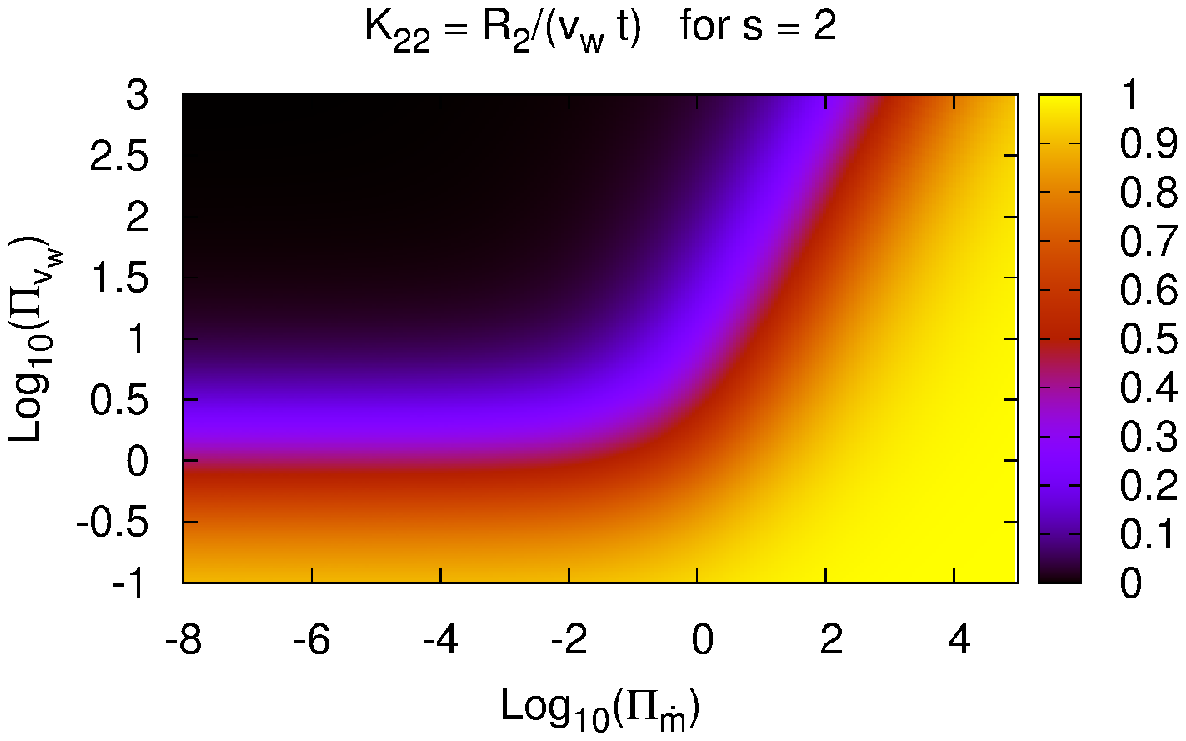}
\caption{Value of $K_{22}$ as a function of $v_w$ and log $K_2 (\Pi_{\dot m}, \Pi_{v_w})$
where $\Pi_{v_w}=(v_{w}-v_{w,1})/v_{w,1})$ and $\Pi_{\dot{m}}=(\dot m/\dot m_1)$. $K_{22} $ is close to constant in the regime of low mass winds running
into high mass loss wind, i.e. $log(\Pi_{\dot m}) < 0$. \vspace{0.5cm}}
\label{R2_lookup2}
\end{figure}

\noindent
Case II (s=2): 
Two $\Pi $ groups exist and, thus, possible solutions are a combination of  $\Pi $ groups with $K_2 (\Pi_{\dot m}, \Pi_{v_w})$ given by equations 27 and 28.
For $r^{-2}$ ambient density profiles, the shell velocity range is sufficient to determine the relation
between $\dot{m}_1$ and $v_{w,1}$ for the prior mass loss. In Figs. \ref{Rc_lookup2}, \ref{R1_lookup2} \& \ref{R2_lookup2}, we show $K_{2\rm{C}}, K_{21}$ and $K_{22}$ as a function of the
wind parameters covering the entire range discussed in this paper.

Regime I: For high mass loss rates, we have no power law relation between the wind and environmental parameters, and the values of
$K_{2\rm{C}}, K_{21}$ and $K_{22}$ need to be interpolated in the figures or can be calculated by SPICE.

\begin{figure}
\begin{center}
$\begin{array}{cc}
\includegraphics[width=0.45\textwidth]{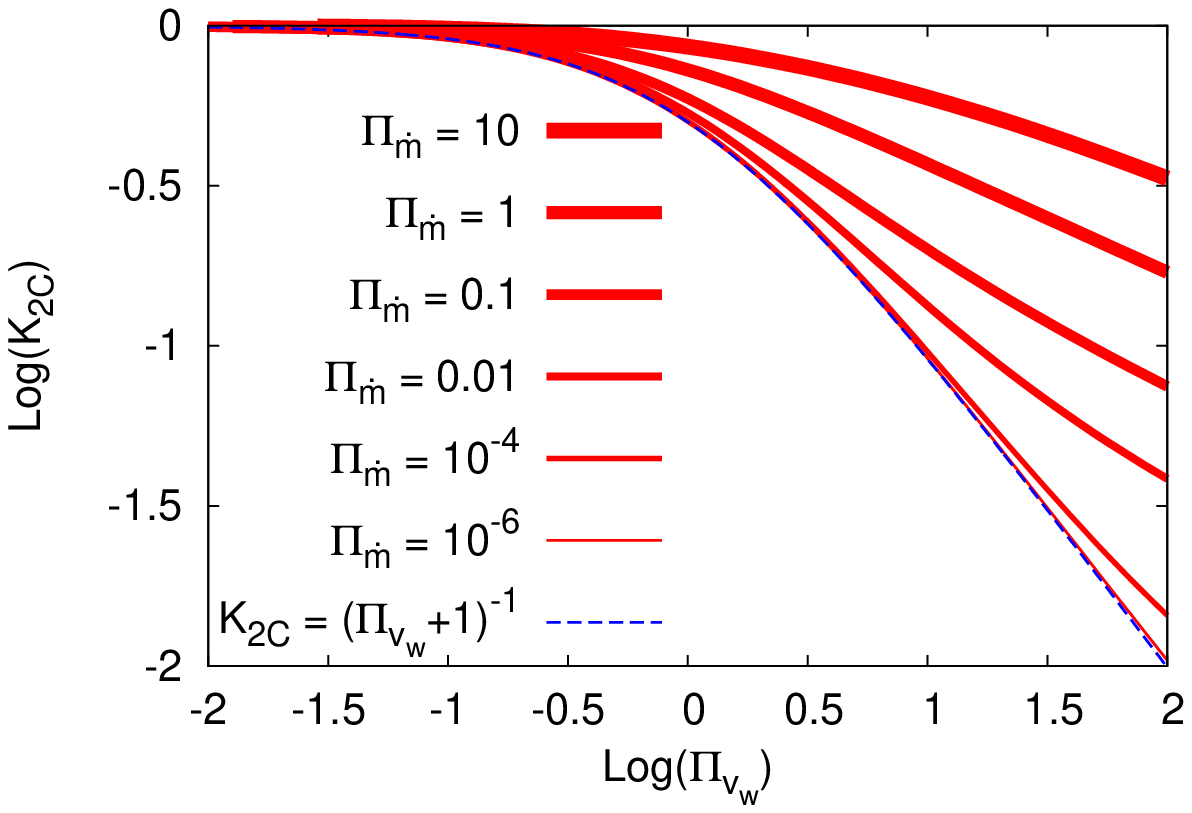} &
\includegraphics[width=0.45\textwidth]{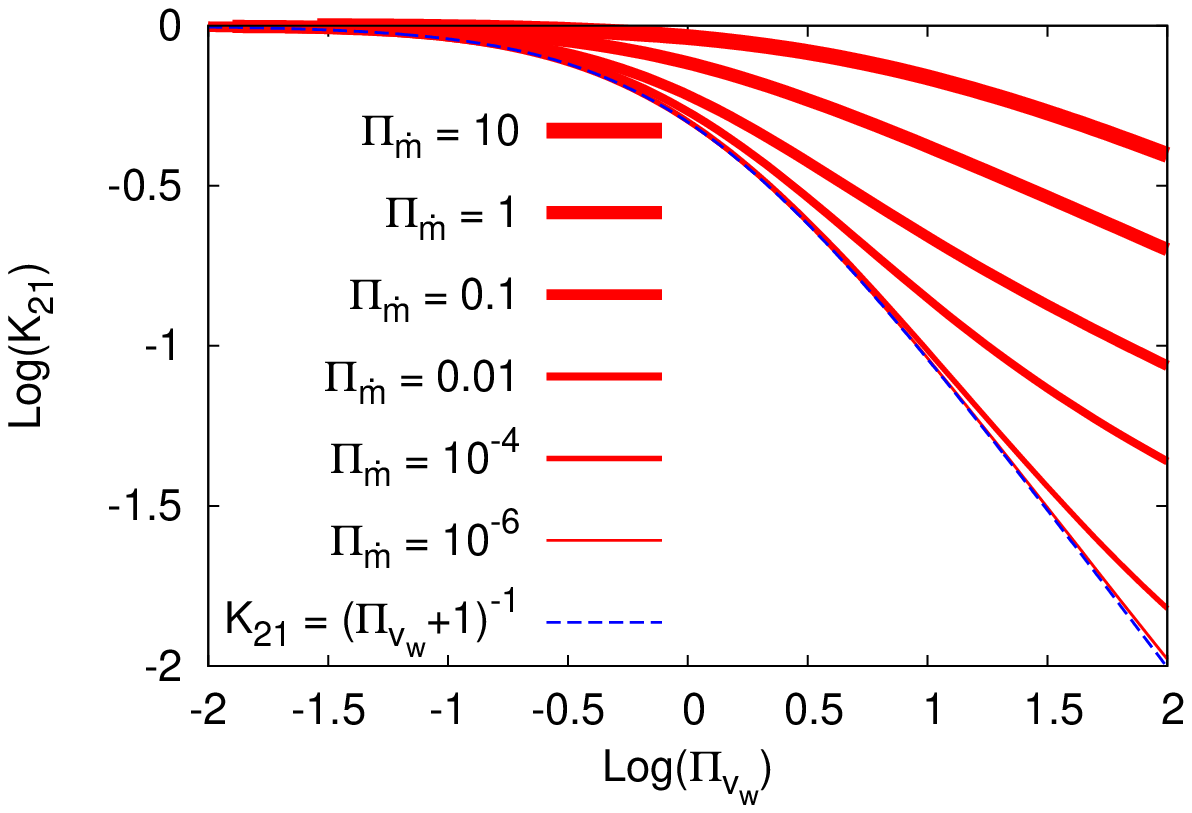} \\
\includegraphics[width=0.45\textwidth]{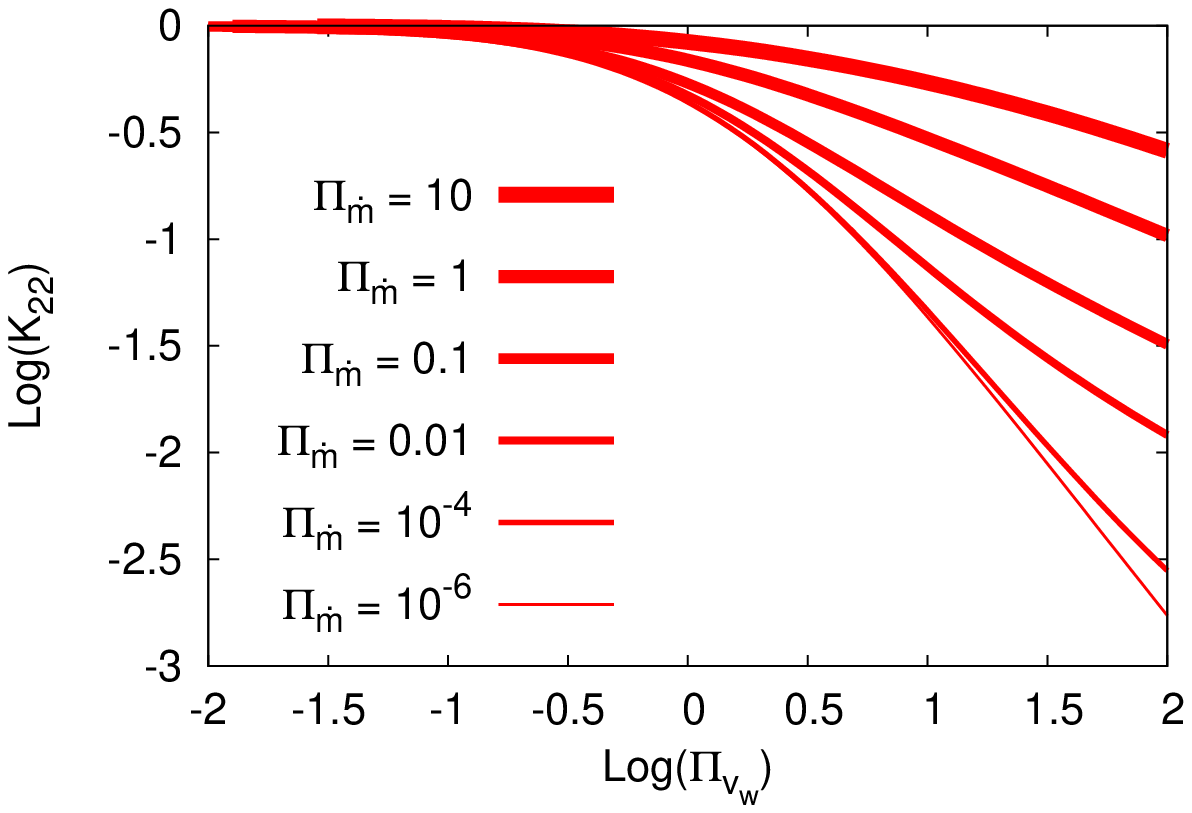} &
\includegraphics[width=0.45\textwidth]{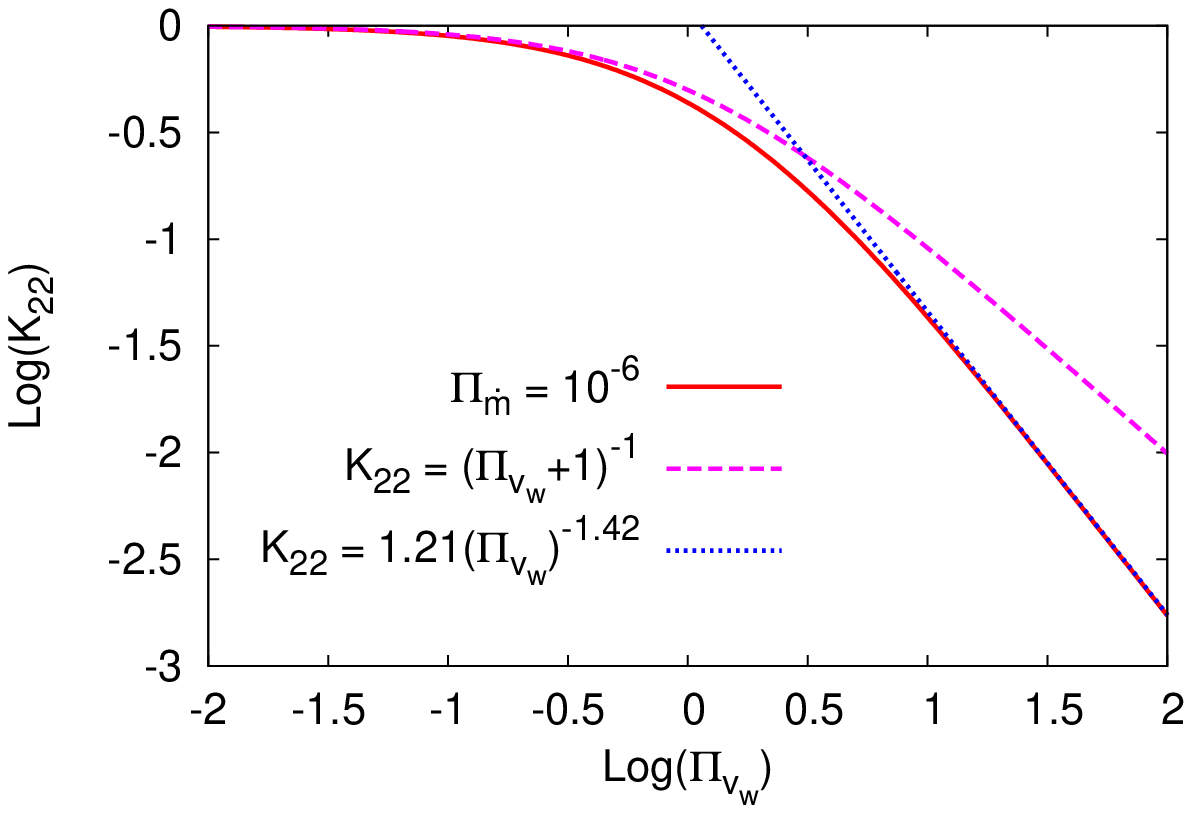}
\end{array}$
\end{center}
\caption{$K_{2(C,1,2)}$ as a function of the ratio between inner and outer wind velocity described by $\Pi_{v,w}$
as obtained by cuts at $log (\Pi_{\dot m})$ of $10^{-6,-4,-2,-1,0,1}$ indicated by progressively thicker lines.
The exact solutions are given in red.
We give the solutions for $K$ in comparison to the fits.
 The approximations are given (magenta and blue, dotted). }
\label{1dR}
\end{figure}

Regime II: If a low mass loss wind runs into a high mass loss wind,  $\Pi_{\dot m} < 0.1 $, $K_{2\rm{C}}, K_{21}, K_{22}$ hardly vary with the ratio $\dot{m}/\dot{m}_1$.
Thus, the contour lines in Figs. \ref{Rc_lookup2}, \ref{R1_lookup2} \& \ref{R2_lookup2} are horizontal.
Their value can be approximated therefore as a function of only the relative wind velocities. In Fig. \ref{1dR}, the variation of the cuts for various $\Pi_{\dot m}$.
$K_{2\rm{C}}$ and $K_{21}$ can be well described by single functions. $K_{22}$ needs two descriptions valid at low and high ratios of $v_{w}/v_{w,1}$ separated
{ at $Log~\Pi_{v_w} \approx 0.5 $}. The resulting power law dependencies are given in Table 2.

\begin{deluxetable}{lcr}
\tablewidth{20pc}
\tablecaption{Same as Table \ref{eqntable1} but for an environment produced by a prior wind (s=2). The index 1 corresponds to
the prior wind. The relations are valid for high velocity winds running into environments produced
by low velocity winds (see Fig. \ref{Rc_lookup2}).
For $R_{\rm{C}}$, $R_1$, and $u_1$, the relations are valid for $\dot{m}/\dot{m}_1 \lesssim 10^{-4}$.
For $R_2$ and $\rho_2$, it is valid for $\dot{m}/\dot{m}_1 \lesssim 10^{-4}$ and 
$v_{w}/v_{w,1}$ between $\sim 10$ and $170$. $\tau_m$ is calculated by formal integraion in SPICE.}

\tablehead{}

\startdata
$R_{\rm{C}}$         & $\approx$ & $v_{w,1}t$  \\ \vspace{0.1cm}
$R_1$         & $\approx$ & $v_{w,1}t$  \\ \vspace{0.1cm}
$R_2$         & $\approx$ & $1.21 \left ( \frac{v_{w,1}}{v_{w}-v_{w,1}}\right)^{1.42} v_{w}t$ \\ \vspace{0.1cm}
$u_1$         & $\approx$ & $\gamma(\gamma +1)^{-1}v_{w,1} $  \\ \vspace{0.1cm}
$\rho_2$      & $\approx$ & $0.385~ \dot{m} {v_{w}}^{-4} \left ( \frac{v_{w,1}}{v_{w}-v_{w,1}}\right)^{-2.84} t^{-2}$ \\ \vspace{0.1cm}
$\tau_{m}$ & $=$ & $\dot{m}_1 {v_{w,1}}^{-1} t^{-1} K_{2,\tau}(\Pi_{\dot{m}},\Pi_{v_w})$ \\

\enddata
\label{eqntable}
\end{deluxetable}

\section{Applications: Environments of Type Ia Supernovae Progenitors}\label{Current Status}
We will explore winds emanating from the progenitor system and interacting with the ISM of mass loss of the system prior to the supernova explosion. 
We consider winds from each source separately, and we address the question of which component is mostly responsible for the 
formation of the environment, and the typical structure to be expected. Subsequently,
we discuss the link between observables and progenitor systems, and  analyze SN 2014J.

We employ our spherical, semi-analytical models constructed by piecewise, scale-free analytic solutions. 
Scales enter the system via the equation of state, the boundary, and the jump conditions.
 The free parameters are: 1) The velocity $v_w$, 2) mass loss rate $\dot{m}$ from the central object, 
and the 3) $n_0=const$ or a mass loss rate $\dot{m}_1$ with $v_{w,1} $, i.e. $n \propto r^{-s}$, and 4) the duration of
the wind interaction $t$.
 As a result, we obtain the density, velocity and pressure  as a function of time, namely $\rho(r,t),v(r,t)$ 
 and $p(r,t)$ which can be linked to observables.
We use typical parameters 
 to discuss the different regimes which may occur in nature. For actual fits of
observations, appropriate solutions can be constructed by tuning these parameters with SPICE. 

 The wind may originate from the AD, the donor star which may be a MS or a RG-, horizontal- and asymptotic-branch star, 
the WD during a phase of over-Eddington accretion, or a combination of AD with a RG-like wind. As shown below, the time scale for the accretion and, thus, the progenitor, is an important factor in formation of the environment.
 The time scales $t$ vary widely depending on the scenario and chemical composition
of the accreted material and the initial mass of the progenitor (e.g. \citep{Sugimoto1980,piersanti2003,Wang2012} and reviews cited in the introduction). To reach $M_{Ch}$, about 0.2 to 0.8 $M_\odot$ of material needs to be accreted.
For hydrogen accretion, the rates for stable hydrogen burning are between   
$\dot{M}\approx 2 \times 10^{-8...-6}~M_\odot/yr$ depending on the metallicity \citep{nomoto82,hachisu10}.
The upper and lower limits for $\dot{M}$ are set by the Eddington limit for the luminosity and the minimum amount of fuel needed for steady burning, respectively.
 However, it is under discussion whether and at which accretion rates
steady H-burning can continue until the WD approaches  $M_{Ch} $.
 It depends on the chemical composition, rotation of the WD, and details of the approximations used 
\citep{nomoto82,starrfield85,vah92,Hachisu99,piersanti2003,Yaron05,Sako08,Hachisu12,Bours13}. 
For a recent review, see \citet{Maoz2014}. In this study, we use the wide range of accretion rates to avoid restricting possible solutions.
Thus, we consider time scales $t$ between $10^5 $ and  $4 \times 10^7$ years. 
 Larger rates of mass overflow result in over-Eddington luminosity and a strong wind from the progenitor WD with 
properties typical of RG winds \citep{Hachisu96,Hachisu08}. Subsequently, we refer to the high-density, low velocity winds as ``RG-like''.
Accretion of He and C/O -rich matter allow much shorter timescales down to the dynamical times of merging WDs.  

 For accretion disk winds the mass loss rate ranges from $10^{-8}$ to $10^{-10}$ solar masses per year; and
the wind velocities originating from the disks are believed to be from 2000 to 5000 $km/s$ 
\citep{Kafka}. 

Mass loss in ``RG-like'' stars are typically between $10^{-6}$ and $10^{-8}~M_\odot /yr$ with wind velocities between $10$ and $60$ $km/s$
 \citep{reimers77a,Judge91,Ramstedt2009}.

 Main sequence star winds are similar to the solar wind \citep{wood2002}.  
 For solar wind the velocity is between  $400~km/s$ and $750~km/s$ and the mass loss is $2...3 \times 10^{-14}~M_\odot/yr$ \citep{noci97,feldman05,March06}. 

\subsection{Parameterized Study}

 In the following, we will assume typical wind parameters as follows:
 In the ``RG-like'' case, we use mass loss rates between $\dot m $ of $10^{-6}$, $10^{-7}$ and $10^{-8}~M_\odot/yr$, and a wind velocity of 30 $km/s$. 
 Similar winds can be expected for WDs with high accretion rates. For the case of a MS star donor, we use mass loss rates between $10^{-13}$, $10^{-14}$, and 
 $10^{-15}~M_\odot/yr$, and $v_w = 500~km/s$. 
 For AD  winds the mass loss rate ranges from $10^{-8}$, $10^{-9}$, and $10^{-10}$ solar masses per year; a 
 typical rate has been measured to be $10^{-9}$ $M_\odot/yr$.
 A wind velocity of 3000 $km/s$ is used. For the duration of the winds, we consider $t$ between $1.5 \times 10^5 $ and $6 \times 10^8$ years
with $ 3 \times 10^5$ years for the references.

 The outer environment of the system depends on its history including the delay time between the formation of the  
WD and the onset of the accretion phase. For long delays, we assume an ISM with constant density, i.e. $s=0$. 
For short delay times and small peculiar velocities of the system, the outer environment may be created 
during the final stage of the progenitor evolution, namely the red giant branch (RGB), horizontal giant branch (HGB) and the asymptotic
giant branch (AGB) phase.

\subsection{Results}\label{Results}
\subsubsection{Case I: Constant ISM density}
\label{CID}

We first consider scenarios where the wind blows out into a medium of constant density for a wide range of parameters
(Tables \ref{s0_table1a}-\ref{s0_table1c}). 
  The structures are characterized by I) an undisturbed, inner layer dominated by the stellar wind,
 II) an inner, shocked region with almost constant, low density and a velocity declining with distance,
III) a slowly expanding shell of high density, swept up material, 
and IV) the ISM. The overall solution is representative for all cases as has been shown in Sect. \ref{Theory}.
 For the estimate of the equivalent width $EW$ of the Na I doublet  at 5890/5896 $\mbox{\normalfont{\AA}} $, we 
 use solar abundances, $X_{\rm{Na}} = 3.34 \times 10^{-5}$. EW is estimated according to
\citet{Spitzer1968,Draine2011}. For potassium lines, the corresponding equations apply.

\begin{figure}
\begin{center}
$\begin{array}{cc}
\includegraphics[width=0.48\textwidth]{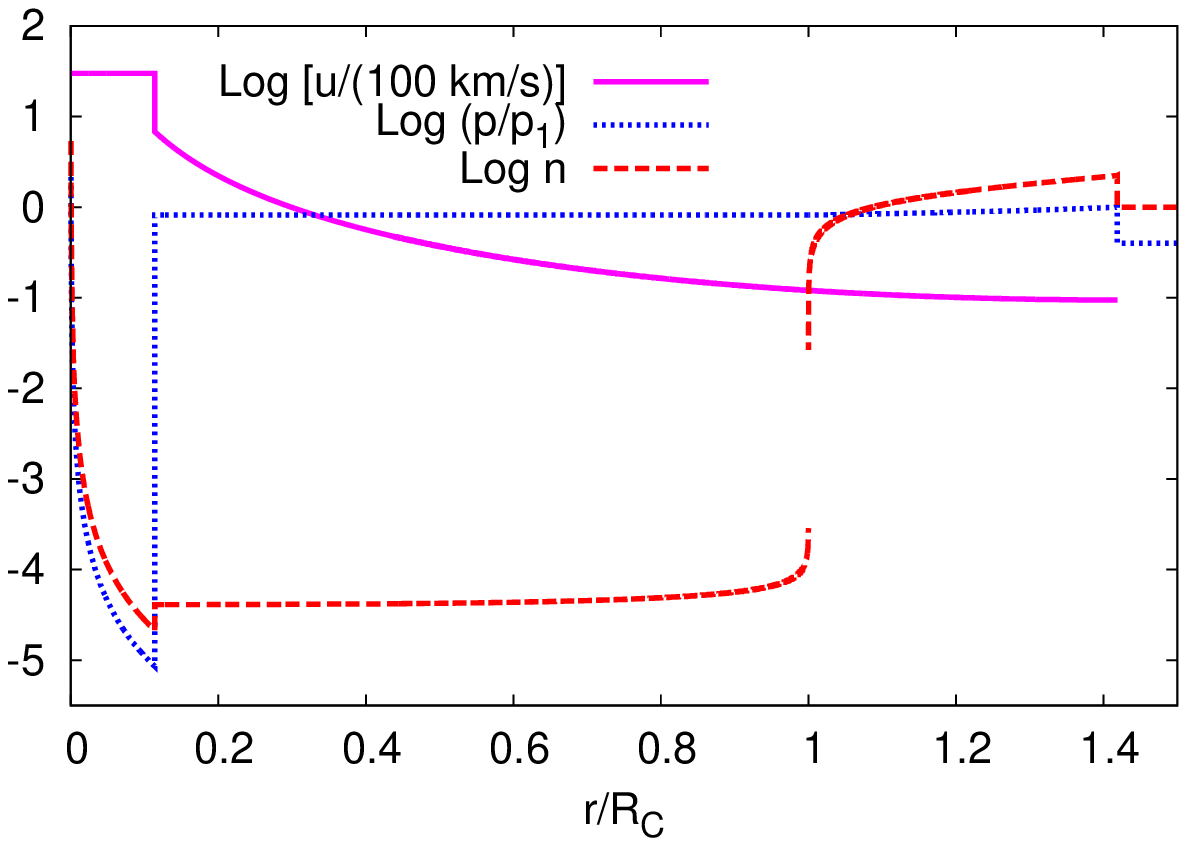}&
\includegraphics[width=0.48\textwidth]{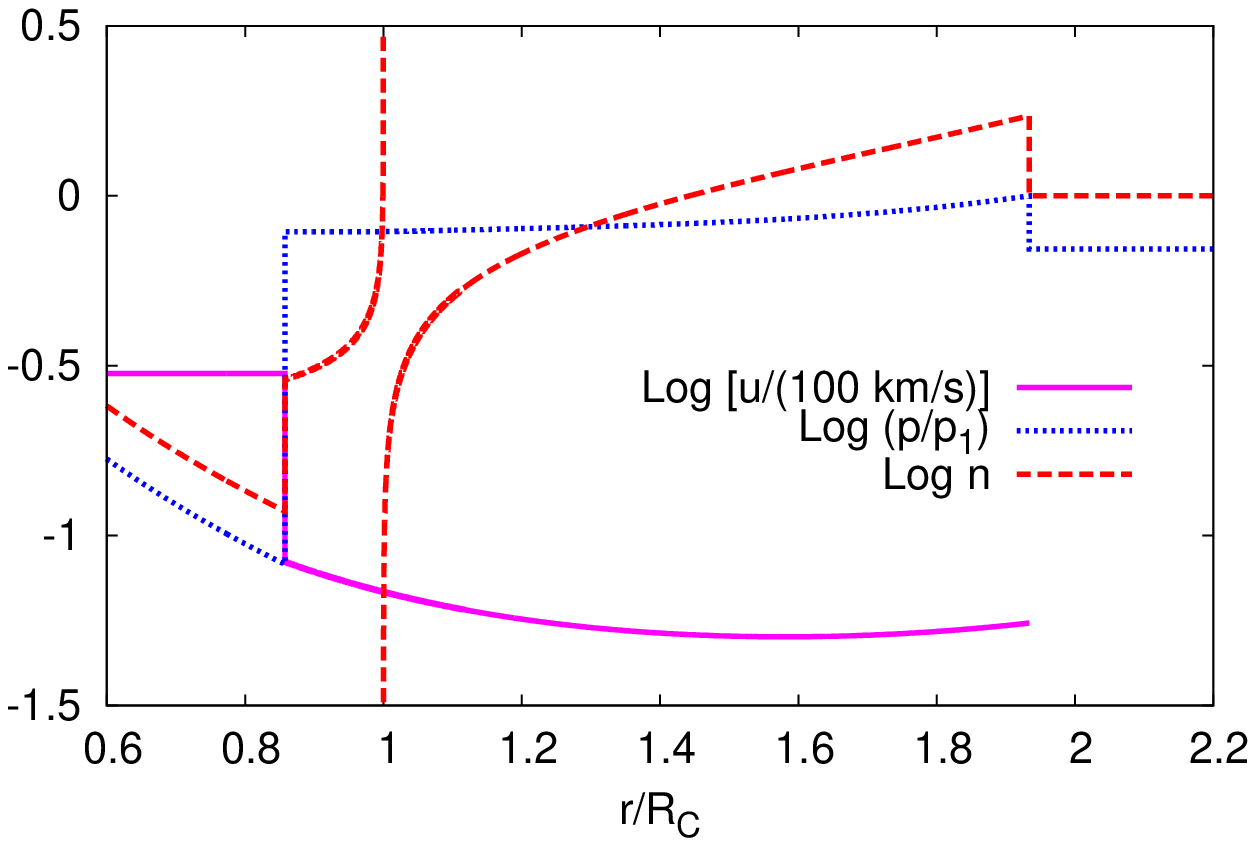} 
\end{array}$
\end{center}
\caption{Hydrodynamic profile for a wind typical for an accretion disk (left), 
$v_w=3000~km/s$ and mass-loss rate of $10^{-8}~M_\odot/yr$,
and a RG-like wind (right), 30 $km/s$ and mass-loss rate of 
$10^{-7}~M_\odot/yr$, running into a constant interstellar medium density of 
1 particle per $cm^3$ after a time of 300,000 years. The contact discontinuity 
is at 21.7 and 5.45 light-years, respectively. Fluid velocity $u$ (magenta) 
is normalized to 100 $km/s$, pressure $p$ (blue) is normalized to the pressure 
$p_1$ just inside the outer shock, and particle density $n$ is unnormalized.}
\label{hydroplot1}
\end{figure}

\noindent
{\bf Case Ia: Fast winds from an accretion disk:}
 Table \ref{s0_table1a} contains calculated results from several cases with different parameters but the same time-scales $t$.
 Our reference, model 2 of Table \ref{s0_table1a}, is shown in Fig. \ref{hydroplot1}. It has a mass loss $\dot{m} $ of $10^{-8}$ $M_\odot/yr$ and  
 a wind velocity $v_w=3,000~km/s$. For the duration of the wind, we choose a duration of $3 \times 10^5$ years which, within the SD scenario,
corresponds to the evolutionary time for a low mass WD to grow to $M_{Ch}$ at an accretion rate of 
 $ \approx 2 \times 10^{-6}~M_\odot /yr$.

 Up to about $2.3~ly$, the environment is dominated by the on-going wind.
  In this region, the particle density drops below 1(100) $cm^{-3}$ at a distance of 
$0.005(0.0005)~ly ~\approx 1000 (100)~AU \approx 10^{16(15)}~cm$ which will be overrun by the SN ejecta within 5(0.5) days. 
 Particle densities below 100 $cm^{-3}$ in this region will hardly affect the light curves or spectra because 
the swept up mass will be small.
Using the same argument, the low densities within $20~ly$ are too low to affect the hydrodynamics of the SN envelope.
 A high density, outer shell expands at a velocity of 11 $km/s$ with a velocity dispersion of 
$ \approx 13 \% \approx 2~km/s $. This shell would produce a narrow line Doppler shifted by about 11 $km/s$.
 For an interstellar medium, the equivalent width would be about
$ 165~m\mbox{\normalfont{\AA}} $ for the NaID line well comparable to values found by \citep{Phillips2013} who
found between 27 and 441 $m\mbox{\normalfont{\AA}}$ in a sample of some 20 SNe~Ia.

\begin{deluxetable*}{lcccccccc}
\tabletypesize{\small}
\tablewidth{0pc}
\tablecaption{Interaction of winds of an accretion disk (AD) with a constant ISM (s=0). Numerical results  are given for parameters typical
for those winds for the FAP model (see text and Table \ref{eqntable1}).  In addition, we give the velocity dispersion of the shell $\sigma_u$ and
the optical depth $\tau$ and equivalent width $EW$ of the $NaID$ line.
Our reference model is marked by $*$.
}
\tablehead{ 
\colhead{No} &
\colhead{$n_0$}      & 
\colhead{$v_w$} &
\colhead{$\dot{m}$} & 
\colhead{$t$} &
\colhead{$t_p$} &
\colhead{$R_{\rm{C}}$} & 
\colhead{$R_1$} &
\colhead{$R_2$} \\ 
\colhead{} &
\colhead{($cm^{-3}$)} &
\colhead{($km/s$)} &
\colhead{($M_\odot/yr$)} &
\colhead{($Myr$)} &
\colhead{($Myr$)} &
\colhead{($ly$)} &
\colhead{($ly$)} &
\colhead{($ly$)} 
}
                                              
\startdata
1  &    0.1  &  3000   & \tenexp{ -8 } &    0.3  &    2.36    &   33.34  &     41.94  &    4.89  \\  
2* &    1.0  &  3000   & \tenexp{ -8 } &    0.3  &    0.745   &   20.11  &     28.53  &    2.29  \\  
3  &   10.0  &  3000   & \tenexp{ -8 } &    0.3  &    0.236   &   11.39  &     22.02  &    0.98  \\ 
4  &    1.0  &  3000   & \tenexp{ -9 } &    0.3  &    0.236   &   11.39  &     22.02  &    0.98  \\  
5  &    1.0  &  3000   & \tenexp{ -10} &    0.3  &    0.0745  &    5.80  &     20.47  &    0.36  \\  
6  &    1.0  &  3000   & \tenexp{ -8 } &    0.15 &    0.745   &   13.70  &     17.84  &    1.82  \\  
7  &    1.0  &  3000   & \tenexp{ -8 } &    0.4  &    0.745   &   23.44  &     35.07  &    2.50  \vspace{0.2cm}\\  
\hline \vspace{-0.40cm} \\  
\hline \vspace{-0.1cm} \\ 

No & $u_1$  & $u_c$  & $\sigma_u$ & $n_2$ & $\tau_{m}$ & $log(\tau)$ & $EW_{NaI}$ \\ 
 & ($km/s$)  & ($km/s$)  & ($km/s$) & ($cm^{-3}$) & ($g/cm^2$) & & ($m\mbox{\normalfont{\AA}}$) \vspace{0.1cm}\\

\hline \vspace{-0.1cm} \\
1  &   16.56  &   19.99  &    0.85   & \xtenexp{ 2.07 }{-5 } &   \xtenexp{2.60}{-6}  &    1.34   & 157 \\
2* &    9.45  &   12.06  &    0.62   & \xtenexp{ 9.45 }{-5 } &   \xtenexp{1.91}{-5}  &    2.34   & 165 \\
3  &    5.53  &    6.83  &    0.32   & \xtenexp{ 5.20 }{-4 } &   \xtenexp{1.71}{-4}  &    3.59   & 192 \\
4  &    5.53  &    6.83  &    0.32   & \xtenexp{ 5.20 }{-5 } &   \xtenexp{1.71}{-5}  &    2.59   & 98.4 \\
5  &    4.50  &    3.48  &    0.62   & \xtenexp{ 3.94 }{-5 } &   \xtenexp{1.84}{-5}  &    2.33   & 163 \\
6  &   13.34  &   16.42  &    0.76   & \xtenexp{ 1.50 }{-4 } &   \xtenexp{1.13}{-5}  &    2.03   & 180 \\
7  &    8.16  &   10.54  &    0.56   & \xtenexp{ 7.95 }{-5 } &   \xtenexp{2.42}{-5}  &    2.49   & 157 \vspace{0.0cm}\\
\enddata

\label{s0_table1a}
\end{deluxetable*}

 The morphology of the structures are hardly affected by variations in the wind parameters as expected from the $\Pi$ theorem.
However, the actual size of the regions and the densities vary (see Table \ref{s0_table1a}) with dependencies as expected from Table \ref{eqntable1}.
Within the framework of SNe~Ia, the ram pressure dominates the ambient pressure which, therefore, hardly affects the solution.
Typically, wind from an accretion disk will produce an inner cavity between 5 and 30 $ly$ surrounded by an expanding shell with 
a velocity of $\approx 5 ... 20~km/s$. EW (NaID) is  $\approx 100$ to $200~m\mbox{\normalfont{\AA}}$.
The combination of line shifts and strength allows us to derive the wind parameters.

In our table, we assumed an accretion rate of $\approx 2 \times 10^{-6}~M_\odot/yr$ which is at the upper limit allowed for stable accretion of 
hydrogen rich matter. Higher mass loss can be expected for over-Eddington accretion or He or C/O accreting WDs (see Sect. 1). Higher and lower 
mass loss rates will increase/decrease the size of the cavity but the dependency is relatively weak, $\propto \dot{m}^{1/5}$.
 
However, the actual size of the region does depend sensitively on the time $t$ of the progenitor evolution as the size of the cavity goes like $\propto t^{3/5}$.
 In hydrogen accreters, the rate of accretion may be smaller by a factor of 100 and, thus, $t$ may be larger by the same factor which, 
in turn, will increase the size of the cavity by 16 and decrease shell velocities by about a factor of 6.
Shorter durations can be realized if we start with a WD of $1.2 M_\odot$, the upper end of mass range for a C/O WD (see Introduction).
 Indeed, there is some evidence and theoretical arguments that the progenitors originate close to the upper end of that mass range
\citep{nomoto06a,hoeflich06,sadler12} which may reduce the amount of accreted material from $\approx 0.8$ to $0.2~M_\odot$. Thus, the 
 duration of the accretion $t$ may be correspondingly shorter which, in turn, 
reduces the size of the cocoon and increases the shell velocities by $\approx 2 $ and $1.5$, respectively.  

On the other hand, rates for He and C/O accretion can be larger than hydrogen accreters by, at least, the same factor of 100 \citep{WI73,piersanti2003,WMC09,WCMH09}, 
reducing the size of the shell and increasing the velocity of the shell lines by the same factors.
 High shell velocities may indicate He- or C/O accreters.
Despite the line shifts, systems with high shell velocities can be expected to have smaller low density regions. The SN ejecta
have velocities up to about $10 $ to $20 \%$ of the speed of light. For such SNe~Ia, we may expect interaction on time scales from
1 to 10 years for our set of parameters.

\noindent
{\bf Case Ib: MS star winds:}
MS star winds produced by a donor star are expected to show low velocities $v_w = 500$ $km/s$ and a very  low mass loss  
$\dot{m} = 10^{-14}$  $M_\odot/yr$. Although the morphology of the shells remain the same as above, the corresponding
regions will be overrun and dominated by wind from the accretion disk. The radius of the reverse shock $R_2$
show little change for time scales much longer than $t_p$ (see Table \ref{s0_table2}). As discussed above, $R_{\rm{C}}$ becomes unphysical
due to mixing in a regime of weak shocks. For SNe~Ia, the MS wind component can be neglected.

\noindent
{\bf Case Ic: Slow, RG-like winds} may be produced during the
RG phase of the donor star or the  progenitor prior to the formation of the progenitor WD, or as a result of
over-Eddington mass overflow (Table \ref{s0_table1b}). We refer to those as RG-like winds. 
   The resulting structures are very similar to Case Ia but the densities     
are higher by an order of magnitude (Fig. \ref{hydroplot1}). In particular,
the density contrast $n_2/n_0 \approx 0.1$ (see also Table \ref{eqntable1}). The cocoons are smaller and less pronounced in this case. 

\begin{deluxetable*}{lcccccccc}
\tabletypesize{\small}
\tablewidth{0pc}
\tablecaption{Same as Table \ref{s0_table1a} but for an ``RG-like'' wind. Our reference model is marked by $*$.
}
\tablehead{ 
\colhead{No} &
\colhead{$n_0$}      & 
\colhead{$v_w$} &
\colhead{$\dot{m}$} & 
\colhead{$t$} &
\colhead{$t_p$} &
\colhead{$R_{\rm{C}}$} & 
\colhead{$R_1$} &
\colhead{$R_2$} \\ 
\colhead{} &
\colhead{($cm^{-3}$)} &
\colhead{($km/s$)} &
\colhead{($M_\odot/yr$)} &
\colhead{($Myr$)} &
\colhead{($Myr$)} &
\colhead{($ly$)} &
\colhead{($ly$)} &
\colhead{($ly$)} 
}

\startdata
8   &    0.1  &    30   & \tenexp{ -7 } &    0.3  &    0.0745  &    5.80  &     20.47  &    3.55  \\
9*  &    1.0  &    30   & \tenexp{ -7 } &    0.3  &    0.0236  &    2.75  &     20.28  &    1.16  \\
10  &    1.0  &    30   & \tenexp{ -5 } &    0.3  &    0.236   &    11.39 &     22.02  &    9.77  \\  
11  &    1.0  &    30   & \tenexp{ -6 } &    0.3  &    0.0745  &    5.80  &     20.47  &    3.55  \\ 
12  &    1.0  &    30   & \tenexp{ -8 } &    0.3  &    0.0075  &    1.28  &     20.27  &    0.37  \\ 
13  &    1.0  &    30   & \tenexp{ -7 } &    0.15 &    0.0236  &    2.16  &     10.17  &    1.15  \\   
14  &    1.0  &    30   & \tenexp{ -7 } &    2.0  &    0.0236  &    5.20  &     135  &    1.16  \\   
15  &    0.1  &    30   & \tenexp{ -7 } &    3    &    0.0745  &    12.8  &     203  &  3.69 \\
16  &    1.0  &    30   & \tenexp{ -7 } &    3    &    0.0236  &    5.96  &     203  &  1.17 \\
17  &    0.1  &    30   & \tenexp{ -7 } &    13   &    0.0745  &    20.9  &     878  &  3.70 \\
18  &    1.0  &    30   & \tenexp{ -7 } &    13   &    0.0236  &    9.71  &     878  &  1.17 \\
19  &    0.1  &    30   & \tenexp{ -7 } &    50   &    0.0745  &    32.8  &    3377  &  3.70 \\
20  &    1.0  &    30   & \tenexp{ -7 } &    50   &    0.0236  &    15.2  &    3382  &  1.17 \vspace{0.2cm}\\
\hline \vspace{-0.40cm} \\  
\hline \vspace{-0.1cm} \\ 

No & $u_1$  & $u_c$  & $\sigma_u$ & $n_2$ & $\tau_{m}$ & $log(\tau)$ & $EW_{NaI}$ \\ 
 & ($km/s$)  & ($km/s$)  & ($km/s$) & ($cm^{-3}$) & ($g/cm^2$) & & ($m\mbox{\normalfont{\AA}}$) \vspace{0.1cm}\\

\hline \vspace{-0.1cm} \\
8  &    4.50  &    3.48  &    0.62   & \xtenexp{ 4.19 }{-2 }  &   \xtenexp{1.84}{-6}  &    1.33   & 114 \\
9* &    4.37  &    1.65  &    0.94   & \xtenexp{ 3.75 }{-1 }  &   \xtenexp{1.94}{-5}  &    2.17   & 231 \\
10 &    5.53  &    6.83  &    0.32   & \xtenexp{ 6.62 }{-1 }  &   \xtenexp{1.71}{-5}  &    2.59   & 98.4 \\
11 &    4.50  &    3.48  &    0.62   & \xtenexp{ 4.19 }{-1 }  &   \xtenexp{1.84}{-5}  &    2.33   & 163 \\
12 &    4.36  &    0.77  &    1.03   & \xtenexp{ 3.66 }{-1 }  &   \xtenexp{1.97}{-5}  &    2.14   & 251 \\
13 &    4.41  &    2.60  &    0.79   & \xtenexp{ 3.92 }{-1 }  &   \xtenexp{9.46}{-6}  &    1.94   & 181 \\
14 &    4.37  &    0.47  &    1.05   & \xtenexp{ 3.65 }{-1 }  &   \xtenexp{1.32}{-4}  &    2.96   & 329 \\
15 &    4.36  &    0.77  &    1.03  &  \xtenexp{ 3.66 }{-2 }  &   \xtenexp{1.97}{-5}  &    2.14   & 251 \\
16 &    4.36  &    0.36  &    1.05  &  \xtenexp{ 3.64 }{-1 }  &   \xtenexp{1.98}{-4}  &    3.13   & 353 \\
17 &    4.36  &    0.29  &    1.05  &  \xtenexp{ 3.64 }{-2 }  &   \xtenexp{8.58}{-5}  &    2.77   & 309 \\
18 &    4.36  &    0.13  &    1.05  &  \xtenexp{ 3.63 }{-1 }  &   \xtenexp{8.58}{-4}  &    3.77   & 500 \\
19 &    4.36  &    0.12  &    1.05  &  \xtenexp{ 3.63 }{-2 }  &   \xtenexp{3.30}{-4}  &    3.35   & 391 \\
20 &    4.38  &    0.05  &    1.06  &  \xtenexp{ 3.64 }{-1 }  &   \xtenexp{3.31}{-3}  &    4.35   & 809  \vspace{0.0cm} \\

\enddata

\label{s0_table1b}
\end{deluxetable*}

For typical properties of the environment, we have to distinguish between winds for an RG donor, RG-like winds from over-
Eddington accretion, and the prior RG-phase of the progenitor.

 In Table \ref{s0_table1b}, models are shown for RG-like winds for various $n_0$, $\dot{m}$ and $t$. As expected, the size of
the cavity decreases with  $n_0$ (model 8 vs. 9). We choose $t$ in the range for stable hydrogen accreters (models 8 to 14).
As duration $t$ increases, the location of $R_2$ ``stalls'' and $n_2$ is hardly affected because the outer and inner pressure equilibrates ($t/t_p \gg 1$) 
as discussed in the Sect. \ref{Boundary}. See also Table \ref{eqntable1}. For the same reason, the size of the cavity increases with $\dot{m}$ but $n_2/n_0$ 
hardly changes.
 Typically, RG-like winds will produce an inner cavity between 1 and 10 $ly$ surrounded by an expanding shell with 
a velocity of $\approx 5~km/s$. EW (NaID) is a few hundred $ m\mbox{\normalfont{\AA}}$. Note that the scales of RG winds
are smaller by an order of magnitude compared to those of AD winds unless those have much lower mass loss than
considered in our example above. An RG wind is more likely to form a combined AD-RG wind as discussed below.

RG winds from the progenitor prior to the WD phase (models 15-20) will produce a 
similar structure as an RG donor in a system with high accretion rates 
($\sim 10^{-6}~M_{\odot}/yr$) but are systematically larger because the longer 
duration $t$.  This wind may form an environment for subsequent winds. 
Progenitor system winds may run into this environment if the delay time between the formation of the WD and the onset 
of the accretion is sufficiently short, and if the peculiar velocity of the system is small. 
The peculiar velocities of stars show a wide range with a typical value of 25-50 $km/s$ for the
Galactic plane (\citep{Griv2009}, and references therein). Here, the duration of the wind is given by the evolutionary time
scale of the RG phase rather than the time to reach $M_{Ch}$. For models 15-18, we use $t$ corresponding to the
post-main sequence life time of 5 and 8 $M_\odot$ stars with a mass loss rates of $10^{-7}~M_\odot$ 
The resulting total mass loss is $1.3$ and $0.3~M_\odot$ for the 5 and 8 $M_\odot$ star, respectively \citep{Schaller1992,Chieffi2001}.

The environment formed by a wind consists of inner region with $s\approx 2$ with a size $R_2$ and a low 
density void ($s \approx 0$) of size $R_{\rm{C}}$. The resulting size of the void,  $R_{\rm{C}}$, is typically 10-20 $ly$.
 The duration $t$ can be increased by lower main sequence masses for the progenitor. However,
then, the  amount of mass lossed prior of forming a WD of $M(WD) \approx 0.6~M_\odot$. Using a mass loss of $10^{-7}~M_\odot$
and $v_w=30~km/s$, the maxima in $R_{\rm{C}}$ and $R_2$ is produced by a $3.6~M_\odot$ star: $\approx 33~ly $ and $\approx 3.7~ly$, 
respectively (models 19, 20, Table \ref{s0_table1b}). Models with durations of 3, 13 and 50 $Myr$ correspond to 
progenitor stars of 8, 5, and 3.6 $M_\odot$. 

We note that long durations may also be produced during the evolution of the progenitor system if the hydrogen accretion rate is close to the lower limit of $\sim 10^{-8}~M_{\odot}/yr$, though this low a rate may not allow for stable accretion for WD close to $M_{Ch}$ \citep{Piersanti2004}. 
Models with RG-like winds may correspond to systems with over-Eddington accretion and a MS donor star (models 15-20).  
For those long duration RG-like winds, the resulting cavities can be $\approx 15$ to $33~ly$ and high EW of $\approx 500~m\mbox{\normalfont{\AA}}$.

\begin{figure}
$\begin{array}{cc}
\includegraphics[width=0.48\textwidth]{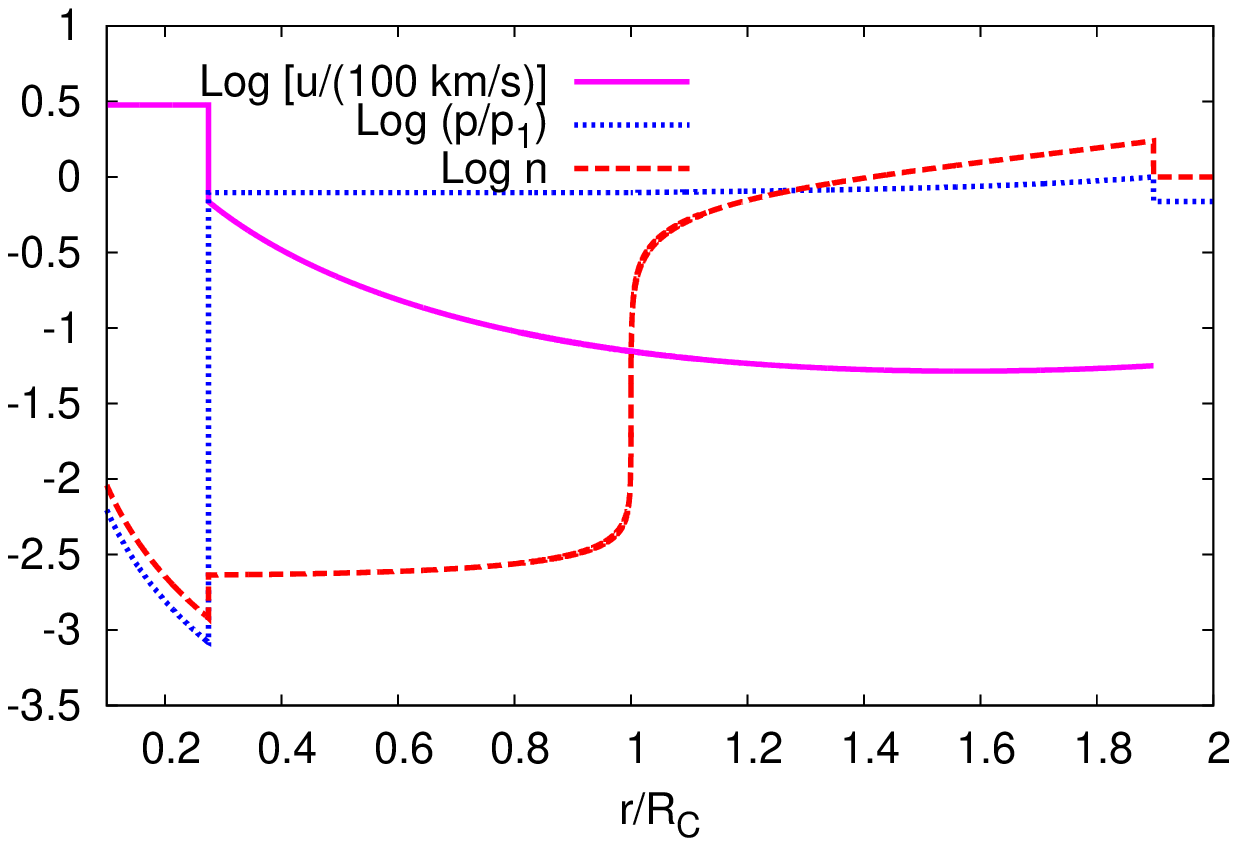}&
\includegraphics[width=0.48\textwidth]{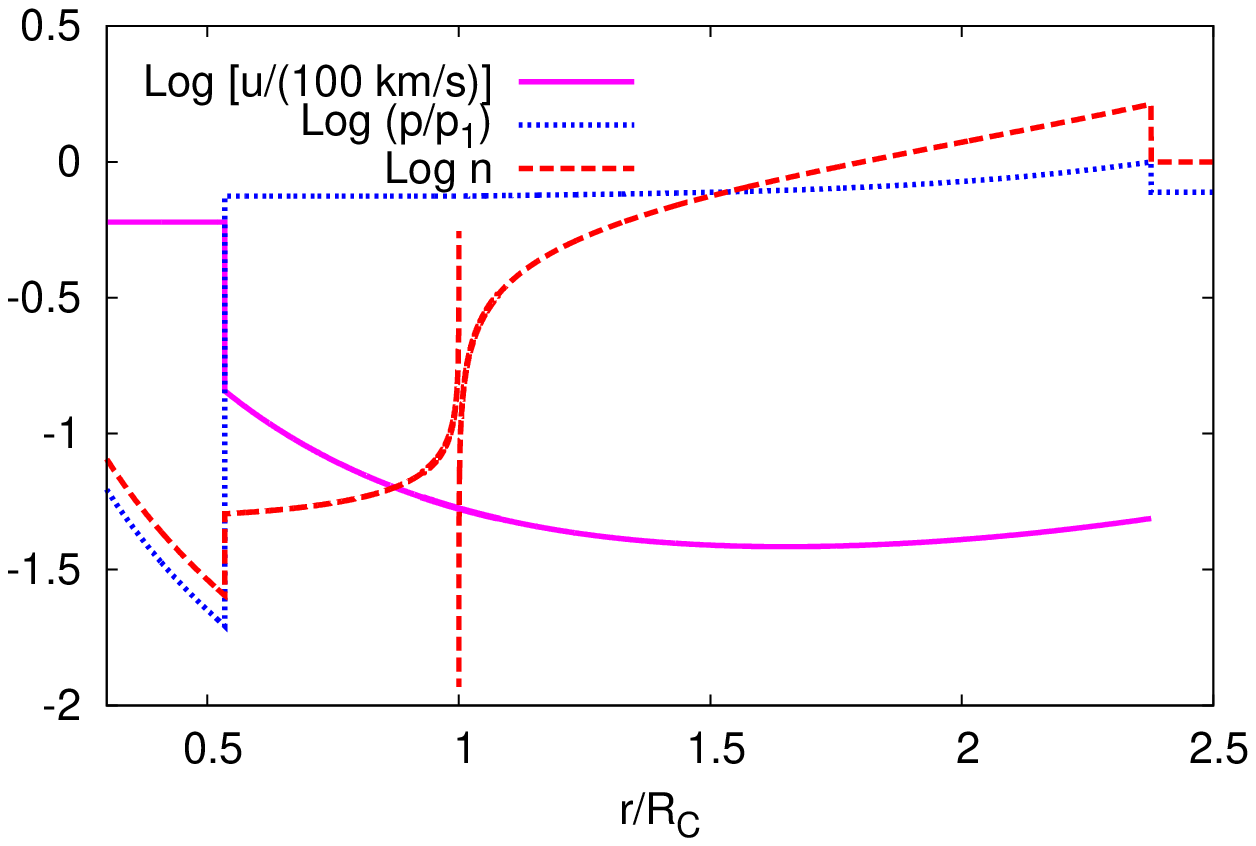} 
\end{array}$
\clearpage
\caption{Same as Fig. \ref{hydroplot1} but for a mixture of AD plus RG-like winds. Left and right are shown models no. 21 and 22 from Table \ref{s0_table1c}, respectively.}
\label{hydroplot3}
\end{figure}

\noindent
{\bf Case Id: Fast wind from an accretion disk combined with mass loss from RG donor star or super-Eddington accretion:}
If we have a system with both a dense RG donor wind and an accretion disk wind, the two will combine. The inner interaction 
region will be RT unstable and mix fast (Fig. \ref{hydroplot3}). They can be expected to form a uniform wind 
with an acceleration region. Assuming momentum conservation and our reference models for the AD wind 
and ``RG-like'' winds, we obtain a total mass loss rate of $\approx 10^{-5,-6,-7}~M_\odot$ and wind velocity of 
$\approx 33, 60 $ and $300~km/s$ (see models 21-23, Table \ref{s0_table1c}).

\begin{deluxetable*}{lcccccccc}
\tabletypesize{\small}
\tablewidth{0pc}
\tablecaption{Same as Table \ref{s0_table1a} but for the combination of an AD- and ``RG-like'' wind (see text). }
\tablehead{ 
\colhead{No} &
\colhead{$n_0$}      & 
\colhead{$v_w$} &
\colhead{$\dot{m}$} & 
\colhead{$t$} &
\colhead{$t_p$} &
\colhead{$R_{\rm{C}}$} & 
\colhead{$R_1$} &
\colhead{$R_2$} \\ 
\colhead{} &
\colhead{($cm^{-3}$)} &
\colhead{($km/s$)} &
\colhead{($M_\odot/yr$)} &
\colhead{($Myr$)} &
\colhead{($Myr$)} &
\colhead{($ly$)} &
\colhead{($ly$)} &
\colhead{($ly$)} 
}

\startdata
21 &    1.0  &    300  & \xtenexp{1.1}{ -7 } &    0.30 &    0.247  &   11.68  &     22.17  &    3.21   \\   
22 &    1.0  &    60   & \xtenexp{1.01}{ -6 } &    0.30 &    0.150 &    8.85  &     21.03  &    4.73   \\   
23 &    1.0  &    33   & \xtenexp{1.001}{ -5 } &    0.30 &    0.082 &    6.15  &     20.50  &    3.70   \vspace{0.2cm}\\
\hline \vspace{-0.40cm} \\  
\hline \vspace{-0.1cm} \\ 

No & $u_1$  & $u_c$  & $\sigma_u$ & $n_2$ & $\tau_{m}$ & $log(\tau)$ & $EW_{NaI}$ \\ 
 & ($km/s$)  & ($km/s$)  & ($km/s$) & ($cm^{-3}$) & ($g/cm^2$) & & ($m\mbox{\normalfont{\AA}}$) \vspace{0.1cm}\\

\hline \vspace{-0.1cm} \\
21 &    5.62  &    7.00  &    0.33   & \xtenexp{ 5.33 }{-3 } &   \xtenexp{1.71}{-5}  &    2.57   & 100 \\
22 &    4.88  &    5.30  &    0.33   & \xtenexp{ 1.17 }{-1 } &   \xtenexp{1.75}{-5}  &    2.59   & 100 \\
23 &    4.52  &    3.69  &    0.58   & \xtenexp{ 3.49 }{-1 } &   \xtenexp{1.83}{-5}  &    2.36   & 154 \vspace{0.0cm}\\

\enddata

\label{s0_table1c}
\end{deluxetable*}

\begin{deluxetable*}{lccccccccc}
\tablewidth{0pc}
\tabletypesize{\small}
\tablecaption{ Same as Table \ref{s0_table1a} but for a MS wind, including the ambient pressure (FAP models). 
 Our reference model is marked by $*$.
 Values are given for the radius of the contact discontinuity and the inner shock using parameters typical for the wind of a main sequence star, 
and using the solution including the ambient pressure (FAP models). For times (t) much larger than $t_p$, 
the location of the reversed shock becomes stationary and close to the distance $R_{2,\rm{PE}}$ expected
from equilibrating the pressure at the reversed shock with the ambient medium. 
 In addition, models without outer pressure are given and marked by ``Rem.''. This shows the importance of the
pressure term for $t/t_p \gg 1 $ in the extreme case. However, 
for winds relevant in SNe, the difference is of the order of several $\%$ (see text, and Fig. \ref{ComparePlots}).
}
  
\tablehead{ 
\colhead{No} &
\colhead{$n_0$}      & 
\colhead{$v_w$} &
\colhead{$\dot{m}$} & 
\colhead{$t$} &
\colhead{$t_p$} &
\colhead{$R_{\rm{C}}$} & 
\colhead{$R_2$} &
\colhead{$R_{2,\rm{PE}}$} \\ 
\colhead{} &
\colhead{($cm^{-3}$)} &
\colhead{($km/s$)} &
\colhead{($M_\odot/yr$)} &
\colhead{($Myr$)} &
\colhead{($yr$)} &
\colhead{($ly$)} &
\colhead{($ly$)} &
\colhead{($ly$)}
}                                          
 
\startdata
1      & 0.1 & 500 & \tenexp{-14} & 40  & 393   & 0.921   &  0.00477    & 0.00390  \\
~~Rem. & "   & "   & "            & "   & -     & 20.0    & 0.481       & \\
2*     & 1   & 500 & \tenexp{-14} & 40  & 124   & 0.427   &  0.00151    & 0.00123  \\
~~Rem. & "   & "   & "            & "   & -     & 12.6    & 0.241       &  \\
3      & 10  & 500 & \tenexp{-14} & 40  & 39.3  & 0.198   &  0.000477   & 0.000390 \\
~~Rem. & "   & "   & "            & "   & -     & 7.94    & 0.121       &  \\
4  & 1   & 500 & \tenexp{-13} & 40  & 393   & 0.921   &  0.00477  & 0.00390  \\
5  & 1   & 500 & \tenexp{-15} & 40  & 3.93  & 0.198   &  0.000477 & 0.00039  \\ 
6  & 1   & 500 & \tenexp{-14} & 2   & 124   & 0.157   &  0.00151  & 0.00123  \\
7  & 1   & 500 & \tenexp{-14} & 600 & 124   & 1.05    &  0.00151  & 0.00123  \\
8  & 0.1 & 500 & \tenexp{-14} & 0.3 & 393   & 0.180   &  0.00477  & 0.00390  \\
9  & 1   & 500 & \tenexp{-14} & 0.3 & 124   & 0.0836  &  0.00151  & 0.00123  \\
10 & 10  & 500 & \tenexp{-14} & 0.3 & 3.93  & 0.0388  &  0.000477 & 0.000390  \\
11 & 1   & 500 & \tenexp{-15} & 0.3 & 3.93  & 0.0388  &  0.000477 & 0.000390 \\ 
12 & 1   & 500 & \tenexp{-14} & 0.4 & 124   & 0.0921  &  0.00151  & 0.00123    

\enddata
\label{s0_table2}
\end{deluxetable*}

This case applies in system with high peculiar velocity. There,  the system has moved out of the environment
formed during the stellar evolution of the progenitor. The cavities are somewhat larger by about a factor of 2
 compared to the RG-like wind cavities, surrounded by an expanding shell of
a similar velocity and EW (NaID). The larger cavity will result in a slower evolution of the narrow lines.

\noindent
{\bf Extremely low Density Environments:}
In case of long delay times of SNe~Ia, the progenitor system may have
wandered away from the galactic disk into the halo, in a regime of very low densities \citet{Graham_Sand2015},
or in a hot-ISM \citet{fesen07}, or in the constant density region II of the cavity from prior mass loss. 
In Table \ref {s0_table3}, some examples are given for constant density with particle densities between $10^{-2}$ to $10^{-3}~cm^{-3}$.
Most likely, the donor star will not be a RG star. The inner, low density region is expected to 
be of the order of or larger than $50-100~ly$. EW(NaID) are between 20 to 100 $m\mbox{\normalfont{\AA}}$ and show 
increased  Doppler shifts of $\approx 30 $ to $50~km/s$ or larger for even lower density 
than found in the vicinity of SN1885 \citet{fesen07}. 
In conclusion, of all of wind components analyzed separately, the AD wind component will dominate the formation of cocoons.

\begin{deluxetable*}{lccccccccccccc}
\tabletypesize{\small}
\tablewidth{0pc}
\tablecaption{Same as tables \ref{s0_table1a} and \ref{s0_table2} but for very low density ISM 
typical for the galactic halo and elliptical galaxies.
}
\tablehead{ 
\colhead{No} &
\colhead{$n_0$}      & 
\colhead{$v_w$} &
\colhead{$\dot{m}$} & 
\colhead{$t$} &
\colhead{$t_p$} &
\colhead{$R_{\rm{C}}$} & 
\colhead{$R_1$} &
\colhead{$R_2$} \\ 
\colhead{} &
\colhead{($cm^{-3}$)} &
\colhead{($km/s$)} &
\colhead{($M_\odot/yr$)} &
\colhead{($Myr$)} &
\colhead{($Myr$)} &
\colhead{($ly$)} &
\colhead{($ly$)} &
\colhead{($ly$)} 
}                                  

\startdata
1 & 0.01  & 3000 & \tenexp{-8}  & 0.4 & 2.36    & 63.8  & 77.2  & 11.21  \\
2 & 0.001 & 3000 & \tenexp{-8}  & 0.4 & 23.6    & 102   & 121   & 22.7   \\
3 & 0.01  & 30   & \tenexp{-7}  & 50  & 0.236   & 70.8  & 3368  & 11.7   \\
4 & 0.001 & 30   & \tenexp{-7}  & 50  & 0.745   & 152   & 3377  & 36.9  \vspace{0.2cm} \\

\hline \vspace{-0.40cm} \\  
\hline \vspace{-0.1cm} \\ 

No & $u_1$  & $u_c$  & $\sigma_u$ & $n_2$ & $\tau_{m}$ & $log(\tau)$ & $EW_{NaI}$ \\ 
 & ($km/s$)  & ($km/s$)  & ($km/s$) & ($cm^{-3}$) & ($g/cm^2$) & & ($m\mbox{\normalfont{\AA}}$) \vspace{0.1cm}\\

\hline \vspace{-0.1cm} \\                                                    

1 & 24.36 & 28.67  &  1.08  &  \xtenexp{1.72}{-6}  &  \xtenexp{4.67}{-7}   &   0.49  & 104  \\
2 & 39.6  & 45.87  &  1.58  &  \xtenexp{4.21}{-7}  &  \xtenexp{7.18}{-8}   &  -0.48  & 24.0 \\
3 & 4.32  & 0.25   &  1.05  &  \xtenexp{3.62}{-3}  &  \xtenexp{3.29}{-5}   &   2.35  & 272  \\
4 & 4.36  & 0.55   &  1.04  &  \xtenexp{3.65}{-4}  &  \xtenexp{3.29}{-6}   &   1.36  & 193 \vspace{0.0cm}\\

\enddata
\label{s0_table3}
\end{deluxetable*}

\subsubsection{Case II: Environments produced by Winds}

Now we consider the scenario where the wind of the progenitor system runs into a nearby environment produced 
by a prior mass loss ($s=2$). ``Nearby'' means that the reverse shock $R_2$ produced by the prior mass loss must be larger than the size of the region produced by the wind of the system. Otherwise the interaction region will move into a region of constant density, shocked RG wind, or the ISM.
Moreover, the speed of the ongoing wind must exceed that of the outer wind.
For our discussion, we disregard the effect of prior mass loss by a MS wind because it produces on 
an cocoon of $< 0.005$ $ly$ Table \ref{s0_table2}  and consider parameters only in which the prior wind
can be produced by an RG star.

\begin{figure}
\includegraphics[width=0.48\textwidth]{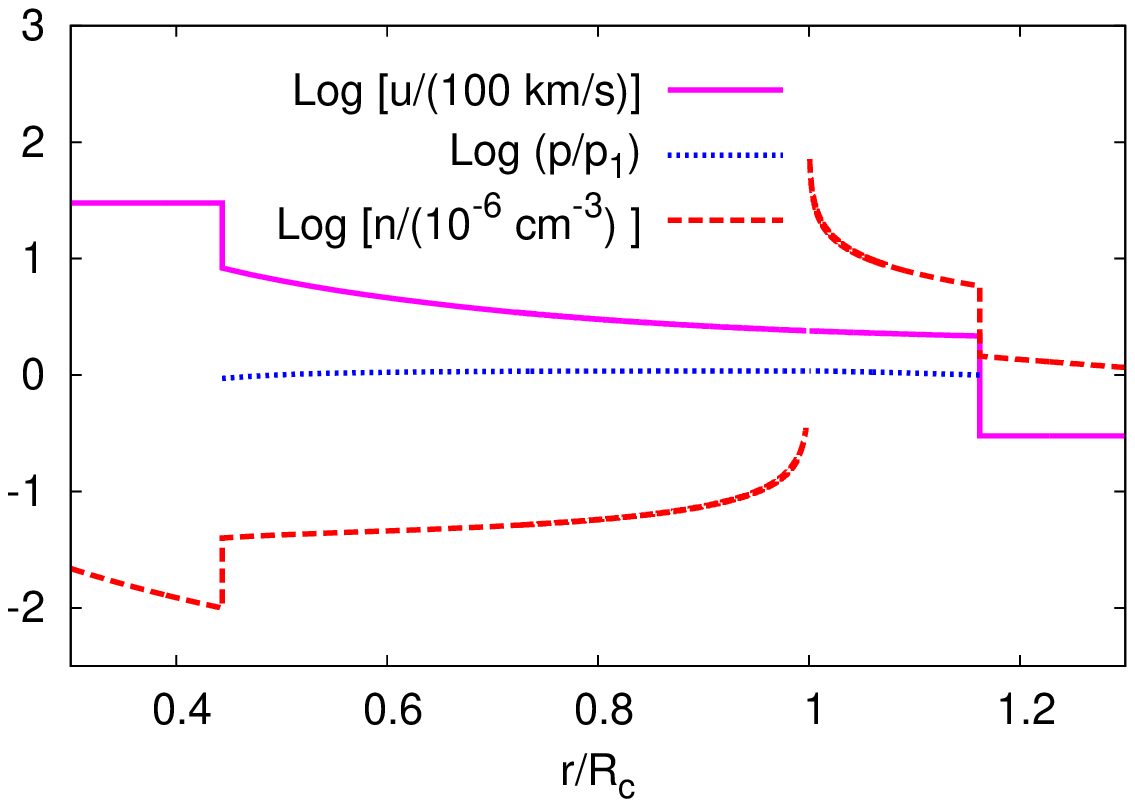}
\includegraphics[width=0.48\textwidth]{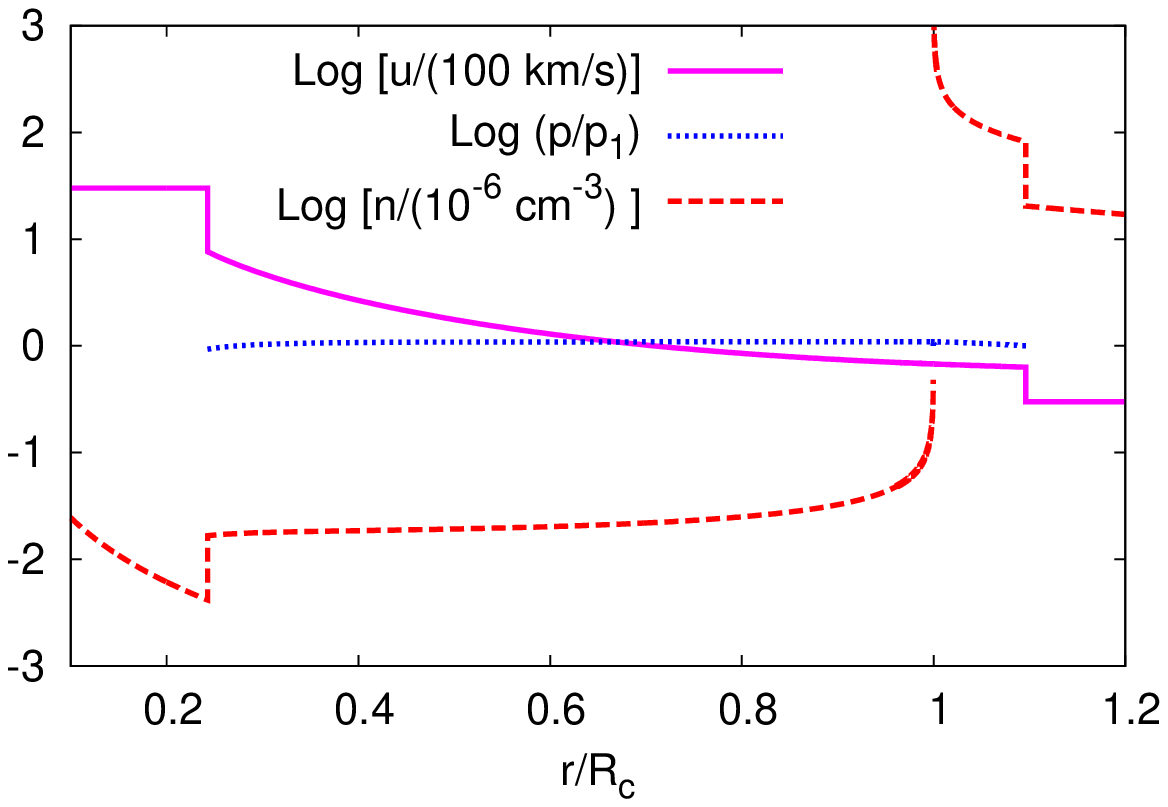}
\caption{Hydrodynamic profile for a wind typical of an accretion disk running into an $r^{-2}$ density profile of prior mass loss. 
  In the left figure, a wind of velocity $v_{w} = 3000~km/s$
 and mass-loss rate of $\dot{m} = 10^{-8}~M_\odot/yr$ runs
 into another wind of mass loss $\dot{m}_1=10^{-7}~M_\odot/yr$
 and $v_{w,1}=30~km/s$ for a time of 300,000 years. The right figure is the same except the ongoing mass loss rate 
 $\dot{m} = 10^{-10}~M_{\odot}/yr$.
  The contact discontinuity is at 240 and 67.7 light years, respectively. Fluid velocity $u$ (magenta) is normalized to 100 $km/s$, 
  pressure $p$ (blue) is normalized to the pressure $p_1$ just inside the outer shock, and particle density $n$ is normalized to $10^{-6}~cm^{-3}$. 
  Note that the particle density normalization used is much smaller than in the constant density case.}
\label{hydroplot2}
\end{figure}

\noindent
{\bf Case IIa: Fast wind from an accretion disk and a non-RG donor star:}
If the donor star is a compact object like a MS or He-star, the AD wind will dominate. Examples are
shown in  Table \ref{s2_table}. As reference, 
some typical parameters are  $\dot{m}_1= 10^{-6}$ $M_\odot/yr$,  $v_{w,1}=30$ $km/s$, $\dot{m}= 10^{-10}$ $ M_\odot/yr$, 
$v_{w}=3000$ $km/s$ and a total run time of  $1.5 \times 10^5$ years.
 Again, we can identify the different zones as above (Fig. \ref{hydroplot2}). The density contrast
between the inner  bubble within $R_{\rm{C}}$ is smaller than in the constant density case, i.e. a factor of $10^{-2}$, but
the density at $R_{\rm{C}}$ is significantly lower than the ISM. This results in an
an ``ultra-low'' density bubble of $n \approx 4 \times 10^{-8}~cm^{-3}$.

\begin{deluxetable*}{lccccccccccccc}
\tablewidth{0pc}
\tabletypesize{\small}
\tablecaption{Interaction of winds with an environment produced by RG wind originating from the progenitor WD
($v_{w,1}= 30$ $km/s$, $s=2$). The index $1$ corresponds to the prior mass loss.
Models 1-7 and 8-9 show the results for an AD-wind and a 
combination of a AD and ``RG-like'' wind, respectively (Tables \ref{s0_table1a} \& \ref{s0_table1c}).}
\tablehead{ 
\colhead{No} &
\colhead{$v_{w,1}$} & 
\colhead{$\dot{m}_1$} & 
\colhead{$v_{w}$} & 
\colhead{$\dot{m}$} & 
\colhead{$t$} & 
\colhead{$R_{\rm{C}}$} &
\colhead{$R_1$} & 
\colhead{$R_2$} & \\          
\colhead{} &
\colhead{($km/s$)} &
\colhead{($M_\odot/yr$)} &
\colhead{($km/s$)} &
\colhead{($M_\odot/yr$)} &
\colhead{($Myr$)} &
\colhead{($ly$)} &
\colhead{($ly$)} &
\colhead{($ly$)} &
}

\startdata
AD &&&&&&&&\\
1  & 30 & \tenexp{-6}  & 3000 & \tenexp{-9}  & 0.3  & 67.7  & 74.2  & 16.4  \\
2  & 30 & \tenexp{-7}  & 3000 & \tenexp{-9}  & 0.3  & 122   & 138   & 39.3  \\
3  & 30 & \tenexp{-8}  & 3000 & \tenexp{-9}  & 0.3  & 240   & 278   & 106   \\
4* & 30 & \tenexp{-7}  & 3000 & \tenexp{-8}  & 0.3  & 240   & 278   & 106   \\
5  & 30 & \tenexp{-7}  & 3000 & \tenexp{-10} & 0.3  & 67.7  & 74.2  & 16.4  \\
6  & 30 & \tenexp{-7}  & 3000 & \tenexp{-9}  & 0.15 & 60.8  & 69.1  & 19.6  \\
7  & 30 & \tenexp{-7}  & 3000 & \tenexp{-9}  & 0.4  & 162   & 184   & 52.4  \\
AD+RG-like &&&&&&&&\\
8  & 30 & \tenexp{-7}  & 300 &  \xtenexp{1.1}{-7}  & 0.3  & 114   & 129   & 91.8  \\
9  & 30 & \tenexp{-7}  & 60 & \xtenexp{1.01}{-6} & 0.3  & 51.3   & 54.9  & 50.0  \vspace{0.2cm}\\

\hline \vspace{-0.40cm} \\
\hline \vspace{-0.1cm} \\

No & \colhead{$u_c$} & \colhead{$n_2$} & \colhead{$\tau_{m}$} & \colhead{$log(\tau)$} & \colhead{$EW_{NaI}$} \\          
 & \colhead{($km/s$)} & \colhead{($cm^{-3}$)} & \colhead{($g/cm^2$)} & \colhead{} & \colhead{($m\mbox{\normalfont{\AA}}$)} \vspace{0.1cm} \\

\hline \vspace{-0.1cm} \\
AD &&&&&&&&\\
1  & 67.7  & \xtenexp{1.7}{-7}   & \xtenexp{1.59}{-8}  & -2.02 & 5.63  \\
2  & 122   & \xtenexp{2.9}{-8}   & \xtenexp{1.18}{-9}  & -3.34 & 0.420  \\
3  & 239   & \xtenexp{4.0}{-9}   & \xtenexp{6.96}{-11} & -4.80 & 0.0248  \\
4* & 239   & \xtenexp{4.0}{-8}   & \xtenexp{6.96}{-10} & -3.80 & 0.248  \\
5  & 67.7  & \xtenexp{1.7}{-8}   & \xtenexp{1.59}{-9}  & -3.02 & 0.566  \\
6  & 122   & \xtenexp{1.2}{-7}   & \xtenexp{2.36}{-9}  & -3.02 & 0.837  \\
7  & 122   & \xtenexp{1.6}{-8}   & \xtenexp{8.86}{-10} & -3.45 & 0.315  \\
AD+RG-like &&&&&&&&\\
8  & 114   & \xtenexp{5.9}{-6}   & \xtenexp{1.59}{-9}  & -2.92 & 0.565  \\
9  & 51.3  & \xtenexp{9.1}{-4}   & \xtenexp{5.05}{-9}  & -1.86 & 1.79 \vspace{0.0cm}\\

\enddata
\label{s2_table}
\end{deluxetable*}

Moreover, this solution shows a qualitative different feature compared to the constant density ISM:
We see a very  thin and dense shell with particle densities $\leq 10^2~cm^{-3}$ and a thickness of 
light-weeks to light-months (Fig. \ref{hydroplot2}). The resulting shell produces a narrow, optically
thick NaID line with small equivalent width EW of 0.24 $m\mbox{\normalfont{\AA}}$, Doppler shifted by about $u_c = 240~km/s$ 
and a width of $\approx 23~km/s$.  We note that, to first order, EW $\propto \dot{m}$ 
(model 1, Table \ref{s2_table}).

\begin{figure}
\includegraphics[width=0.48\textwidth]{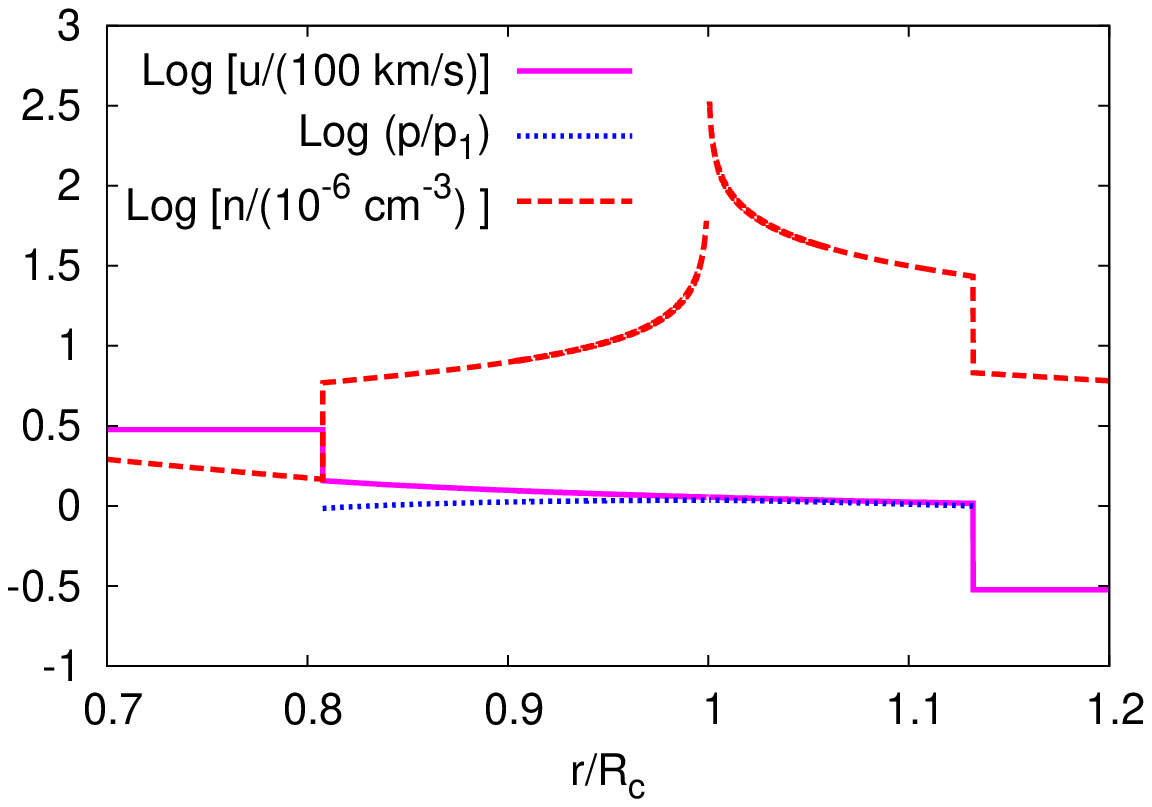}
\includegraphics[width=0.48\textwidth]{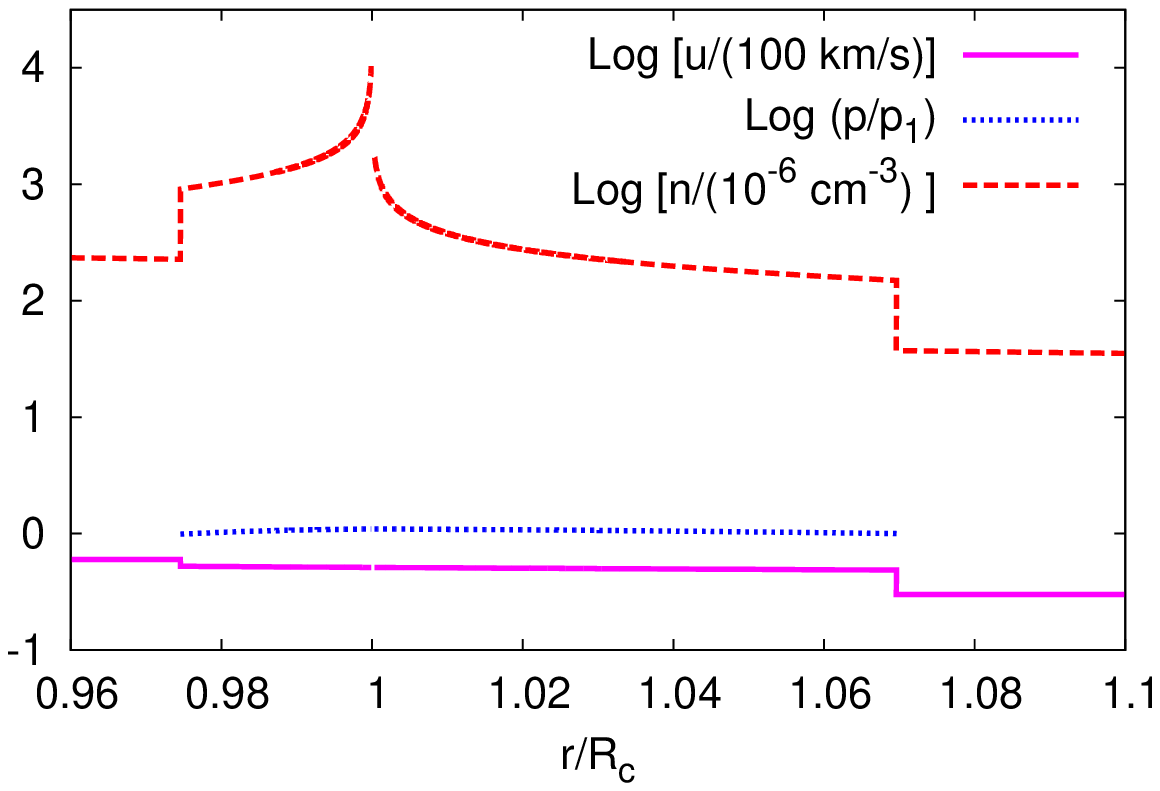}
\caption{Same as Fig. \ref{hydroplot2} but for a mixture of AD plus RG-like winds. Left and right are shown models 8 and 9 from Table \ref{s2_table}, respectively.}
\label{hydroplot4}
\end{figure}

\noindent
{\bf Case IIb: Fast wind from an accretion disk combines with a ``RG-like'' wind:}
If we have a system with a dense RG donor and wind plus an accretion disk 
wind, they would form a uniform wind with an acceleration region (Fig. \ref{hydroplot4}). We use the same parameters
as in Sect. \ref{CID}  but omit the mixed wind with the highest mass loss because it's velocity
will be comparable to the wind velocity of the surrounding medium. As expected, 
the cavities are smaller than the AD-wind case, and EW(NaID) and its Doppler shift are larger. 

Comparing interaction regions of environments created by winds with constant density ISM, the 
typical scales are larger in the former and the resulting narrow lines
show a larger blue-shift with smaller EW.

\subsection{Application to SN 2014J}\label{SN 2014J}
We will now explore the application of our semi-analytic models to the case of SN 2014J.

\noindent
{\bf Background:} This object was discovered in January of 2014 in a high density region of the nearby M82. 
The higher than average ISM densities and the small proximity of the host make SN 2014J an excellent candidate for investigating the interaction of the 
SN with the environment. This gives us a premier opportunity to implement our models and make predictions. 

Observational constraints for the environment and progenitor are obtained from searches for X-rays \citep{Margutti}, 
time-dependence of high-resolution spectra 
of narrow potassium lines formed in the environment\citep{Graham}, IR imaging \citep{Kelly2014},
 and radio \citep{Perez-Torres}. X-rays and radio provided the most stringent constraints on the
average density of ions in the environment. 
In the case of SN~2014J, X-rays and radio luminosities at maximium light  
probe the wind from the progenitor system (region I in Fig. \ref{example}). 

When the SN material propagates into the circumstellar surrounding, a shock is formed 
and leads to the acceleration of partially or fully relativistic electrons with a power-law distribution
$n_e(\gamma)= n_0 \gamma^{-p}$ with $p$ between 2 and 3, which produce X-ray and radio emission.
 From radio observations of SNe~Ib/c, \citet{CF06} find $p\approx 3$.
Because the outer layers of a SNe~Ia and SN~Ibc have similar structure, abundances and velocities,
we use this value in the following. 

\noindent
For X-rays \citep{xraystudy}, $L_\nu$ is given by

\begin{equation} 
L_\nu = 16500\, \epsilon_{\rm{e}}^2  \left(\frac{\dot m/v_w}
{M_\odot\,yr^{-1}\,km^{-1}\,s}\right)^{0.64} \left(\frac{t}{d}\right)^{-1.36}\nu^{-1}
\left(\frac{L_{\rm{bol}}}{erg\,s^{-1}}\right)\frac{erg}{s\,Hz}.
\label{mdvwx}
\end{equation}


\noindent
For the radio \citep{Perez-Torres}, $L_\nu$ is given by

\begin{equation} 
L_\nu = 5.81 \times 10^{-9}\, \epsilon_{\rm{e}}^{1.71}\,
\epsilon_{\rm{B}}^{1.07} \left(\frac{\dot m/v_w}{M_\odot\,yr^{-1}\,km^{-1}\,s}
\right)^{1.37} \left(\frac{t}{d}\right)^{-1.55}T_{\rm{bright}}
~\nu^{-1}\frac{erg}{s\,Hz}.
\label{mdvwr}
\end{equation}


\noindent
The parameters are the fraction of relativistic electrons $\epsilon_{\rm{e}}$, 
the energy fraction in the magnetic field $\epsilon_{\rm{B}}$, and the brightness temperature
$T_{\rm{bright}}$ which, based on observations, can be expected to be $\approx 10^{11} K$
\citep{readhead94}. \citet{Margutti} and \citet{Perez-Torres} use a value  of 0.1 for
$\epsilon_{\rm{e}}$ and $\epsilon_{\rm{B}}$.    

If the supernova shell runs into a constant density environment,
the X-ray luminosity is  given by

\begin{equation}
L_\nu = 6.44 \times 10^{-5}\, \epsilon_{e}^2 \left(\frac{n_0}
{cm^{-3}}\right)^{0.5} \left(\frac{t}{d}\right)^{-0.45}\nu^{-1}
 \left(\frac{L_{\rm{bol}}}{erg\,s^{-1}}\right)\frac{erg}{s\,Hz}.
\label{s0xray}
\end{equation}


For late times when the SN shock runs into region II (Fig.  \ref{example}), 
the radio luminosity is given by

\begin{equation}
L_\nu \propto T_{\rm{bright}}\,\epsilon_{\rm{e}}^{0.86}\,
\epsilon_{\rm{B}}^{1.07}\,n_{\rm{ISM}}^{1.27}\,t^{0.88}.
\label{s0radio}
\end{equation}

We note that region II has a constant density. Thus, we expect the radio luminosity to increase
with time. Densities in our models of region II (see Fig. \ref{example}) 
are two to four orders of magnitude smaller than limits discussed above. However, radio must be expected
when the SN shockfront hits the shell after 10 to 100 years after the explosion.

In the case of SN 2014J running into a constant density medium and from radio observations, \citet{Perez-Torres}  found an upper limit $n_0 \lesssim 1.3 $ particles
per $cm^{3}$. The same luminosity can be produced in a wind with $\dot{m} \lesssim 7.0 \times 10^{-10} \times v_w/(100~km/s)$ $M_\odot/yr$. 
The neutral lines and the IR emission indicate shells at distances
between 10 and 20 ly expanding at velocities between $120-140~km/s$. 
\citet{Margutti} and \citet{Perez-Torres} concluded from their X-ray and radio observations that a DD system was the likely 
progenitor. However the findings of \citep{Graham,Graham} led them to favor a SD progenitor, although 
excluding a Red Giant as donor.

In light of the different conclusions, we will apply our analytic description to explore the wide range of parameters
within the observational limits from the radio- and X-rays. For the outer environment, we consider both an interstellar medium
with constant density, s=0, and one consisting of a wind, s=2, produced during the stellar evolution history.

The X-ray and radio observations provide limits on the far-inside region (region I) within the contact discontinuity (Figs. \ref{hydroplot1} and \ref{hydroplot2}).
In order to apply this constraint to our parameter space we will consider our three cases: RG-like wind, MS star wind, 
and wind originating from the accretion disk. Case I: If the environment is produced by an ongoing RG-like wind with a velocity
of $30~km/s$, the mass loss would have to be $\dot{m} \lesssim $ $2.1 \times 10^{-10}~M_\odot/yr$ which is much too low for RG wind. 
Ongoing RG-like wind from the donor star is also very unlikely to be consistent with
the IR imaging \citep{Kelly2014}. If the environment is formed by a RG-like wind, it must originate from the progenitor
star prior to the formation of the WD.
Case II: If the wind originates from a MS star with $500$ $km/s$, radio limits would allow 
 $\dot{m} \lesssim $ $3.5 \times 10^{-9}~M_\odot/yr$.
Case III: For an AD wind with  $3000~km/s$,  this means $\dot{m} \lesssim $ $2.1 \times 10^{-8}~M_\odot/yr$, well
within limits discussed in the introduction.

Narrow circumstellar lines have also been used to probe the environment close 
to the contact discontinuity (e.g. Fig. \ref{example}).
Using ionization models, \citet{Graham} attributed the blueshifted K~I and NaI~D absorption features to circumstellar shells at 
$\sim 10 - 20~ly$ at velocities of $\sim 120 - 140~km/s$. A further reduction of the parameter space
comes from the limits on X-rays and radio.  Using our models, we can infer what kind 
of progenitor wind (and system) could have produced such shells expanding at the proper velocity and distance
with the duration of the wind as a free parameter.

\noindent
{\bf Analysis of SN 2014J:}
 In the following, we want to apply our analysis by combining the tools of the last section
with the observations of SN 2014J. For finding allowed parameters, we use the given distance and velocity of the shells
in combination with wind parameters for the AD-, MS-, RG-like and ``RG-like plus AD'' winds.

\begin{figure}
\begin{center}
$\begin{array}{cc}
\includegraphics[width=0.45\textwidth]{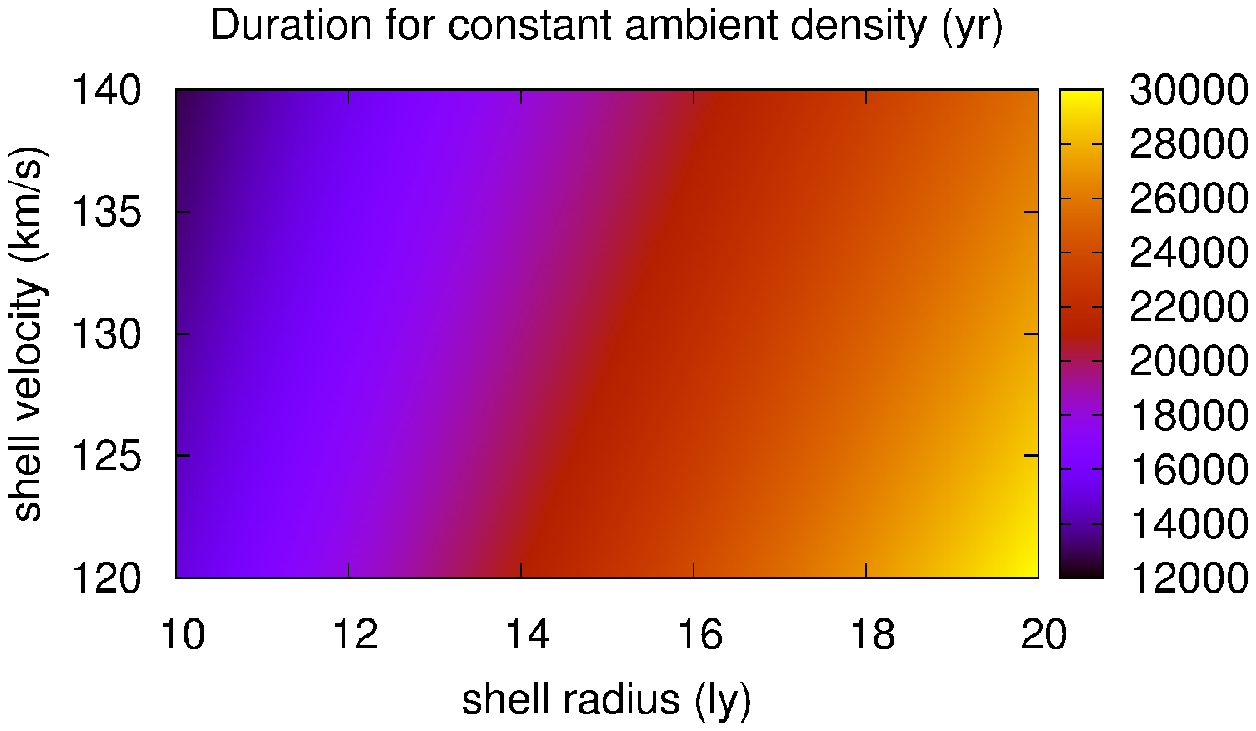} &
\includegraphics[width=0.45\textwidth]{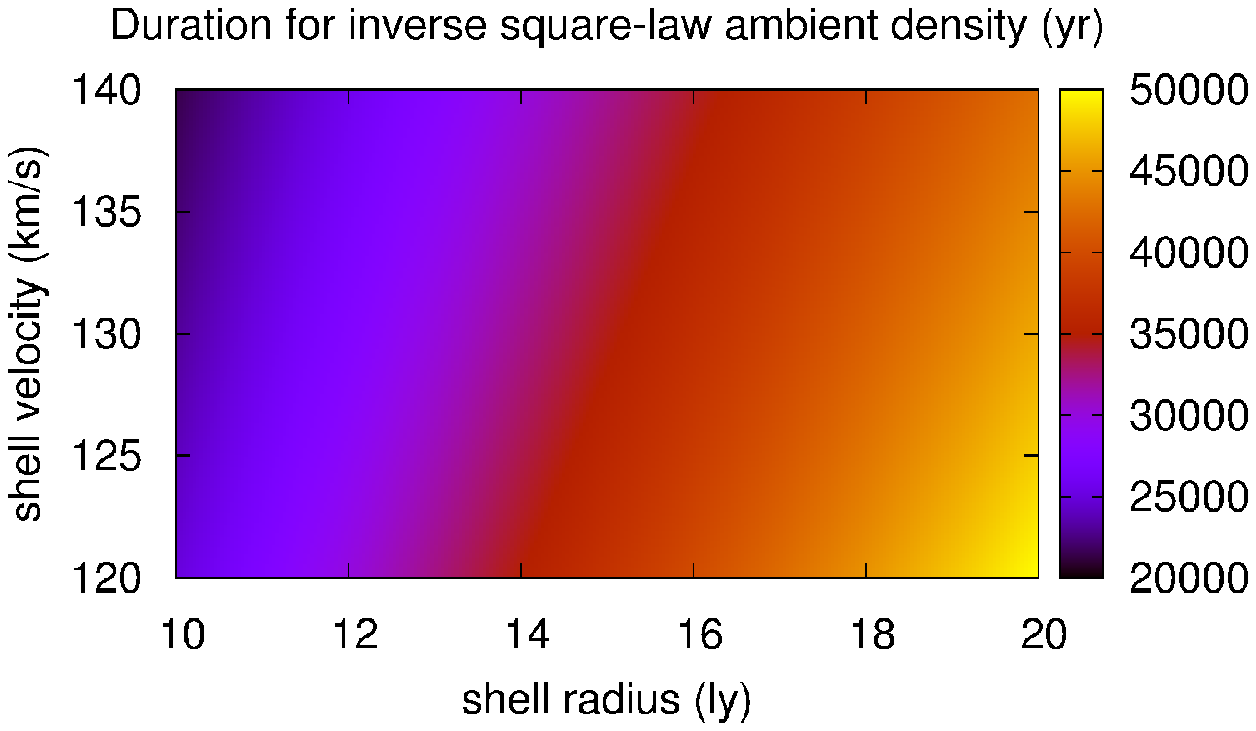}
\end{array}$
\caption{Allowed duration of the wind for SN 2014J as a function of $R_c$ and $\dot{R_c}$, for 
constant density environment, s=0 (left), and wind environment, $s=2$ (right).}
\label{times}
\end{center}
\end{figure}

1.) As a first step, we want to determine the allowed duration of the wind $t$ 
using the shell velocities and distance as given by \citep{Graham}. 
 Assuming the shells are at $R_{\rm{C}}$ and the expansion velocity at $\dot{R_{\rm{C}}}$, we can use equation  
8 to estimate the allowed range for the duration of the wind (Fig. \ref{times}).

2.) In a second step, we make use of the $\Pi$-theorem to determine combinations of $n_0$, $\dot{m}$ and $v_w$. 
Because the solutions depend on one and two $\Pi$ groups for constant ISM and wind-environments, respectively, 
we have to consider two cases.
  
\begin{figure}
\includegraphics[width=1\textwidth]{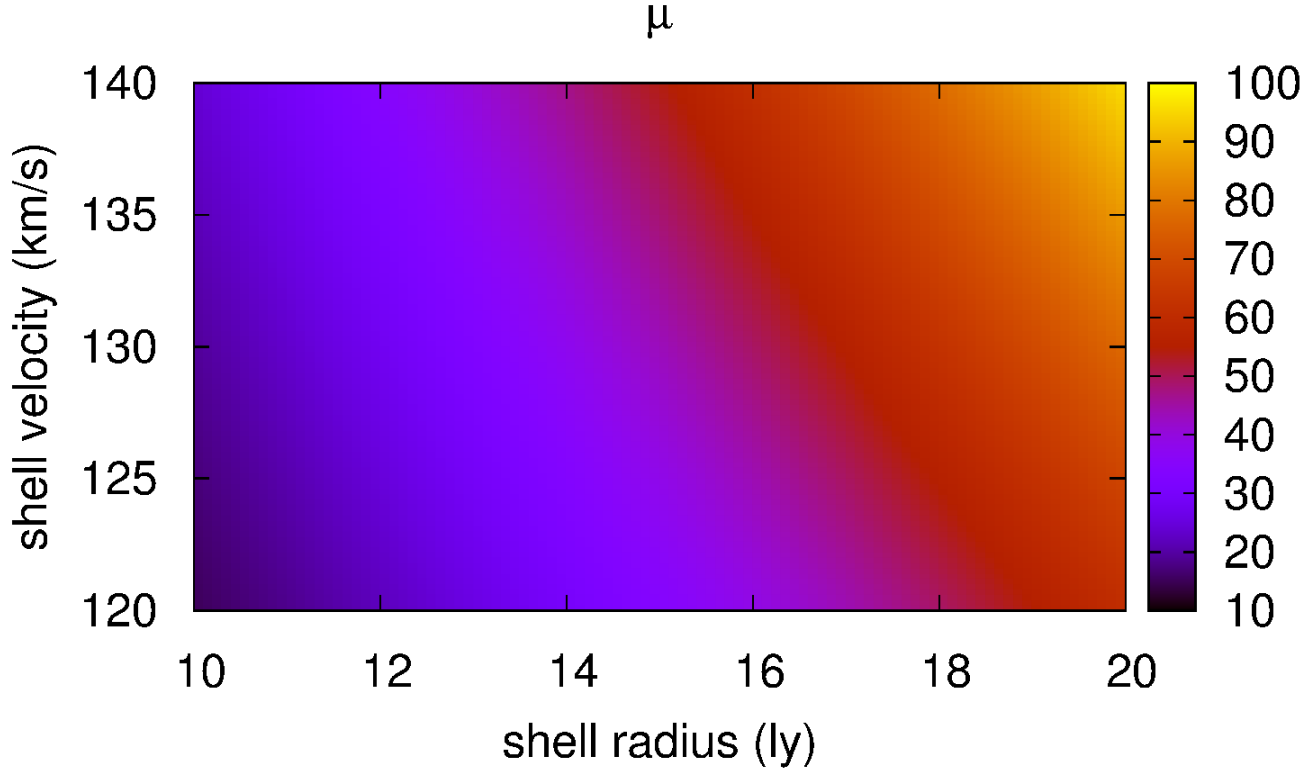}
\caption{Determination of $\mu = \dot{m}{v_w}^2/n_0$ by observations.
$\mu $ is given as a function of the velocity of the shell and the shell radius in the range
given by \citep{Graham}. We assume particle densities of the ISM  between 0.1 and 10 $cm^{-3}$.
We find a range of possible solutions $\mu({SN 2014J})$  between $15.11$ and $95.85$ in units of 
$(M_{\odot}/yr)(km^2/s^2)\div(cm^{-3})$.
Here, the range of $\mu$ is given by the extrema in shell velocity and radius. \vspace{0.25cm}}
\label{mu_lookup}
\end{figure}

\begin{figure}
\includegraphics[width=1\textwidth]{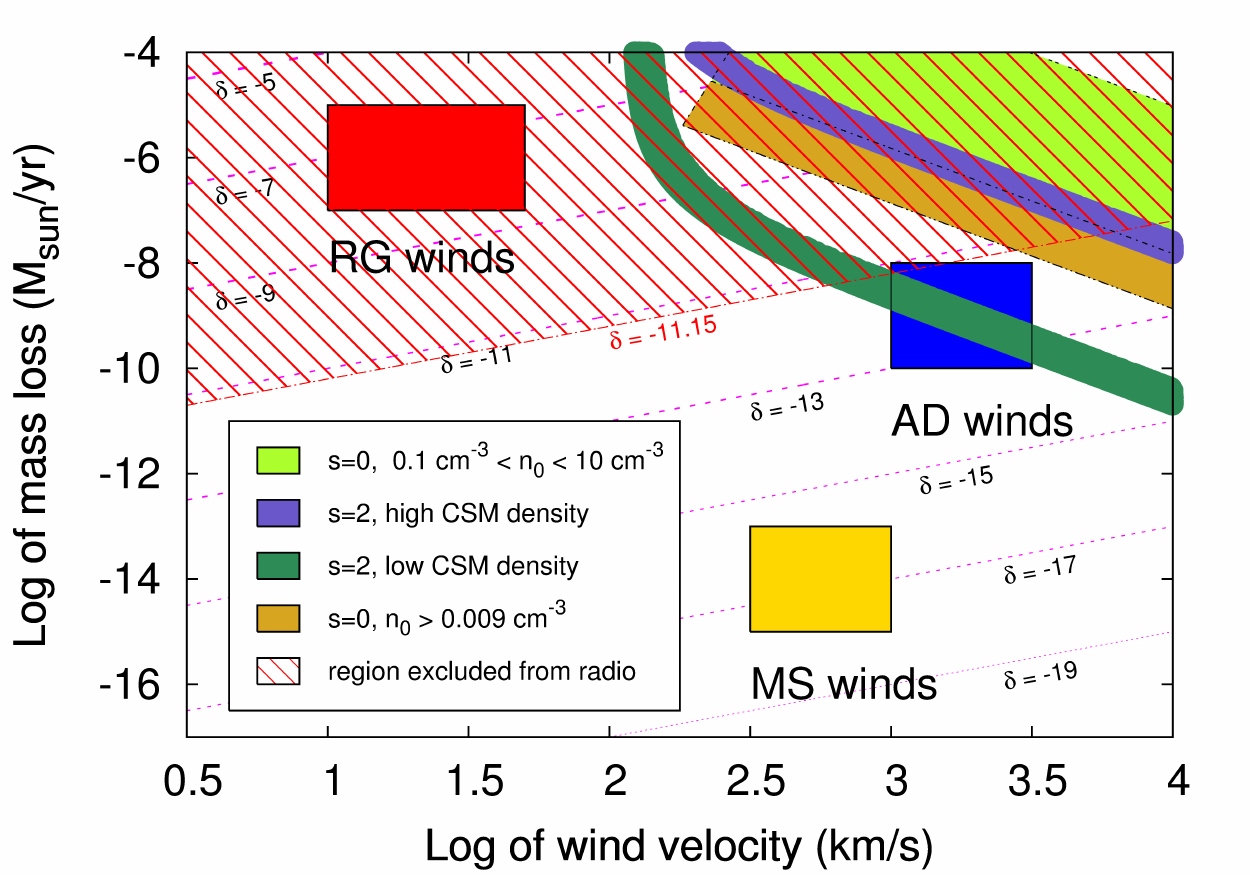}
\caption{{
Parameters for SN~2014J in the $\dot{m}$-$v_w$-space
as constrained from the spectra and the X-ray and radio luminosities.
We give the lines of constant luminosities $L_\nu$ for $\delta :=log_{10}(\dot{m}/v_w)$ which
can be obtained by eqs. 51 and 52.
 The red, striped area is the forbidden region based on radio and X-ray observations of SN 2014J ($\delta \gtrsim -11.15$).
 Note that the radio limit from 2011fe gives $\delta \gtrsim -11.22$ \citep{Chomiuk11fe}.
The bright green area indicates the allowed range for SN 2014J for constant density
environments with particles densities between 0.1
and 10 $cm^{-3}$. A RG wind prior to the formation of the WD may produce a
$r^{-2}$ power law (region I) and a pre-existing constant density surrounding (region II) 
as low as $0.009~cm^{-3}$ (see Sect. \ref{SN 2014J} and Fig. \ref{example}).
 The thick, curved lines correspond progenitor winds running into region I of a prior RG 
wind with $(\dot{m}/v_w)$ in ($M_\odot/yr)/(km/s)$ of ($10^{-7}/60$) and ($10^{-5}/20$) which
produce low and high CSM, respectively. These lines bracket the range of possible prior RG winds.  
Their widths indicate the range of solutions of possible AD or RG-like progenitor winds (see text).
 The boxes indicate the range observed of ongoing RG, MS and AD winds as discussed in
Chapter \ref{Current Status}. Note that the variations in wind properties are not
well known.} \vspace{0.25cm}
}
\label{PS_allowed}
\end{figure}

Case I: For the constant density environment (s=0), the solutions can be described in the 
$\Pi$ space by a unique parameter $\mu = \frac{\dot{m}v_w^2}{n_0}$. We we have to find $\mu $ from the observations and 
step one as described above (Fig. \ref{mu_lookup}). Assuming a particle density of the ISM between 0.1 and 10 $cm^{-3}$, 
 the allowed range for $s=0$ in the $\dot{M}$-$v_w$-space 
is shown in Fig. \ref{PS_allowed}. Note that the allowed range has linear boundaries because the exponential dependence of the free parameter (see Table 2). 
The minimum and maximum of $\mu $ are found by the values in the lower left and upper right of Fig. \ref{mu_lookup}, respectively. 
In Fig. \ref{PS_allowed}, the allowed range of $\mu$ in the $\dot{m}-v_w$ space is shown in bright green. 
A constant density may also be produced in region II of prior winds (Fig. \ref{example}). 

Case II: For ISM environments produced by a wind (s=2, region I), the allowed solutions depend on two $\Pi$ groups, $\Pi_{\dot{m}}$ 
and $\Pi_{v_w}$ (eq. \ref{Pi_m} and \ref{Pi_v}) which relate the mass loss rates and velocities of the interacting winds. 
Possible solutions are linear in time (eq. 8). Thus, $K_{2\rm{C}}= R_{\rm{C}}/(v_{w} t) = \dot{R_{\rm{C}}}/v_{w}$.
Possible solutions depend only on the shell velocity which can be measured directly from the spectra \citet{Graham}. 
Note that the velocity $\dot R_{\rm{C}}$ equals the matter velocity. Because it varies slowly in this region (see Fig. \ref{hydroplot2}), 
the Doppler shift of lines is a good measure of $\dot{R_{\rm{C}}}$. Knowing $v_{w}$, the original wind
velocity is given for any $\Pi_{v_w}$. $\Pi_{\dot{m}}$ gives the ratio between the ongoing and the outer
wind mass loss rates. We assume that the progenitor wind has the parameters given by a RG star as discussed in Sect. \ref{Current Status}. To solve for the implicit condition imposed given by the observations and possible model solutions,
a standard rejection technique is applied.
 We use a Monte Carlo Scheme to sample the parameter space of the ongoing wind for 
acceptable solutions for $v_{w,1}$. For a given wind parameter of the outer wind, we test each position in Fig.
 \ref{Rc_lookup2} (or the lookup table) for whether the resulting wind velocity $v_{w,1}$ is consistent with an RG wind.
If it is, we accept the possible solution. The lines of solutions found in this manner are shown in Fig. \ref{PS_allowed}. 
The line widths are given by the range of RG wind properties accepted to be consistent with a solution. 

3.) In a third step, finally, we compare the allowed range in parameter space with the ranges consistent with our three cases.
The ranges of $\dot{m}$ and $v_w$ for each type of wind is presented as a box.

A comparison between the predicted ranges for various winds and the allowed range for SN 2014J
constrains the possible progenitor properties. In addition, we marked the parameter range excluded
from the lack of X-ray and radio detections.  
The possible wind parameters for SN~2014J and the location of 
typical RG, MS and AD winds are shown in Fig. \ref{PS_allowed}. In the figure, we use the
quantity $\delta \equiv log_{10}(\dot{m}/v_w)$ as an abbreviation.
 In addition, 
we show the allowed region for an environment produced by an RG wind of the progenitor
prior to the WD formation.

 For constant ISM environments (Case I), there is obviously no overlap between regions allowed from the observations
and the parameter space of solutions. Constant particle density environments
ranging from 0.1 to 10 $cm^{-3}$ can be ruled out. However, a very low density ISM
of $\lesssim 0.01~cm^{-3}$ is consistent. 

 We find two other possible solutions both of which require an environment formed by 
a prior RG-phase of the progenitor (Case II). In either case, we will need a short delay
time between the formation of the WD and the explosion.

As discussed above, a wind from the progenitor prior to the WD phase will 
produce an inner structure with a density close to $\propto r^{-2}$ (region I, Fig. 
\ref{example}), and an outer cocoon of constant, low density (s=0, region II) 
which is lower than the ISM by about an order of magnitude. The lower limit for
the particle density in region II is $\approx 0.009~cm^{-3}$ which widens the 
regime of solutions for s=0.

 The AD wind may either interact in the $r^{-2}$ or constant density region.
In either case, we can exclude RG or RG-like winds from the progenitor system. 
A RG star as  the donor star can be ruled out, too, by the X-rays and radio. 
This finding is also consistent with the lack of a RG remnant at the location of 
SN~2014J. 

Within the scenario of $M_{Ch} $ explosions, AD winds are consistent with 
both interaction in region I and II. AD winds are also supported by the high
Doppler shifts of the narrow lines reported by \citep{Kelly2014}. 

For further constraints, we want to discuss the formation of the necessary wind environment formed 
from prior mass loss which created the RG-cocoon to, at least, 10 to 20 ly.
As discussed above, we use $v_w = 20$ to $60~km/s$, mass loss rates between $10^{-5}$ and $10^{-7}~km/s$, and particle densities
$n_0 $ between 0.1 and 10 $cm^{-3}$ for the initial ISM. Using the relation for
$R_{\rm{C}}$ of Table \ref{eqntable1} and the reference model for normalization, we obtain
for the duration needed for the wind: 
$t_{RG}= const\,[ \,R_{\rm{C}}^{5} \,n_0 \,v_w^{-2} \,\dot{m}^{-1}\, ]^{1/3}$.

\noindent
We obtain wind durations between $1.3 \times 10^4 $  and $1.4 \times 10^6 $ $yr$ prior to the explosion.
This provides an upper limit for the duration of the AD disk wind and hence the accretion.
Within the picture of an $M_{Ch}$ mass explosion starting from an $0.6$  to $1.1 M_\odot$ C/O WD,
some $0.7...0.25 M_\odot$ must be accreted. As discussed above, hydrogen is only stable between
$2 \times 10^{-8...-6} M_\odot /yr$. Thus, only the long time
scale is consistent with a hydrogen accreter. This makes it more likely that the donor was a  
He star or a C/O accretion disk. Assuming an ISM density of 0.1 to 1 $cm^{-3}$, between
0.5 and 10 $M_\odot $ would have been swept up by the ISM.
We note that, with this constraint, the region within $R_2$, with $s=2$ is much too small in all our 
models (see Table \ref{s0_table1b}). The most likely solution is therefore an AD wind interaction within 
region II. 

 For SN 2014J, the following picture emerges as our most likely solution:
An enviroment is produced during the RG phase of the WD progenitor, consisting of a low density
cavity surrounded by a shell, or a very low density ISM which would obviously not be expected for 
the starburst galaxy M82. Wind from the 
accretion disk (AD-wind) runs into this environment and produces a shell 
structure responsible for the variable narrow lines observed. 

 Consistent solutions can be obtained if the AD wind interacts either with
region I (case A, s=2) or region II (case B, s=0) of the prior RG wind. 
This wind runs into an ISM of $0.03$ and $0.1~cm^{-3}$, respectively.
The RG winds have a mass loss rate, wind velocity and duration which are $8 \times 10^{-7}~M_\odot/yr$,
 $30~km/s$, $10^6~yr$; and $10^{-7}~M_\odot/yr$, $60~km/s$
 and $2 \times 10^6~yr$, respectively. For case A, the location of the reversed
 shock is larger than 20 ly. For case B, the densities in region
II are about $0.01~cm^{-3}$. Obviously, a lower ISM density by a factor
of 10 is a possible solution without the need for an RG wind. 

 The most likely solutions for case A and B is an AD wind with ($\dot m
[M_\odot/yr],v_w[km/s]$) of ($10^{-8},3000$) and ( $1.5 \times 10^{-8},5000$) with a duration
 $t$ of $35000~yr$ and $20000~yr$, respectively. 
 The inner corresponding void contains $\approx 3.5 \times 10^{-4}~M_\odot$
and $\approx 2.9 \times 10^{-4}~M_\odot$ of material at a density of 
$\approx 4 \times 10^{-5}$ and $  3 \times 10^{-5}~cm^{-3}$. 
The corresponding density is $\sim 4 \times 10^{-5}~cm^{-3}$, 
the outer shell width is $\approx 2~ly$ and its density is $\sim 0.02~cm^{-3}$. 
For case B, the shell has a thickness of about $2~ly$ and a density of $\approx 0.03~cm^{-3}$. 
For case A, a thin shell is formed (Fig. \ref{example}) with densities decreasing from $10 $ to
$0.1~cm^{-3}$ over some $0.015~ly$. The equivalent width for Na I D is 
$EW = 27~m\mbox{\normalfont{\AA}}$ with a velocity dispersion of $14~km/s$ for case A, and 
$ 27~m\mbox{\normalfont{\AA}}$ and $4.2~km/s$ for case B. 

{Other analyses of SN 2014J include that of \citet{Soker14J} who concluded that the core-degenerate 
model alone could account for the observations discussed. 
Obviously, winds of CDs are a solution among many which may be more favorable 
because the consistency with other observational properties of SN2014J.} 

{In this context, it may be interesting to apply our method to SN2011fe,  one
of the best observed SN~Ia at a distance of ($\approx 6.4~Mpc$) \citep{nug_11fe_11}. 
This is another nearby, well-observed, very normal SN~Ia.
Similar to SN 2014J, environmental density
constraints have been set for SN2011fe using X-rays by \citet{Margutti11fe} and
radio by \citet{Chomiuk11fe}. They find upper limits, for a wind CSM environment, of
$\dot{m} \lesssim 6 \times 10^{-10} \times v_w/(100~km/s)~M_\odot/yr$. For
an AD wind of $3000~km/s$ this limit corresponds to $\dot{m} \lesssim 1.8
\times 10^{-8}~M_\odot/yr$.
These parameters are consistent with the wind from an accretion disk with a SD progenitors 
as discussed above.
For constant CSM environment, \citet{Chomiuk11fe} found $n_0 \lesssim 6~cm^{-3}$ which
is above typical density found in our models within the bubble.

  Unlike for SN 2014J, no time varying narrow lines have been observed for this event \citep{Patat11fe}, which prevents us
from specifying shell velocities or distances in conducting our analysis.
The narrow lines may either attributed to the ISM or distances significantly larger
than in SN 2014J which remains unaffected by the light front 
(see e.g. Fig. 5 in \citet{Graham}).   
If there is a shell produced in a CSM, the bubble must be larger than 
$\gtrsim 40~ly$. For typical AD wind parameters and a constant density environment of 
$1~cm^{-3}$, this requires durations of $t \gtrsim 10^6~yr$ or an environment produced by 
prior mass loss and a duration larger than in SN02014J by a factor of 2 to 4.}

\section{Discussion and Conclusions}\label{Discussion}

We presented theoretical, semi-analytic models for the interaction of stellar winds with the ISM 
and its implementation in our code SPICE \footnote{The code SPICE can be obtained by request}.  We assumed  spherical symmetry and power-law 
ambient density profiles.   Our free parameters are: the
a) mass loss $\dot{m}$, b) wind velocity $v_w$, c) the particle density $n_0$ of the in case of a constant density ISM or the mass loss and 
wind velocity for environments produced by a prior stellar wind, 
and d) the duration of the wind from the progenitor system.

Our approach provides an efficient method to study a wide range of parameters well beyond what is feasible with complex numerical simulations.
 Using the $\Pi$-theorem allows us to test a wide variety of configurations, properties of the solutions along with their sensitivity and  
 dependencies on the wind and environment parameters.  Using these dependencies, we showed how to use observations to find possible solutions.
The formalism presented and SPICE may be used for a wide variety of objects, including stellar winds. The speed of the semi-analytic approach produces solutions with low computational overhead. This allows us to evaluate a large parameter space for individual objects, and to include realistic feedback by many objects in star-formation and galactic large-scale simulations. 

Here, the formalism has been applied to study SNe~Ia and to constrain the properties of progenitor systems. 
As discussed in the introduction, SNe~Ia may originate from a wide range of scenarios and progenitor channels which
lead to an ongoing discussion about the nature of these objects, and their diversity.  Most of the channels are not well 
understood and, thus, our range may be wider than realized in nature.

We studied a variety of winds within the scenario of $M_{Ch}$ explosions. The winds may originate from  
the accretion disk, MS and RG donor star, and over-Eddington accretion of H/He rich matter within this scenario.

The environment considered may be produced by the ISM or may be produced by 
winds during the stages of the evolution of the progenitor prior to becoming a WD for both double or single degenerate systems.  
 Within the $M_{Ch}$ scenarios, we studied wind from the accretion disk, MS and RG donor stars,
and over-Eddington accretion of H/He rich matter within parameters suggested from literature as discussed in Sect. \ref{Current Status}.

Within the $M_{Ch} $ explosions, we find that the wind from the accretion disk dominates the environment, or the combined wind from an RG donor 
and the AD. Such wind leads to a low-density ``cocoon'' of the order of light years in size. The actual size depends on the duration of the progenitor evolution.  
The calculations reveal that these cocoons are characterized by interior regions with particle densities often as low as $10^{-4} cm^{-3}$ and which are surrounded by a thin shell. This
 explains why most SNe~Ia appear to explode in low density environment, although SN~Ia are observed in the galactic
halo, the disk and the bulge. 
The lack of ongoing interaction in SNe~Ia may well be understood in the framework of $M_{Ch}$ mass explosions whether
they originate from accretion from a MS, RG, He-star or a tidally disrupted WD.
 If the wind of the progenitor system interacts with a constant, interstellar medium, we expect narrow lines produced by the shell. 
Our calculations show shell velocities ranging from 10 to 100 $km/s$.
 For the $s=0$ models, the narrow lines are expected to have an equivalent width of  $\approx 100 m\mbox{\normalfont{\AA}}$ and are Doppler shifted  
 by about 10 to 20 $km/s$ (see Tables \ref{s0_table1a}, \ref{s0_table1b} \& \ref{s0_table1c}). 
 In contrast, an environment dominated by a prior stellar wind will result in weaker lines with EW lower by one to two orders 
of magnitude, i.e. $0.5$ to $5~m\mbox{\normalfont{\AA}}$ (see Table \ref{s0_table2}), which is beyond current observational limits for most SNe~Ia.
 For small cocoons, the narrow lines may show variations in strength and velocity on time scales of months due to the radiation from a SNe~Ia.
Radiation pressure of the SNe~Ia light may accelerate nearby shells (as seen in, e.g. SN1993J).

 As a separate effect, the SN ejecta may interact with the shell as discussed in Sect. 1. The outer layers of 
a SNe~Ia expands with velocities of 10 to 30 \% of the speed of light and we may expect some interaction 
on  time scales between years and several decades (see Tables \ref{s0_table1a}, \ref{s0_table1c} \& \ref{s0_table2}).
For more details, see \citet{hoeflich2013,paul}. Note the possible implications for SN-remnants and their evolution.

We studied the effect of winds in ultra-low density environments.
 If there is a long delay time between the formation of the WD and the explosion, the progenitor system may have  
moved into a low density environment. Within $M_{Ch}$ mass explosions, the size of the ``cocoon'' will be larger by
an order of magnitude and we would neither expect narrow lines nor interaction (Sect. 3.2.1).
  In general, dynamical mergers are expected to often explode in low density environments due to their long delay times but, also, 
  are unlikely to produce a long-duration wind. Thus, some of these objects are expected to show ongoing interaction for sufficiently large samples of SNe~Ia.

A step-by-step scheme has been developed to use the narrow lines for the analysis of SNe~Ia progenitors.  We applied our models to SN 2014J using both the limits from the X-ray and radio, and the observation
of narrow lines (Sect. \ref{SN 2014J}), and discussed the allowed range of progenitor properties using the analytic relations (Fig. \ref{PS_allowed}).  
We found the observations to be consistent with an environment produced by a stellar wind.  
We applied both our analytic solutions to produce the environment and the wind from the progenitor system.
We require an environment which has been created recently with times between $1.3 \times 10^4$  and $1.4 \times 10^6~yr$ prior to the explosion.
Within the picture of an $M_{Ch}$ explosion starting from an initial $0.6$  to $1.1~M_\odot$ C/O WD,
some $0.7...0.25~M_\odot$ must be accreted. As discussed in the reviews quoted in the introduction, hydrogen is only stable between
$2 \times 10^{-8...-6}~M_\odot /yr$. Thus, only the upper limit for the evolutionary times 
is consistent with a hydrogen accreter. This makes it more likely that the donor was a  
He star or a C/O accretion disk as a result of a tidally disrupted WD in a DD system. 
{ As discussed in the introduction, He-triggered or double degenerate explosions may show winds similar to AD and the $M_{Ch}$ models. However,
to build up a sufficiently large a He-layer seems to require low accretion rates inconsistent with the short delay times between the RG phase 
and the explosion. We note that a future generation of double degenerate models may void this argument.} 

 Our analysis of SN 2014J is consistent with other constraints: No RG donor star has been found \citep{Kelly2014},
 optical to MIR light curves and spectra are consistent with an Branch-normal $M_{Ch} $ mass explosion and $0.6 M_\odot $ of $^{56}\rm{Ni}$
\citep{Marion2015,Telesco2015,Churazov2015,Diamond2015,Isern2015}, and the lack of polarization \citep{patat2015}.  

 However, within an increasing number of well observed SNe~Ia, it also becomes increasingly obvious that SNe~Ia are not 
 ``all the same''. Our conclusion on SN~2014J should not be generalized.
Better observation of narrow line systems are a key to the environment of SNe~Ia and the diversity.

 Within the class of single degenerate  systems we found configurations for SN~2014J  which are consistent
 with both the limits from X-rays and radio and the narrow lines. 
Our solutions invoke a wind from the accretion disk running into an unusually low density ISM environment, $\lesssim 0.01~cm^{-3}$, or 
a low density cavity created by a RG wind some $20,000$ to $35,000~yr$ prior to the explosion (see Sect. \ref{SN 2014J}).
 
We want to put our findings in context of the possible $H_\alpha$ emission discussed in the introduction.
Its detectability depends on the amount of hydrogen, the mechanism of ionization, and 
the sensitivity and timing of the observations.  Possible mechanisms include hard radiation from  the shock breakout, $^{56}\rm{Ni}$-decay,
surface burning of the WD prior to the explosion, interaction in the forward shock with the CSM, and the reversed shock 
as a result of the interaction between SN and the CSM. The results are very model dependent with various regimes.
Based on ionization calculations and the deflagration model W7 \citep{Thielemoto84} , \citet{Cumming} studied in detail pre-ionization of the 
progenitor winds by supersoft X-ray sources \citep{greiner91,rap94,kah97} and the $H_\alpha$ emission in the outer supernovae ejecta ionized by 
the reversed  shock interacting with the progenitor wind. They parameterized their study in terms of $\dot M/v_w~[{M}_\odot \,yr^{-1}\,km^{-1}\,s]$ with 
$v_w$ being the progenitor wind velocity. \citet{Cumming} put their results into context for the observability  
of $H_\alpha$ emission for local SNe~Ia at distances similar to SN~2014J. For $\dot M /v_{w} \lesssim 1.9 \times 10^{-8}$,
supersoft X-ray sources dominate the reversed shock mechanism. 
 We note that the actual flux from supersoft X-ray sources may be significantly lower because 
 of material in the progenitor system \citep{Gerardy}. The $H_\alpha$ emission in the reversed shock region depends 
 on the ionization fraction of hydrogen.  \citet{Cumming} found the ionization fraction less than 0.01, at which the emission becomes inefficient
 for  $\lesssim \dot{M}/v_{w}=1.5 \times10^{-8}$, and full ionization for $\dot{M}/v_{w} = 1.5 \times 10^{-6}$.
  For local supernovae with a reversed shock in a H-rich region, $H_\alpha $ should be observable in high quality spectra
for a RG wind with  $ \dot{M} \sim 10^{-5}~M_\odot/yr$ and $v_w=10~km/s$. Note that the result of \citet{Cumming} is consistent with the    
finding of \citet{Gerardy}. Using the observed light curves, \citet{Gerardy} put strong 
limits on the ongoing interaction. Without a strong interaction, there is no reversed shock or too weak of one to ionize hydrogen. 
 We note that the dominance of the ionization mechanism will depend on the class of explosion. For example, excitation by $^{56}\rm{Ni}$-decay or
the shock breakout can be expected to be small for deflagration models such as W7 or pulsating delayed-detonation models with a significant 
C/O shell reducing the $\gamma$ emission, however the role of $^{56}\rm{Ni}$ may become important in double-detonation/HeD explosions 
which show some $^{56}\rm{Ni}$ in the outer layers and may produce He-lines instead \citep{hk96}. 
 For the progenitor wind and environment we found for SN~2014J, we do not expect early, observational $H_\alpha$  emission.
The mass loss from ADs is smaller by some 3 to 4 orders of magnitude compared to the limits by \citet{Cumming} for local SNe~Ia,
 and the forward shock runs into low density material of some  $10^{-3...-4} M_\odot$ in the cavity or the ISM.  

 However, we may be able to see late-time, narrow $H_\alpha$ emission if the the supernova ejecta runs into the shell of the cavity.
Due to the low masses of the cavity, the SN ejecta will remain largely unmodified as it travels through the void and produce 
$H_\alpha$ emission on impact by both the forward and the reversed shock if we have H in the outer layers of the SN.  
The velocities of the outer layers are about  $1/3~c$ and, for SN~2014J, we may expect emission in about 50~yr.
 The impact will be earlier for higher density, more compact shells or shells with inner clumps. While we wait to see 
hydrodynamic interaction for SN 2014J, we could observe young SN remnants for this signature. One such example is the Branch-normal SNe~Ia SN1972E in 
NGC 5253, $3.3~Mpc$ away \citep{kirshner75}.

Here, our method has been applied to SNe~Ia but studies of other types of stellar explosions, 
namely core-collapse SNe, and SNe interacting with Wolf-Rayet star mass loss, are under way. SPICE may also be used in modeling other hydrodynamic phenomena and and has the potential to include feedback from stellar winds 
on a subgrid scale in starformation and large scale galactic evolution simulations.

 This brings us also to the limit of our analysis.  Although our analytic models provide a practical tool for individual shells or shells from well separated phases of mass loss. In reality, 
nature may be more complicated. The mass loss can be expected to be time dependent. 
If an environment is formed by this wind, it will deviate from an $r^{-2}$ law.
 The mass loss may come in phases, namely brief periods of ``superwinds'' during the RG phase.
Moreover, mass loss from the progenitor system may change over time. 
 In fact, multiple narrow lines have been observed e.g. for SN~2014J \citep{Graham}. 
 However, the narrow  lines may be produced both in the vicinity of the SN or at any distance by unrelated clouds in the ISM.  Without time 
 evolution of the narrow lines or measurements of dust components, the origin of the systems of narrow lines remain unclear. 
  Obviously, detailed calculations for spectra and light curves
are needed to quantify the intensity of the narrow lines.

Moreover, multi-dimensional effects need to be considered for a full analysis, for example in instabilities
and mixing, non-symmetric winds, and motion of the progenitor system through the ambient medium. 
A combination of semi-analytic solutions provide a good starting point for more complex, multi-D calculations
with more detailed physics, which are under way, to be presented in forthcoming papers. What role and influence dust may have, especially in the outer 
layers, is another question deserving close attention.

 In light of the wide range of progenitor scenarios and properties of the resulting environment, our approach has been justified in order to find the right 
 ballpark in the vast sea of parameters. In reality, however, 
 multi-dimensional affects such as asymmetric winds will become important, including variable winds.
Moreover, proper cooling functions and equation of states for the gas are to be taken into account for 
detailed analysis of high quality data such as SN 2014J.

\acknowledgments 
\noindent{\bf Acknowledgments:} 
We would like to thank many colleges, collaborators and the referee for helpful discussions and comments.
 The work presented in this paper has been supported by
the NSF projects AST-1008962,
 ``Interaction of Type Ia Supernovae with their Environment'', and
AST-0708855, ``Three-Dimensional Simulations of Type Ia Supernovae: Constraining
Models with Observations''. In parts, the results presented have been obtained 
as part of the PhD thesis of Paul Dragulin at Florida State University.

\bibliography{MS.bbl}

\begin{thebibliography}{146}
\expandafter\ifx\csname natexlab\endcsname\relax\def\natexlab#1{#1}\fi


\bibitem[{{Aldering} {et~al.}(2006){Aldering}, {Antilogus}, {Bailey}, {Baltay},
  {Bauer}, {Blanc}, {Bongard}, {Copin}, {Gangler}, {Gilles}, {Kessler},
  {Kocevski}, {Lee}, {Loken}, {Nugent}, {Pain}, {P{\'e}contal}, {Pereira},
  {Perlmutter}, {Rabinowitz}, {Rigaudier}, {Scalzo}, {Smadja}, {Thomas},
  {Wang}, {Weaver}, \& {Nearby Supernova Factory}}]{aldering06}
{Aldering}, G., {Antilogus}, P., {Bailey}, S., {et~al.} 2006, \apj, 650, 510

\bibitem[{{Blondin} {et~al.}(2009){Blondin}, {Prieto}, {Patat}, {Challis},
  {Hicken}, {Kirshner}, {Matheson}, \& {Modjaz}}]{Blondin09}
{Blondin}, S., {Prieto}, J.~L., {Patat}, F., {et~al.} 2009, \apj, 693, 207

\bibitem[{{Bours} {et~al.}(2013){Bours}, {Toonen}, \& {Nelemans}}]{Bours13}
{Bours}, M.~C.~P., {Toonen}, S., \& {Nelemans}, G. 2013, \aap, 552, A24

\bibitem[{{Branch} {et~al.}(1995){Branch}, {Livio}, {Yungelson}, {Boffi}, \&
  {Baron}}]{Branch1995}
{Branch}, D., {Livio}, M., {Yungelson}, L.~R., {Boffi}, F.~R., \& {Baron}, E.
  1995, \pasp, 107, 1019

\bibitem[{{Buckingham}(1914)}]{Buckingham1914}
{Buckingham}, E. 1914, Physical Review, 4, 345

\bibitem[{{Burns} {et~al.}(2014){Burns}, {Stritzinger}, {Phillips}, {Hsiao},
  {Contreras}, {Persson}, {Folatelli}, {Boldt}, {Campillay}, {Castell{\'o}n},
  {Freedman}, {Madore}, {Morrell}, {Salgado}, \& {Suntzeff}}]{burns2014}
{Burns}, C.~R., {Stritzinger}, M., {Phillips}, M.~M., {et~al.} 2014, \apj, 789,
  32

\bibitem[{{Cardelli} {et~al.}(1989){Cardelli}, {Clayton}, \&
  {Mathis}}]{cardelli89}
{Cardelli}, J.~A., {Clayton}, G.~C., \& {Mathis}, J.~S. 1989, \apj, 345, 245

\bibitem[{{Chandra} {et~al.}(2012){Chandra}, {Chevalier}, {Irwin}, {Chugai},
  {Fransson}, \& {Soderberg}}]{Chandra2012}
{Chandra}, P., {Chevalier}, R.~A., {Irwin}, C.~M., {et~al.} 2012, \apjl, 750,
  L2

\bibitem[{{Chevalier}(1982)}]{Chevalier}
{Chevalier}, R.~A. 1982, \apj, 258, 790

\bibitem[{{Chevalier} \& {Fransson}(2006)}]{CF06}
{Chevalier}, R.~A., \& {Fransson}, C. 2006, \apj, 651, 381

\bibitem[{{Chevalier} \& {Imamura}(1983)}]{Chevalier&Imamura}
{Chevalier}, R.~A., \& {Imamura}, J.~N. 1983, \apj, 270, 554

\bibitem[{{Chevalier} \& {Irwin}(2012)}]{Chevalier2012}
{Chevalier}, R.~A., \& {Irwin}, C.~M. 2012, \apjl, 747, L17

\bibitem[{{Chieffi} {et~al.}(2001){Chieffi}, {Dom{\'{\i}}nguez}, {Limongi}, \&
  {Straniero}}]{Chieffi2001}
{Chieffi}, A., {Dom{\'{\i}}nguez}, I., {Limongi}, M., \& {Straniero}, O. 2001,
  \apj, 554, 1159

\bibitem[{{Chomiuk} {et~al.}(2012{\natexlab{a}}){Chomiuk}, {Soderberg}, {Moe},
  {Chevalier}, {Rupen}, {Badenes}, {Margutti}, {Fransson}, {Fong}, \&
  {Dittmann}}]{Chomiuk12}
{Chomiuk}, L., {Soderberg}, A.~M., {Moe}, M., {et~al.} 2012{\natexlab{a}},
  \apj, 750, 164

\bibitem[{{Chomiuk} {et~al.}(2012{\natexlab{b}}){Chomiuk}, {Soderberg}, {Moe},
  {Chevalier}, {Rupen}, {Badenes}, {Margutti}, {Fransson}, {Fong}, \&
  {Dittmann}}]{Chomiuk11fe}
---. 2012{\natexlab{b}}, \apj, 750, 164

\bibitem[{{Chugai}(1986)}]{Chugai86}
{Chugai}, N.~N. 1986, \sovast, 30, 563

\bibitem[{{Chugai} \& {Danziger}(1994)}]{Chugai1994}
{Chugai}, N.~N., \& {Danziger}, I.~J. 1994, \mnras, 268, 173

\bibitem[{{Churazov} {et~al.}(2014){Churazov}, {Sunyaev}, {Isern},
  {Kn{\"o}dlseder}, {Jean}, {Lebrun}, {Chugai}, {Grebenev}, {Bravo}, {Sazonov},
  \& {Renaud}}]{Churazov2015}
{Churazov}, E., {Sunyaev}, R., {Isern}, J., {et~al.} 2014, \nat, 512, 406

\bibitem[{{Crotts} \& {Yourdon}(2008)}]{crotts08}
{Crotts}, A.~P.~S., \& {Yourdon}, D. 2008, \apj, 689, 1186

\bibitem[{{Cumming} {et~al.}(1996){Cumming}, {Lundqvist}, {Smith}, {Pettini},
  \& {King}}]{Cumming}
{Cumming}, R.~J., {Lundqvist}, P., {Smith}, L.~J., {Pettini}, M., \& {King},
  D.~L. 1996, \mnras, 283, 1355

\bibitem[{{Di Stefano} \& {Kilic}(2012)}]{Di_Stefano2012}
{Di Stefano}, R., \& {Kilic}, M. 2012, \apj, 759, 56

\bibitem[{{Di Stefano} {et~al.}(2011){Di Stefano}, {Voss}, \&
  {Claeys}}]{Di_Stefano2011}
{Di Stefano}, R., {Voss}, R., \& {Claeys}, J.~S.~W. 2011, \apjl, 738, L1

\bibitem[{{Diamond} {et~al.}(2014){Diamond}, {Hoeflich}, \&
  {Gerardy}}]{Diamond2015}
{Diamond}, T., {Hoeflich}, P., \& {Gerardy}, C.~L. 2014, ArXiv e-prints

\bibitem[{{D'Odorico} {et~al.}(1989){D'Odorico}, {di Serego Alighieri},
  {Pettini}, {Magain}, {Nissen}, \& {Panagia}}]{Odorico89}
{D'Odorico}, S., {di Serego Alighieri}, S., {Pettini}, M., {et~al.} 1989, \aap,
  215, 21

\bibitem[{{Dragulin}(2015)}]{paul}
{Dragulin}, P. 2015, PhD thesis, Florida State University, Tallahassee

\bibitem[{{Draine}(2011)}]{Draine2011}
{Draine}, B.~T. 2011, Physics of the Interstellar and Intergalactic Medium

\bibitem[{{Dwarkadas} \& {Chevalier}(1998)}]{Dwarkadas1998}
{Dwarkadas}, V.~V., \& {Chevalier}, R.~A. 1998, \apj, 497, 807

\bibitem[{{Eggleton}(1983)}]{eggleton83}
{Eggleton}, P.~P. 1983, \apj, 268, 368

\bibitem[{{Elias-Rosa} {et~al.}(2006){Elias-Rosa}, {Benetti}, {Cappellaro},
  {Turatto}, {Mazzali}, {Patat}, {Meikle}, {Stehle}, {Pastorello}, {Pignata},
  {Kotak}, {Harutyunyan}, {Altavilla}, {Navasardyan}, {Qiu}, {Salvo}, \&
  {Hillebrandt}}]{elias66}
{Elias-Rosa}, N., {Benetti}, S., {Cappellaro}, E., {et~al.} 2006, \mnras, 369,
  1880

\bibitem[{{Feldman} {et~al.}(2005){Feldman}, {Landi}, \&
  {Schwadron}}]{feldman05}
{Feldman}, U., {Landi}, E., \& {Schwadron}, N.~A. 2005, Journal of Geophysical
  Research (Space Physics), 110, 7109

\bibitem[{Ferri\`ere(2001)}]{Ferri01}
Ferri\`ere, K.~M. 2001, Rev. Mod. Phys., 73, 1031

\bibitem[{{Fesen} {et~al.}(2007){Fesen}, {H{\o}flich}, {Hamilton}, {Hammell},
  {Gerardy}, {Khokhlov}, \& {Wheeler}}]{fesen07}
{Fesen}, R.~A., {H{\o}flich}, P.~A., {Hamilton}, A.~J.~S., {et~al.} 2007, ApJ,
  658, 396

\bibitem[{{Fisher}(2000)}]{fisher00}
{Fisher}, A.~K. 2000, PhD thesis, THE UNIVERSITY OF OKLAHOMA

\bibitem[{{Folatelli} {et~al.}(2010){Folatelli}, {Phillips}, {Burns},
  {Contreras}, {Hamuy}, {Freedman}, {Persson}, {Stritzinger}, {Suntzeff},
  {Krisciunas}, {Boldt}, {Gonz{\'a}lez}, {Krzeminski}, {Morrell}, {Roth},
  {Salgado}, {Madore}, {Murphy}, {Wyatt}, {Li}, {Filippenko}, \&
  {Miller}}]{folatelli2014}
{Folatelli}, G., {Phillips}, M.~M., {Burns}, C.~R., {et~al.} 2010, \aj, 139,
  120

\bibitem[{{Folatelli} {et~al.}(2013){Folatelli}, {Morrell}, {Phillips},
  {Hsiao}, {Campillay}, {Contreras}, {Castell{\'o}n}, {Hamuy}, {Krzeminski},
  {Roth}, {Stritzinger}, {Burns}, {Freedman}, {Madore}, {Murphy}, {Persson},
  {Prieto}, {Suntzeff}, {Krisciunas}, {Anderson}, {F{\o}rster}, {Maza},
  {Pignata}, {Rojas}, {Boldt}, {Salgado}, {Wyatt}, {Olivares E.}, {Gal-Yam}, \&
  {Sako}}]{folatelli2013a}
{Folatelli}, G., {Morrell}, N., {Phillips}, M.~M., {et~al.} 2013, \apj, 773, 53

\bibitem[{{Foley} {et~al.}(2012){Foley}, {Simon}, {Burns}, {Gal-Yam}, {Hamuy},
  {Kirshner}, {Morrell}, {Phillips}, {Shields}, \& {Sternberg}}]{Foley11}
{Foley}, R.~J., {Simon}, J.~D., {Burns}, C.~R., {et~al.} 2012, \apj, 752, 101

\bibitem[{{Fransson} {et~al.}(2014){Fransson}, {Ergon}, {Challis}, {Chevalier},
  {France}, {Kirshner}, {Marion}, {Milisavljevic}, {Smith}, {Bufano},
  {Friedman}, {Kangas}, {Larsson}, {Mattila}, {Benetti}, {Chornock}, {Czekala},
  {Soderberg}, \& {Sollerman}}]{Fransson2014}
{Fransson}, C., {Ergon}, M., {Challis}, P.~J., {et~al.} 2014, \apj, 797, 118

\bibitem[{Gerardy {et~al.}(2003)}]{gerardy03du04}
Gerardy, C., {et~al.} 2003, 607, 391

\bibitem[{{Gerardy} {et~al.}(2004){Gerardy}, {Hoeflich}, {Fesen}, {Marion},
  {Nomoto}, {Quimby}, {Schaefer}, {Wang}, \& {Wheeler}}]{Gerardy}
{Gerardy}, C.~L., {Hoeflich}, P., {Fesen}, R.~A., {et~al.} 2004, \apj, 607, 391

\bibitem[{{Goobar}(2008)}]{goobar2008}
{Goobar}, A. 2008, \apjl, 686, L103

\bibitem[{{Graham} {et~al.}(2015{\natexlab{a}}){Graham}, {Sand}, {Zaritsky}, \&
  {Pritchet}}]{Graham_Sand2015}
{Graham}, M.~L., {Sand}, D.~J., {Zaritsky}, D., \& {Pritchet}, C.~J.
  2015{\natexlab{a}}, ArXiv e-prints

\bibitem[{{Graham} {et~al.}(2015{\natexlab{b}}){Graham}, {Valenti}, {Fulton},
  {Weiss}, {Shen}, {Kelly}, {Zheng}, {Filippenko}, {Marcy}, {Howell}, {Burt},
  \& {Rivera}}]{Graham}
{Graham}, M.~L., {Valenti}, S., {Fulton}, B.~J., {et~al.} 2015{\natexlab{b}},
  \apj, 801, 136

\bibitem[{{Greiner} {et~al.}(1991){Greiner}, {Hasinger}, \&
  {Kahabka}}]{greiner91}
{Greiner}, J., {Hasinger}, G., \& {Kahabka}, P. 1991, \aap, 246, L17

\bibitem[{{Griv} {et~al.}(2009){Griv}, {Gedalin}, \& {Eichler}}]{Griv2009}
{Griv}, E., {Gedalin}, M., \& {Eichler}, D. 2009, \aj, 137, 3520

\bibitem[{{Hachisu} {et~al.}(1996){Hachisu}, {Kato}, \& {Nomoto}}]{Hachisu96}
{Hachisu}, I., {Kato}, M., \& {Nomoto}, K. 1996, \apjl, 470, L97

\bibitem[{{Hachisu} {et~al.}(1999){Hachisu}, {Kato}, \& {Nomoto}}]{Hachisu99}
---. 1999, \apj, 522, 487

\bibitem[{{Hachisu} {et~al.}(2008){Hachisu}, {Kato}, \& {Nomoto}}]{Hachisu08}
---. 2008, \apj, 679, 1390

\bibitem[{{Hachisu} {et~al.}(2010){Hachisu}, {Kato}, \& {Nomoto}}]{hachisu10}
{Hachisu}, I., {Kato}, M., \& {Nomoto}, K.-i. 2010, in Progenitors and
  Environments of Stellar Explosions, 56

\bibitem[{{Hachisu} {et~al.}(2012){Hachisu}, {Kato}, {Saio}, \&
  {Nomoto}}]{Hachisu12}
{Hachisu}, I., {Kato}, M., {Saio}, H., \& {Nomoto}, K. 2012, \apj, 744, 69

\bibitem[{{Hamuy} {et~al.}(2000){Hamuy}, {Trager}, {Pinto}, {Phillips},
  {Schommer}, {Ivanov}, \& {Suntzeff}}]{Hamy00}
{Hamuy}, M., {Trager}, S.~C., {Pinto}, P.~A., {et~al.} 2000, \aj, 120, 1479

\bibitem[{Han \& Webbink(1999)}]{hanweb99}
Han, Z., \& Webbink, R. 1999, 349, L17

\bibitem[{{Hatano} {et~al.}(2000){Hatano}, {Branch}, {Lentz}, {Baron},
  {Filippenko}, \& {Garnavich}}]{hatano99}
{Hatano}, K., {Branch}, D., {Lentz}, E.~J., {et~al.} 2000, ApJl, 543, L49

\bibitem[{{Hoeflich}(2006)}]{hoeflich06}
{Hoeflich}, P. 2006, Nuclear Physics A, 777, 579

\bibitem[{{Hoeflich} {et~al.}(2013){Hoeflich}, {Dragulin}, {Mitchell},
  {Penney}, {Sadler}, {Diamond}, \& {Gerardy}}]{hoeflich2013}
{Hoeflich}, P., {Dragulin}, P., {Mitchell}, J., {et~al.} 2013, Frontiers of
  Physics, 8, 144

\bibitem[{{Hoeflich} \& {Khokhlov}(1996)}]{hk96}
{Hoeflich}, P., \& {Khokhlov}, A. 1996, \apj, 457, 500

\bibitem[{{H{\"o}flich} {et~al.}(1997){H{\"o}flich}, {Khokhlov}, \&
  {M\"uller}}]{het97plp}
{H{\"o}flich}, P.~A., {Khokhlov}, A., \& {M\"uller}, E. 1997, in NATO ASIC
  Proc. 486: Thermonuclear Supernovae, ed. P.~{Ruiz-Lapuente}, R.~{Canal}, \&
  J.~{Isern}, 659

\bibitem[{{Hoyle} \& {Fowler}(1960)}]{hf60}
{Hoyle}, F., \& {Fowler}, W.~A. 1960, \apj, 132, 565

\bibitem[{{Isern} {et~al.}(2014){Isern}, {Knoedlseder}, {Jean}, {Lebrun},
  {Renaud}, {Bravo}, {Churazov}, {Grebenev}, {Sunyaev}, {Soldi}, {Domingo},
  {Kuulkers}, {Hoeflich}, {Elias-Rosa}, {Hartmann}, {Hernanz}, {Badenes},
  {Dominguez}, {Garcia-Senz}, {Jordi}, {Lichti}, {Vedrenne}, \& {Von
  Ballmoos}}]{Isern2015}
{Isern}, J., {Knoedlseder}, J., {Jean}, P., {et~al.} 2014, The Astronomer's
  Telegram, 6099, 1

\bibitem[{{Judge} \& {Stencel}(1991)}]{Judge91}
{Judge}, P.~G., \& {Stencel}, R.~E. 1991, \apj, 371, 357

\bibitem[{{Kafka} \& {Honeycutt}(2004)}]{Kafka}
{Kafka}, S., \& {Honeycutt}, R.~K. 2004, \aj, 128, 2420

\bibitem[{{Kahabka} \& {van den Heuvel}(1997)}]{kah97}
{Kahabka}, P., \& {van den Heuvel}, E.~P.~J. 1997, \araa, 35, 69

\bibitem[{{Kasen}(2010)}]{kasen10}
{Kasen}, D. 2010, \apj, 708, 1025

\bibitem[{{Kashi} \& {Soker}(2011)}]{Soker}
{Kashi}, A., \& {Soker}, N. 2011, \mnras, 417, 1466

\bibitem[{{Kawara} {et~al.}(2011){Kawara}, {Hirashita}, {Nozawa}, {Kozasa},
  {Oyabu}, {Matsuoka}, {Shimizu}, {Sameshima}, \& {Ienaka}}]{Kawara2014}
{Kawara}, K., {Hirashita}, H., {Nozawa}, T., {et~al.} 2011, \mnras, 412, 1070

\bibitem[{{Kelly} {et~al.}(2014){Kelly}, {Fox}, {Filippenko}, {Cenko}, {Prato},
  {Schaefer}, {Shen}, {Zheng}, {Graham}, \& {Tucker}}]{Kelly2014}
{Kelly}, P.~L., {Fox}, O.~D., {Filippenko}, A.~V., {et~al.} 2014, \apj, 790, 3

\bibitem[{{Kirshner} \& {Oke}(1975)}]{kirshner75}
{Kirshner}, R.~P., \& {Oke}, J.~B. 1975, \apj, 200, 574

\bibitem[{{Krisciunas} {et~al.}(2000){Krisciunas}, {Hastings}, {Loomis},
  {McMillan}, {Rest}, {Riess}, \& {Stubbs}}]{kevin00}
{Krisciunas}, K., {Hastings}, N.~C., {Loomis}, K., {et~al.} 2000, \apj, 539,
  658

\bibitem[{{Krisciunas} {et~al.}(2003){Krisciunas}, {Suntzeff}, {Candia},
  {Arenas}, {Espinoza}, {Gonzalez}, {Gonzalez}, {H{\o}flich}, {Landolt},
  {Phillips}, \& {Pizarro}}]{kevin2013}
{Krisciunas}, K., {Suntzeff}, N.~B., {Candia}, P., {et~al.} 2003, \aj, 125, 166

\bibitem[{{Kromer} {et~al.}(2010){Kromer}, {Sim}, {Fink}, {R{\"o}pke},
  {Seitenzahl}, \& {Hillebrandt}}]{Kromer2010}
{Kromer}, M., {Sim}, S.~A., {Fink}, M., {et~al.} 2010, \apj, 719, 1067

\bibitem[{{Kr{\"u}gel}(2015)}]{kruegel15}
{Kr{\"u}gel}, E. 2015, \aap, 574, A8

\bibitem[{{Landau} \& {Lifshitz}(1971)}]{landau71}
{Landau}, L.~D., \& {Lifshitz}, E.~M. 1971, {The classical theory of fields}

\bibitem[{{Livne}(1990)}]{livne1990}
{Livne}, E. 1990, \apjl, 354, L53

\bibitem[{{Mannucci} {et~al.}(2006){Mannucci}, {Della Valle}, \&
  {Panagia}}]{Mannucci2006}
{Mannucci}, F., {Della Valle}, M., \& {Panagia}, N. 2006, \mnras, 370, 773

\bibitem[{{Maoz} {et~al.}(2014){Maoz}, {Mannucci}, \& {Nelemans}}]{Maoz2014}
{Maoz}, D., {Mannucci}, F., \& {Nelemans}, G. 2014, \araa, 52, 107

\bibitem[{{Margutti} {et~al.}(2014){Margutti}, {Parrent}, {Kamble},
  {Soderberg}, {Foley}, {Milisavljevic}, {Drout}, \& {Kirshner}}]{Margutti}
{Margutti}, R., {Parrent}, J., {Kamble}, A., {et~al.} 2014, \apj, 790, 52

\bibitem[{{Margutti} {et~al.}(2012{\natexlab{a}}){Margutti}, {Soderberg},
  {Chomiuk}, {Chevalier}, {Hurley}, {Milisavljevic}, {Foley}, {Hughes},
  {Slane}, {Fransson}, {Moe}, {Barthelmy}, {Boynton}, {Briggs}, {Connaughton},
  {Costa}, {Cummings}, {Del Monte}, {Enos}, {Fellows}, {Feroci}, {Fukazawa},
  {Gehrels}, {Goldsten}, {Golovin}, {Hanabata}, {Harshman}, {Krimm}, {Litvak},
  {Makishima}, {Marisaldi}, {Mitrofanov}, {Murakami}, {Ohno}, {Palmer},
  {Sanin}, {Starr}, {Svinkin}, {Takahashi}, {Tashiro}, {Terada}, \&
  {Yamaoka}}]{xraystudy}
{Margutti}, R., {Soderberg}, A.~M., {Chomiuk}, L., {et~al.} 2012{\natexlab{a}},
  \apj, 751, 134

\bibitem[{{Margutti} {et~al.}(2012{\natexlab{b}}){Margutti}, {Soderberg},
  {Chomiuk}, {Chevalier}, {Hurley}, {Milisavljevic}, {Foley}, {Hughes},
  {Slane}, {Fransson}, {Moe}, {Barthelmy}, {Boynton}, {Briggs}, {Connaughton},
  {Costa}, {Cummings}, {Del Monte}, {Enos}, {Fellows}, {Feroci}, {Fukazawa},
  {Gehrels}, {Goldsten}, {Golovin}, {Hanabata}, {Harshman}, {Krimm}, {Litvak},
  {Makishima}, {Marisaldi}, {Mitrofanov}, {Murakami}, {Ohno}, {Palmer},
  {Sanin}, {Starr}, {Svinkin}, {Takahashi}, {Tashiro}, {Terada}, \&
  {Yamaoka}}]{Margutti11fe}
---. 2012{\natexlab{b}}, \apj, 751, 134

\bibitem[{{Marietta} {et~al.}(2000){Marietta}, {Burrows}, \&
  {Fryxell}}]{Marietta}
{Marietta}, E., {Burrows}, A., \& {Fryxell}, B. 2000, \apjs, 128, 615

\bibitem[{{Marion} {et~al.}(2015){Marion}, {Sand}, {Hsiao}, {Banerjee},
  {Valenti}, {Stritzinger}, {Vink{\'o}}, {Joshi}, {Venkataraman}, {Ashok},
  {Amanullah}, {Binzel}, {Bochanski}, {Bryngelson}, {Burns}, {Drozdov},
  {Fieber-Beyer}, {Graham}, {Howell}, {Johansson}, {Kirshner}, {Milne},
  {Parrent}, {Silverman}, {Vervack}, \& {Wheeler}}]{Marion2015}
{Marion}, G.~H., {Sand}, D.~J., {Hsiao}, E.~Y., {et~al.} 2015, \apj, 798, 39

\bibitem[{{Marsch}(2006)}]{March06}
{Marsch}, E. 2006, Living Reviews in Solar Physics, 3, 1

\bibitem[{{Moore} {et~al.}(2013){Moore}, {Townsley}, \& {Bildsten}}]{Moore2013}
{Moore}, K., {Townsley}, D.~M., \& {Bildsten}, L. 2013, \apj, 776, 97

\bibitem[{{Noci} {et~al.}(1997){Noci}, {Kohl}, {Antonucci}, {Tondello},
  {Huber}, {Fineschi}, {Gardner}, {Korendyke}, {Nicolosi}, {Romoli}, {Spadaro},
  {Maccari}, {Raymond}, {Siegmund}, {Benna}, {Ciaravella}, {Giordano},
  {Michels}, {Modigliani}, {Naletto}, {Panasyuk}, {Pernechele}, {Poletto},
  {Smith}, \& {Strachan}}]{noci97}
{Noci}, G., {Kohl}, J.~L., {Antonucci}, E., {et~al.} 1997, in ESA Special
  Publication, Vol. 404, Fifth SOHO Workshop: The Corona and Solar Wind Near
  Minimum Activity, ed. A.~{Wilson}, 75

\bibitem[{{Nomoto}(1982{\natexlab{a}})}]{n82}
{Nomoto}, K. 1982{\natexlab{a}}, ApJ, 257, 780

\bibitem[{{Nomoto}(1982{\natexlab{b}})}]{nomoto82}
---. 1982{\natexlab{b}}, ApJ, 253, 798

\bibitem[{{Nomoto} {et~al.}(2006){Nomoto}, {Saio}, {Kato}, \&
  {Hachisu}}]{nomoto06a}
{Nomoto}, K., {Saio}, H., {Kato}, M., \& {Hachisu}, I. 2006, ArXiv Astrophysics
  e-prints

\bibitem[{{Nomoto} {et~al.}(1984){Nomoto}, {Thielemann}, \&
  {Yokoi}}]{Thielemoto84}
{Nomoto}, K., {Thielemann}, F.-K., \& {Yokoi}, K. 1984, \apj, 286, 644

\bibitem[{{Nomoto} {et~al.}(2003){Nomoto}, {Uenishi}, {Kobayashi}, {Umeda},
  {Ohkubo}, {Hachisu}, \& {Kato}}]{Nomoto2003}
{Nomoto}, K., {Uenishi}, T., {Kobayashi}, C., {et~al.} 2003, in From Twilight
  to Highlight: The Physics of Supernovae, ed. W.~{Hillebrandt} \&
  B.~{Leibundgut}, 115

\bibitem[{{Nordin} {et~al.}(2011){Nordin}, {{\O}stman}, {Goobar}, {Balland},
  {Lampeitl}, {Nichol}, {Sako}, {Schneider}, {Smith}, {Sollerman}, \&
  {Wheeler}}]{nordin11}
{Nordin}, J., {{\O}stman}, L., {Goobar}, A., {et~al.} 2011, \apj, 734, 42

\bibitem[{{Nugent} {et~al.}(2011)}]{nug_11fe_11}
{Nugent}, P.~E., {et~al.} 2011, Nature, in press, astro-ph/1110.6201

\bibitem[{{Osterbrock} \& {Ferland}(2006)}]{Osterbrock}
{Osterbrock}, D.~E., \& {Ferland}, G.~J. 2006, {Astrophysics of gaseous nebulae
  and active galactic nuclei}

\bibitem[{{Parker}(1963)}]{Parker}
{Parker}, E.~N. 1963, Interplanetary dynamical processes.

\bibitem[{{Pastorello} {et~al.}(2011){Pastorello}, {Benetti}, {Bufano},
  {Kankare}, {Mattila}, {Turatto}, \& {Cupani}}]{pastorello11}
{Pastorello}, A., {Benetti}, S., {Bufano}, F., {et~al.} 2011, Astronomische
  Nachrichten, 332, 266

\bibitem[{{Patat} {et~al.}(2007{\natexlab{a}}){Patat}, {Chandra}, {Chevalier},
  {Justham}, {Podsiadlowski}, {Wolf}, {Gal-Yam}, {Pasquini}, {Crawford},
  {Mazzali}, {Pauldrach}, {Nomoto}, {Benetti}, {Cappellaro}, {Elias-Rosa},
  {Hillebrandt}, {Leonard}, {Pastorello}, {Renzini}, {Sabbadin}, {Simon}, \&
  {Turatto}}]{Patat07b}
{Patat}, F., {Chandra}, P., {Chevalier}, R., {et~al.} 2007{\natexlab{a}},
  Science, 317, 924

\bibitem[{{Patat} {et~al.}(2007{\natexlab{b}}){Patat}, {Benetti}, {Justham},
  {Mazzali}, {Pasquini}, {Cappellaro}, {Della Valle}, {-Podsiadlowski},
  {Turatto}, {Gal-Yam}, \& {Simon}}]{patat07}
{Patat}, F., {Benetti}, S., {Justham}, S., {et~al.} 2007{\natexlab{b}}, \aap,
  474, 931

\bibitem[{{Patat} {et~al.}(2013){Patat}, {Cordiner}, {Cox}, {Anderson},
  {Harutyunyan}, {Kotak}, {Palaversa}, {Stanishev}, {Tomasella}, {Benetti},
  {Goobar}, {Pastorello}, \& {Sollerman}}]{Patat11fe}
{Patat}, F., {Cordiner}, M.~A., {Cox}, N.~L.~J., {et~al.} 2013, \aap, 549, A62

\bibitem[{{Patat} {et~al.}(2015{\natexlab{a}}){Patat}, {Taubenberger}, {Cox},
  {Baade}, {Clocchiatti}, {H{\o}flich}, {Maund}, {Reilly}, {Spyromilio},
  {Wang}, {Wheeler}, \& {Zelaya}}]{patat15}
{Patat}, F., {Taubenberger}, S., {Cox}, N.~L.~J., {et~al.} 2015{\natexlab{a}},
  \aap, 577, A53

\bibitem[{{Patat} {et~al.}(2015{\natexlab{b}}){Patat}, {Taubenberger}, {Cox},
  {Baade}, {Clocchiatti}, {H{\"o}flich}, {Maund}, {Reilly}, {Spyromilio},
  {Wang}, {Wheeler}, \& {Zelaya}}]{patat2015}
---. 2015{\natexlab{b}}, \aap, 577, A53

\bibitem[{{P{\'e}rez-Torres} {et~al.}(2014){P{\'e}rez-Torres}, {Lundqvist},
  {Beswick}, {Bj{\o}rnsson}, {Muxlow}, {Paragi}, {Ryder}, {Alberdi},
  {Fransson}, {Marcaide}, {Mart{\'{\i}}-Vidal}, {Ros}, {Argo}, \&
  {Guirado}}]{Perez-Torres}
{P{\'e}rez-Torres}, M.~A., {Lundqvist}, P., {Beswick}, R.~J., {et~al.} 2014,
  \apj, 792, 38

\bibitem[{{Phillips} {et~al.}(2013){Phillips}, {Simon}, {Morrell}, {Burns},
  {Cox}, {Foley}, {Karakas}, {Patat}, {Sternberg}, {Williams}, {Gal-Yam},
  {Hsiao}, {Leonard}, {Persson}, {Stritzinger}, {Thompson}, {Campillay},
  {Contreras}, {Folatelli}, {Freedman}, {Hamuy}, {Roth}, {Shields}, {Suntzeff},
  {Chomiuk}, {Ivans}, {Madore}, {Penprase}, {Perley}, {Pignata}, {Preston}, \&
  {Soderberg}}]{Phillips2013}
{Phillips}, M.~M., {Simon}, J.~D., {Morrell}, N., {et~al.} 2013, \apj, 779, 38

\bibitem[{{Piersanti} {et~al.}(2003{\natexlab{a}}){Piersanti}, {Gagliardi},
  {Iben}, \& {Tornamb{\'e}}}]{piersanti2003}
{Piersanti}, L., {Gagliardi}, S., {Iben}, Jr., I., \& {Tornamb{\'e}}, A.
  2003{\natexlab{a}}, \apj, 598, 1229

\bibitem[{{Piersanti} {et~al.}(2003{\natexlab{b}}){Piersanti}, {Gagliardi},
  {Iben}, \& {Tornamb{\'e}}}]{Piersanti2004}
---. 2003{\natexlab{b}}, \apj, 598, 1229

\bibitem[{{Piersanti} {et~al.}(2009){Piersanti}, {Tornamb{\'e}}, {Straniero},
  \& {Dom{\`i}nguez}}]{PiersantiTornambe09}
{Piersanti}, L., {Tornamb{\'e}}, A., {Straniero}, O., \& {Dom{\`i}nguez}, I.
  2009, in American Institute of Physics Conference Series, Vol. 1111, American
  Institute of Physics Conference Series, ed. G.~{Giobbi}, A.~{Tornambe},
  G.~{Raimondo}, M.~{Limongi}, L.~A. {Antonelli}, N.~{Menci}, \& E.~{Brocato},
  259--266

\bibitem[{{Quimby} {et~al.}(2007{\natexlab{a}}){Quimby}, {Hoeflich}, \&
  {Wheeler}}]{q07}
{Quimby}, R., {Hoeflich}, P., \& {Wheeler}, J.~C. 2007{\natexlab{a}}, \apj,
  666, 1083

\bibitem[{{Quimby} {et~al.}(2007{\natexlab{b}}){Quimby}, {Hoeflich}, \&
  {Wheeler}}]{Quimby}
---. 2007{\natexlab{b}}, \apj, 666, 1083

\bibitem[{{Ramstedt} {et~al.}(2009){Ramstedt}, {Sch{\o}ier}, \&
  {Olofsson}}]{Ramstedt2009}
{Ramstedt}, S., {Sch{\o}ier}, F.~L., \& {Olofsson}, H. 2009, \aap, 499, 515

\bibitem[{{Rappaport} {et~al.}(1994){Rappaport}, {Chiang}, {Kallman}, \&
  {Malina}}]{rap94}
{Rappaport}, S., {Chiang}, E., {Kallman}, T., \& {Malina}, R. 1994, ApJ, 431,
  237

\bibitem[{{Raymond} {et~al.}(2013){Raymond}, {Ghavamian}, {Williams}, {Blair},
  {Borkowski}, {Gaetz}, \& {Sankrit}}]{raymond13}
{Raymond}, J.~C., {Ghavamian}, P., {Williams}, B.~J., {et~al.} 2013, \apj, 778,
  161

\bibitem[{{Readhead}(1994)}]{readhead94}
{Readhead}, A.~C.~S. 1994, \apj, 426, 51

\bibitem[{{Reimers}(1977)}]{reimers77a}
{Reimers}, D. 1977, \aap, 61, 217

\bibitem[{{Rest} {et~al.}(2005){Rest}, {Suntzeff}, {Olsen}, {Prieto}, {Smith},
  {Welch}, {Becker}, {Bergmann}, {Clocchiatti}, {Cook}, {Garg}, {Huber},
  {Miknaitis}, {Minniti}, {Nikolaev}, \& {Stubbs}}]{rest05}
{Rest}, A., {Suntzeff}, N.~B., {Olsen}, K., {et~al.} 2005, \nat, 438, 1132

\bibitem[{{Rest} {et~al.}(2008){Rest}, {Matheson}, {Blondin}, {Bergmann},
  {Welch}, {Suntzeff}, {Smith}, {Olsen}, {Prieto}, {Garg}, {Challis}, {Stubbs},
  {Hicken}, {Modjaz}, {Wood-Vasey}, {Zenteno}, {Damke}, {Newman}, {Huber},
  {Cook}, {Nikolaev}, {Becker}, {Miceli}, {Covarrubias}, {Morelli}, {Pignata},
  {Clocchiatti}, {Minniti}, \& {Foley}}]{rest08}
{Rest}, A., {Matheson}, T., {Blondin}, S., {et~al.} 2008, ApJ, 680, 1137

\bibitem[{{Ruiter} {et~al.}(2014){Ruiter}, {Belczynski}, {Sim}, {Seitenzahl},
  \& {Kwiatkowski}}]{Ruiter2014}
{Ruiter}, A.~J., {Belczynski}, K., {Sim}, S.~A., {Seitenzahl}, I.~R., \&
  {Kwiatkowski}, D. 2014, \mnras, 440, L101

\bibitem[{{Sadler}(2012)}]{sadler12}
{Sadler}, B. 2012, PhD thesis, Florida State University

\bibitem[{{Sako} {et~al.}(2008){Sako}, {Bassett}, {Becker}, {Cinabro},
  {DeJongh}, {Depoy}, {Dilday}, {Doi}, {Frieman}, {Garnavich}, {Hogan},
  {Holtzman}, {Jha}, {Kessler}, {Konishi}, {Lampeitl}, {Marriner}, {Miknaitis},
  {Nichol}, {Prieto}, {Riess}, {Richmond}, {Romani}, {Schneider}, {Smith},
  {SubbaRao}, {Takanashi}, {Tokita}, {van der Heyden}, {Yasuda}, {Zheng},
  {Barentine}, {Brewington}, {Choi}, {Dembicky}, {Harnavek}, {Ihara}, {Im},
  {Ketzeback}, {Kleinman}, {Krzesi{\'n}ski}, {Long}, {Malanushenko},
  {Malanushenko}, {McMillan}, {Morokuma}, {Nitta}, {Pan}, {Saurage}, \&
  {Snedden}}]{Sako08}
{Sako}, M., {Bassett}, B., {Becker}, A., {et~al.} 2008, \aj, 135, 348

\bibitem[{{Schaller} {et~al.}(1992){Schaller}, {Schaerer}, {Meynet}, \&
  {Maeder}}]{Schaller1992}
{Schaller}, G., {Schaerer}, D., {Meynet}, G., \& {Maeder}, A. 1992, \aaps, 96,
  269

\bibitem[{{Schlegel}(1995)}]{schlegel95}
{Schlegel}, E.~M. 1995, Reports on Progress in Physics, 58, 1375

\bibitem[{{Schlegel} \& {Petre}(1993)}]{Schlegel93}
{Schlegel}, E.~M., \& {Petre}, R. 1993, \apjl, 412, L29

\bibitem[{{Schmidt} {et~al.}(1994){Schmidt}, {Kirshner}, {Leibundgut}, {Wells},
  {Porter}, {Ruiz-Lapuente}, {Challis}, \& {Filippenko}}]{Schmidt}
{Schmidt}, B.~P., {Kirshner}, R.~P., {Leibundgut}, B., {et~al.} 1994, \apjl,
  434, L19

\bibitem[{{Sedov}(1959)}]{Sedov1959}
{Sedov}, L.~I. 1959, Similarity and Dimensional Methods in Mechanics

\bibitem[{{Shiga}(2007)}]{Shiga2007}
{Shiga}, D. 2007, New Scientist, 1

\bibitem[{{Silverman} {et~al.}(2015){Silverman}, {Vink{\'o}}, {Marion},
  {Wheeler}, {Barna}, {Szalai}, {Mulligan}, \& {Filippenko}}]{Silverman2015}
{Silverman}, J.~M., {Vink{\'o}}, J., {Marion}, G.~H., {et~al.} 2015, \mnras,
  451, 1973

\bibitem[{{Slavin} {et~al.}(2015){Slavin}, {Dwek}, \& {Jones}}]{slavin15}
{Slavin}, J.~D., {Dwek}, E., \& {Jones}, A.~P. 2015, \apj, 803, 7

\bibitem[{{Soker}(2015)}]{Soker14J}
{Soker}, N. 2015, \mnras, 450, 1333

\bibitem[{{Spitzer}(1968)}]{Spitzer1968}
{Spitzer}, Jr., L. 1968, Dynamics of Interstellar Matter and the Formation of
  Stars, ed. B.~M. {Middlehurst} \& L.~H. {Aller} (the University of Chicago
  Press), 1

\bibitem[{{Starrfield} {et~al.}(1985){Starrfield}, {Sparks}, \&
  {Truran}}]{starrfield85}
{Starrfield}, S., {Sparks}, W.~M., \& {Truran}, J.~W. 1985, \apj, 291, 136

\bibitem[{{Sternberg} {et~al.}(2011){Sternberg}, {Gal-Yam}, {Simon}, {Leonard},
  {Quimby}, {Phillips}, {Morrell}, {Thompson}, {Ivans}, {Marshall},
  {Filippenko}, {Marcy}, {Bloom}, {Patat}, {Foley}, {Yong}, {Penprase},
  {Beeler}, {Allende Prieto}, \& {Stringfellow}}]{Sternberg2011}
{Sternberg}, A., {Gal-Yam}, A., {Simon}, J.~D., {et~al.} 2011, Science, 333,
  856

\bibitem[{{Sugimoto} \& {Nomoto}(1980)}]{Sugimoto1980}
{Sugimoto}, D., \& {Nomoto}, K. 1980, \ssr, 25, 155

\bibitem[{{Telesco} {et~al.}(2015){Telesco}, {H{\"o}flich}, {Li},
  {{\'A}lvarez}, {Wright}, {Barnes}, {Fern{\'a}ndez}, {Hough}, {Levenson},
  {Mari{\~n}as}, {Packham}, {Pantin}, {Rebolo}, {Roche}, \&
  {Zhang}}]{Telesco2015}
{Telesco}, C.~M., {H{\"o}flich}, P., {Li}, D., {et~al.} 2015, \apj, 798, 93

\bibitem[{{Tsebrenko} \& {Soker}(2015)}]{SNIPS}
{Tsebrenko}, D., \& {Soker}, N. 2015, \mnras, 447, 2568

\bibitem[{{van den Heuvel} {et~al.}(1992){van den Heuvel}, {Bhattacharya},
  {Nomoto}, \& {Rappaport}}]{vah92}
{van den Heuvel}, E.~P.~J., {Bhattacharya}, D., {Nomoto}, K., \& {Rappaport},
  S.~A. 1992, \aap, 262, 97

\bibitem[{{Wang} {et~al.}(2009{\natexlab{a}}){Wang}, {Chen}, {Meng}, \&
  {Han}}]{WCMH09}
{Wang}, B., {Chen}, X., {Meng}, X., \& {Han}, Z. 2009{\natexlab{a}}, \apj, 701,
  1540

\bibitem[{{Wang} \& {Han}(2012)}]{Wang2012}
{Wang}, B., \& {Han}, Z. 2012, \nar, 56, 122

\bibitem[{{Wang} {et~al.}(2009{\natexlab{b}}){Wang}, {Meng}, {Chen}, \&
  {Han}}]{WMC09}
{Wang}, B., {Meng}, X., {Chen}, X., \& {Han}, Z. 2009{\natexlab{b}}, \mnras,
  395, 847

\bibitem[{{Wang}(2014)}]{Wang2014}
{Wang}, L. 2014, {Polarimetry of SN 2014J in M82 as a Probe of Its Dusty
  Environment}, HST Proposal

\bibitem[{{Wang} {et~al.}(1997){Wang}, {Hoeflich}, \& {Wheeler}}]{Wang}
{Wang}, L., {Hoeflich}, P., \& {Wheeler}, J.~C. 1997, \apjl, 483, L29

\bibitem[{{Wang} {et~al.}(2003){Wang}, {Baade}, {H{\o}flich}, {Khokhlov},
  {Wheeler}, {Kasen}, {Nugent}, {Perlmutter}, {Fransson}, \&
  {Lundqvist}}]{wang2003}
{Wang}, L., {Baade}, D., {H{\o}flich}, P., {et~al.} 2003, ApJ, 591, 1110

\bibitem[{{Wang} {et~al.}(2008){Wang}, {Li}, {Filippenko}, {Foley}, {Smith}, \&
  {Wang}}]{wang08}
{Wang}, X., {Li}, W., {Filippenko}, A.~V., {et~al.} 2008, \apj, 677, 1060

\bibitem[{{Weaver} {et~al.}(1977){Weaver}, {McCray}, {Castor}, {Shapiro}, \&
  {Moore}}]{Weaver}
{Weaver}, R., {McCray}, R., {Castor}, J., {Shapiro}, P., \& {Moore}, R. 1977,
  \apj, 218, 377

\bibitem[{{Whelan} \& {Iben}(1973)}]{WI73}
{Whelan}, J., \& {Iben}, I.~J. 1973, 186, 1007

\bibitem[{{Wood} {et~al.}(2002){Wood}, {M{\"u}ller}, {Zank}, \&
  {Linsky}}]{wood2002}
{Wood}, B.~E., {M{\"u}ller}, H.-R., {Zank}, G.~P., \& {Linsky}, J.~L. 2002,
  \apj, 574, 412

\bibitem[{{Woosley} \& {Kasen}(2011)}]{WK2011}
{Woosley}, S.~E., \& {Kasen}, D. 2011, \apj, 734, 38

\bibitem[{{Woosley} \& {Weaver}(1994)}]{woosley94}
{Woosley}, S.~E., \& {Weaver}, T.~A. 1994, ApJ, 423, 371

\bibitem[{{Yaron} {et~al.}(2005){Yaron}, {Prialnik}, {Shara}, \&
  {Kovetz}}]{Yaron05}
{Yaron}, O., {Prialnik}, D., {Shara}, M.~M., \& {Kovetz}, A. 2005, \apj, 623,
  398

\bibitem[{{Zhou} {et~al.}(2014){Zhou}, {Wang}, \& {Zhao}}]{Zhou2014}
{Zhou}, W.-H., {Wang}, B., \& {Zhao}, G. 2014, Research in Astronomy and
  Astrophysics, 14, 1146

\end{thebibliography}

\end{document}